\documentclass[12pt,preprint]{aastex}

\shorttitle{Acrminute CMB Anisotropies with Bolocam}
\shortauthors{Sayers et al.}

\begin{document}

\title{A Search for Cosmic Microwave Background Anisotropies
  on Arcminute Scales with Bolocam}

\author{J.~Sayers\altaffilmark{1,6}, S.~R.~Golwala\altaffilmark{1},
  P.~Rossinot\altaffilmark{1}, P.~A.~R.~Ade\altaffilmark{2},
  J.~E.~Aguirre\altaffilmark{3,4}, J.~J.~Bock\altaffilmark{5},
  S.~F.~Edgington\altaffilmark{1}, J.~Glenn\altaffilmark{4},
  A.~Goldin\altaffilmark{5}, D.~Haig\altaffilmark{2},
  A.~E.~Lange\altaffilmark{1}, G.~T.~Laurent\altaffilmark{4},  
  P.~D.~Mauskopf\altaffilmark{2}, and~H.~T.~Nguyen\altaffilmark{5}}
\altaffiltext{1}
  {Division of Physics, Mathematics, \& Astronomy,
  California Institute of Technology, 
  Mail Code 59-33, Pasadena, CA 91125}
\altaffiltext{2}
  {Physics and Astronomy, Cardiff University, 5 The Parade,
  P. O. Box 913, Cardiff CF24 3YB, Wales, UK}
\altaffiltext{3}
  {Jansky Fellow, National Radio Astronomy Observatory}
\altaffiltext{4}
  {Center for Astrophysics and Space Astronomy \& Department of 
  Astrophysical and Planetary Sciences, 
  University of Colorado, 389 UCB, Boulder, CO 80309}
\altaffiltext{5}
  {Jet Propulsion Laboratory, California Institute of Technology, 
  4800 Oak Grove Drive, Pasadena, CA 91109}
\altaffiltext{6}
  {jack@caltech.edu}

\begin{abstract}
  We have surveyed two science fields totaling one square degree with
  Bolocam at 2.1 mm to search for secondary CMB anisotropies caused by
  the Sunyaev-Zel'dovich effect (SZE).  The fields are in the Lynx and
  Subaru/XMM SDS1 fields.  Our survey is sensitive to angular scales
  with an effective angular multipole of $\ell_{eff} = 5700$ with
  FWHM$_{\ell} = 2800$ and has an angular resolution of 60 arcseconds
  FWHM.  Our data provide no evidence for anisotropy.  We are able to
  constrain the level of total astronomical anisotropy, modeled as a
  flat band power in $\mathcal{C}_\ell$, with frequentist 68\%, 90\%, and
  95\% CL upper limits of 590, 760, and 830 $\mu K_{CMB}^2$.  We
  statistically subtract the known contribution from primary CMB
  anisotropy, including cosmic variance, to obtain constraints on the
  SZE anisotropy contribution.  Now including flux calibration
  uncertainty, our frequentist 68\%, 90\% and 95\% CL upper limits on a
  flat band power in $\mathcal{C}_\ell$ are 690, 960, and 1000 $\mu
  K_{CMB}^2$.  When we instead employ the analytic spectrum suggested by
  \citet{komatsu02}, and account for the non-Gaussianity of the
  SZE anisotropy signal, we obtain upper limits on the average amplitude of
  their spectrum weighted by our transfer function of 790, 1060, and 1080
  $\mu K_{CMB}^2$.  We obtain a 90\% CL upper limit on $\sigma_8$, which
  normalizes the power spectrum of density fluctuations, of 1.57.  These
  are the first constraints on anisotropy and $\sigma_8$ from survey
  data at these angular scales at frequencies near 150~GHz.

\end{abstract}

\section{Introduction}

  \subsection{Background}


    The SZE\footnote{Throughout this paper SZE refers
    to the thermal SZE.}
    is the inverse Compton scattering of CMB photons
    with a distribution of hot electrons, causing a net increase in
    the energy of the photons \citep{sunyaev72}.
    Since the background CMB is redshifted along with the 
    SZE-induced distortion, the relative amplitude of the 
    distortion, $\Delta T_{CMB} / T_{CMB}$, is 
    independent of redshift.
    The distortion caused by the SZE is proportional
    to the Comptonization parameter $y$, which is 
    a measure of the integral of the electron thermal energy density
    along the line of sight and is given by
    \begin{displaymath}
      y  = 
      \frac{\sigma_T}{m_e c^2} \int dl \mbox{ } n_e k_B T_e,
    \end{displaymath}
    where $\sigma_T$ is the Thomson cross section,
    $m_e$ is the electron mass, $c$ is the speed of light,
    $k_B$ is Boltzmann's constant, $T_e$ is the temperature of 
    the electrons, and  
    $n_e$ is the number density of electrons.
    Since the scattering process conserves photon number, the 
    thermal spectrum of the CMB is distorted by the SZE;
    there is a negative temperature shift at low
    frequency and a positive temperature shift at high
    frequency.
    The cross-over point where there is no distortion
    of the CMB occurs at approximately 218~GHz.
    The temperature shift caused by the SZE, $\Delta T_{CMB}$, is
    \begin{displaymath}
      \frac{\Delta T_{CMB}}{T_{CMB}} = f(x) y,
    \end{displaymath}
    where
    \begin{displaymath}
      f(x) = x \frac{e^x+1}{e^x-1} - 4,
    \end{displaymath}
    and $x = h \nu / k_B T_{CMB}$, $h$ is Planck's constant, 
    $\nu$ is the frequency, and $T_{CMB} = 2.73$~K is the 
    temperature of the CMB.
    For reference, excellent reviews of the SZE and its relevance to
    cosmology are given by \citet{birkinshaw99} and 
    \citet{carlstrom02}.

  \subsection{Untargeted SZE Surveys}


    To date, there have been no detections of previously unknown
    clusters using the SZE.
    However, 
    unresolved objects in SZE surveys will produce anisotropies
    in the CMB that are expected to dominate the CMB power
    spectrum at small angular scales corresponding
    to angular multipoles above $\ell \simeq 2500$.
    The overall normalization of these SZE-induced CMB anisotropies
    is extremely sensitive to $\sigma_8$ and can
    be used to constrain the value of this cosmological 
    parameter \citep{komatsu02}.
    Several experiments have conducted SZE surveys
    that have produced tentative detections of the SZE-induced 
    anisotropies in the CMB.
    At 30~GHz CBI has measured an excess CMB power between
    $\ell = 2000$ and $\ell = 3500$ at a significance of 
    3.1$\sigma$ \citep{mason03}.
    Also at 30~GHz, BIMA/OVRO has measured
    a CMB anisotropy of $220^{+140}_{-120}$~$\mu$K$^2_{CMB}$
    at an angular multipole of $\ell = 5237$ \citep{dawson06}.
    ACBAR, at 150~GHz and $2000 < \ell < 3000$, has measured an
    excess power of $34 \pm 20$~$\mu$K$_{CMB}^2$ \citep{reichardt08}.
    A joint analysis of the CBI and ACBAR excesses
    shows that they are six times more likely to be caused
    by the SZE than primordial fluctuations \citep{reichardt08}.

    Additionally, these tentative anisotropy detections have  
    been used to constrain cosmological parameters;
    the CBI data are consistent with 
    $\sigma_8 \simeq 1$ \citep{bond05},
    and the BIMA/OVRO data measure 
    $\sigma_8 = 1.03^{+0.20}_{-0.29}$ \citep{dawson06}.
    \citet{reichardt08} combine various data sets to place 
    constraints on $\sigma_8$ via an excess contribution
    to anisotropy at high $\ell$, $\ell > 1950$.
    ACBAR and WMAP3,
    plus CBI and BIMA/OVRO data at high $\ell$ and lower
    frequency, combine to indicate
    $\sigma_8^{SZ} = 0.95^{+0.03}_{-0.04}$
    when the amplitude of the SZE contribution is not
    slaved to its contribution at low $\ell$.
    Since this result is inconsistent with other
    constraints on $\sigma_8$, including those from the
    lower $\ell$ portions of the CMB spectrum, 
    \citet{reichardt08} also consider a case in which
    the high-$\ell$ SZE contribution is slaved to to
    the contribution at lower $\ell$ via
    $\sigma_8^{SZ} = \sigma_8$.
    In this case the excess power
    is produced by point sources.
    Fitted to the CMBall data set, which excludes
    CBI high-$\ell$ and BIMA/OVRO data,
    this model results in $\sigma_8$ values consistent
    with other measurements,
    $\simeq 0.80-0.81 \pm 0.03-0.04$ depending on the
    assumptions and data sets included.
    No attempt is made to explain the CBI and BIMA/OVRO
    excesses. 
    Overall, the current results suggest two possibilities:
    there are point source contributions to all the 
    high-$\ell$ data (ACBAR, CBI, BIMA/OVRO) that have
    not been properly included;
    or the SZ contribution calculated from theory is
    underestimated.

    The survey presented here is the first such survey
    at 150~GHz and at $\ell \simeq 6000$. 
    As we shall explain, contributions from primary
    CMB anisotropies, SZE, radio, and submillimeter point
    sources are all expected to be comparable,
    each at a level of $\mathcal{C}_{\ell} \simeq
    50$~$\mu$K$^2_{CMB}$.


\section{Observations}

  \subsection{Instrument Description}

    Bolocam is a large format, 144 detector, millimeter-wave camera designed
    to be operated at the Caltech Submillimeter Observatory (CSO).
    For these observations the array was comprised of 115 optical
    and 6 dark detectors.
    Each detector is housed within its own integrating cavity,
    formed by a frontshort plate and a backshort plate \citep{glenn02}.
    Smooth-walled conical feedhorns separated by 0.7 (f/\#)$\lambda$, 
    a cold (4~K) 
    high-density polyethylene (HDPE) lens,
    and a room-temperature ellipsoidal mirror are used to
    couple the detectors to the CSO optics.
    Each feedhorn terminates into a cylindrical waveguide,
    which defines the low-frequency cutoff of the system;
    the final filter in 
    a series of six cold metal-mesh filters determines the
    high-frequency cutoff.
    The resulting passband is centered at 143~GHz, and has
    an effective width of 21~GHz.
    A cold (4~K) Lyot stop is used to define
    the illumination of the 10.4~m primary mirror, 
    and the resulting far-field beams
    have FWHMs of 60~arcseconds.
    Bolocam can also observe at 270~GHz, and has been
    used in this mode for several types of observations,
    including surveys for submillimeter
    galaxies and protostellar cores \citep{laurent05, enoch06, young06}.
    
    The detector array has a hexagonal geometry, and utilizes
    silicon nitride micromesh (spider-web) bolometers 
    \citep{mauskopf97} which
    are cooled to 260~mK using a three-stage $^4$He/$^3$He/$^3$He
    sorption refrigerator \citep{bhatia00, bhatia02}.
    JFETs located near the array and operated at 140~K are used
    to buffer the high-impedance bolometer signals from sources
    of current noise.
    In order to avoid the $1/f$ noise from the JFETs, the bolometers
    are biased at 130~Hz and read out using room-temperature
    lockin amplifiers.
    More details of the Bolocam instrument can be found 
    in \citet{golwala08}, \citet{glenn98}, \citet{glenn03}, 
    and \citet{haig04}.

  \subsection{Observing Strategy}

    The data described in this paper was collected during a forty
    night observing run in late 2003.
    During the first half of each night we observed a 
    0.5~deg$^2$ region centered at 02h18m00s, \mbox{-5d00m00s} (J2000),
    which coincides with the Subaru/XMM Deep Survey
    (SXDS or SDS1); and 
    during the second half of each night we observed
    a 0.5~deg$^2$ region centered on the Lynx field
    at 08h49m12s, +44d50m24s (J2000).
    These fields were selected because they have extremely
    low dust emission and a large amount of optical/X-ray
    data that could be used to follow up any
    SZE cluster candidates found in the maps.

    Two nights at the start of the run were used to analyze
    different scan strategies for mapping the science fields.
    The maps were made by repeatedly raster scanning across the field, 
    stepping perpendicular to the scan, then
    rastering across the field in the opposite direction
    until the entire field has been covered.
    Our studies showed that the time-stream noise
    is independent of the angle of the raster scan and the
    turnaround time between scans, so we chose to scan
    parallel to RA or dec and turnaround as quickly
    as the telescope would allow ($\simeq 10$~seconds).
    Additionally, we found that our sensitivity to 
    astronomical signals is maximized when we raster scan
    at a speed of 240~arcseconds/second.\footnote{
      Faster speeds were not attempted due to fears
      that the CSO would not function properly and/or
      would be damaged.}
    At this speed it takes approximately 12.5 seconds to
    complete one scan across the field, which
    means we were on-source approximately 56\% of the time
    during an observation.
    Although scanning at this relatively quick speed
    reduces our time on-source
    because a larger fraction of time is spent on turnarounds
    between scans, it also puts a larger amount of
    our signal band above the $1/f$ atmospheric noise.
    Given the scan speed and turnaround time mentioned above,
    along with our step size of 162~arcseconds 
    ($\simeq 1/3$ of the field of view) 
    a complete map of the field was made in
    approximately eight minutes.

\section{Data Reduction}

  \subsection{Initial Processing}
  \label{sec:merge}

    After merging the bolometer time-streams recorded by
    the data acquisition system with the pointing information
    recorded by the telescope, 
    we parse the data into files that
    contain a single observation.
    Each single observation contains a set of scans that
    completely map the astronomical field or object, and they
    are typically around ten minutes in length.
    Parsing the data by observation is useful because individual
    observations are statistically independent, have a small
    enough number of data samples to be easily manageable
    from an analysis standpoint, and provide a convenient
    division of the data for the sake of bookkeeping.

\begin{figure}
  \plotone{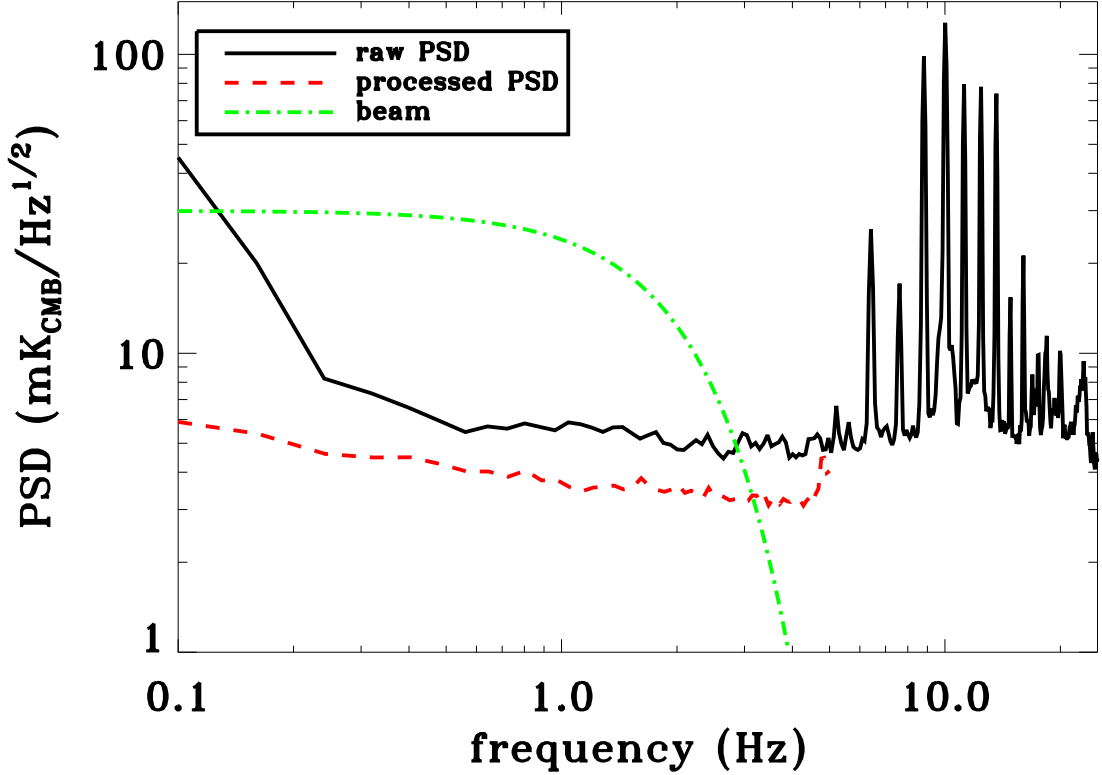}
  \caption{The solid black line represents
    a pre-down-sampled time-stream PSD,
    which has 60~Hz pickup at
    frequencies above $\simeq 10$~Hz.
    The dashed red line shows the time-stream PSD
    after downsampling and processing, including
    removal of an atmospheric noise template.  
    The sharp increase in this PSD near 5~Hz is
    caused by the small amount of noise that is
    aliased in from frequencies just above 5~Hz
    during the downsampling process.
    Since there is approximately no astronomical
    signal at the frequencies 
    where this noise increase occurs, this 
    noise does not have a noticeable effect on 
    our sensitivity.
    Overlaid as a dot-dashed green line is the beam profile, 
    showing that very little astronomical
    signal will be present above a few Hz.}
  \label{fig:pickup_60hz}
\end{figure}

    Once the initial merging and parsing of the data is complete,
    we begin the process of refining the data.
    The first step in this process is to remove the effects
    of the lockin amplifier electronics filters and to
    down-sample the data from 50~Hz to 10~Hz.
    We down-sample the data because essentially no
    astronomical signal is lost, while a large amount
    of 60-Hz pickup noise is removed.
    See Figure~\ref{fig:pickup_60hz} for an illustration
    of the noise spectrum and the shape of the
    beam in frequency space.

  \subsection{Noise Removal}

    There are several forms of correlated noise present in the 
    raw bolometer data which contaminate the astronomical signal
    and therefore must be modeled and removed.
    First, the emission from the atmosphere changes as a 
    function of telescope elevation angle due to the
    changing path length through the atmosphere.
    The path length through the atmosphere relative to the
    zenith path length is called the airmass, $A$, and is described by
    \begin{displaymath}
      A = 1 / \sin(\epsilon),
    \end{displaymath}
    where $\epsilon$ is the elevation angle.
    For a typical observation the range of elevation angles is
    approximately one degree, which corresponds to a 
    change in airmass between 0.005 and 0.060 for elevation
    angles between 75 and 30 degrees.
    For reference, a change of 0.060 in airmass corresponds to
    a change of approximately 0.5~K of optical loading
    from the atmosphere, or a change in surface brightness
    of a little less than 1~K$_{CMB}$.
    To remove this elevation-dependent signal, we calculate
    a linear fit of bolometer signal versus airmass.
    We build up a 
    fit using each 12.5-second-long scan within the observation, 
    after removing the mean signal level and airmass for the scan.
    This process yields one set of linear fit coefficients for 
    each bolometer for the entire observation,
    which is used to create a template that is removed from
    the bolometer time-streams.

    Next, we create a template from the 
    bias voltage monitors to account for the small amount of noise from the
    bias electronics.
    Note that the bias applied to the bolometers is monitored through amplifier
    electronics identical in design to those used to monitor the
    bolometer signals.
    This template is then correlated and removed
    from each of the the bolometer time-streams.
    A template is also created from the dark bolometer signals
    and removed from the bolometer time-streams.
    Note that both the bias template and dark bolometer template
    have an RMS of $\lesssim 1$~mK$_{CMB}$.

    Finally, and most importantly, we remove a template describing
    the fluctuations in emission from the atmosphere
    (\emph{i.e.}, the atmospheric noise).
    Since atmospheric noise is the dominant signal in our data,
    and the beams from the individual detectors overlap to 
    a high degree while passing through the atmosphere,
    a template for the atmospheric signal is created 
    by averaging the signals from all the bolometers.\footnote{
    To quantify the degree of overlap,
    note that the far-field distance for Bolocam at the CSO is
    approximately 30~km; at a height of 20~km,
    well above most of the water vapor in the atmosphere,
    the ray bundles from nearby detectors have only
    separated by one half-width.}
    Three different algorithms are used to construct this
    template, one for which the atmospheric signal is assumed to
    be constant over the array, one for which the atmospheric
    signal is allowed to vary linearly with bolometer
    location on the array, and one for which the atmospheric
    signal is allowed to vary quadratically with
    bolometer location on the array.
    
    For the most basic case of an average template, the
    algorithm proceeds as follows.
    Initially, a template is constructed according to
    \begin{equation}
      T_n = \frac{\sum_{i=1}^{i=N_{b}}
      c_i d_{in}}{\sum_{i=1}^{i=N_{b}} c_i}
      \label{eqn:avg_template}
    \end{equation}
    where $n$ is the sample number, $N_{b}$ is the number
    of bolometers, $c_i$ is the relative responsivity
    of bolometer $i$, $d_{in}$ is the signal recorded by
    bolometer $i$ at time sample number $n$, and $T_n$ is the 
    template.
    The template generally has
    an RMS between 10 and 100~mK$_{CMB}$, depending on
    the observing conditions.
    A separate template is computed for each 12.5-second-long scan.
    After the template is computed, it is correlated with
    the signal from each bolometer to determine the 
    correlation coefficient, with
    \begin{equation}
      \tilde{c_i} = \frac{\sum_{j=1}^{j=N_{s}}
      T_n d_{in}}{\sum_{j=1}^{j=N_{s}} T_n^2}.
      \label{eqn:skysub_corr}
    \end{equation}
    $\tilde{c_i}$ is the correlation coefficient
    of bolometer $i$ and $N_s$ is the number of samples
    in the 12.5-second-long scan.\footnote{
      The best fit correlation coefficients change
      from one scan to the next, typically by a 
      couple percent.}
    Next, the $c_i$ in Equation~\ref{eqn:avg_template} are
    set equal to the values of $\tilde{c_i}$ found from
    Equation~\ref{eqn:skysub_corr}, and a new template is
    computed.
    The process is repeated until the values of $c_i$ 
    stabilize.
    We generally iterate until the average fractional change
    in the $c_i$s is less than $1 \times 10^{-8}$, which
    usually takes five to ten iterations.
    If the $c_i$s fail to converge after 100 iterations,
    then the scan is discarded from the data.
    For the more advanced planar and quadratic algorithms,
    the process proceeds in the same way except linear
    and quadratic variations with bolometer position are
    allowed when the template is constructed.
    These algorithms, along with adaptive PCA
    and time-lagged average template subtraction, 
    are described and compared in more
    detail in \citet{sayers08}. 

    Each of the three different atmospheric noise
    removal algorithms,
    average, planar, and quadratic template
    removal,  was applied to each observation.
    Therefore, three different atmospheric-noise-cleaned
    time-streams are generated for
    each observation.
    A figure of merit is calculated for each of the three
    files for each observation, based on the noise level of the
    data and the expected astronomical signal shape.
    Details of the calculation of this figure of merit
    are given in Section~\ref{sec:opt_skysub}.
    For each observation, the file with the best figure of
    merit value will be the one used to create the final map
    of the data.
    Weather is the main criteria that determines which algorithm will
    be selected as optimal for a given observation;
    more aggressive algorithms (planar or quadratic) are selected
    in poor weather conditions and more benign algorithms (average or planar)
    are selected in good weather conditions.
    However, there is some dependence on the profile of the source,
    and observations of compact objects tend to be
    optimally processed using more aggressive algorithms
    than observations of extended objects.

\section{Calibration}

  \subsection{Pointing reconstruction}

    Pointing reconstruction consists of determining the
    location of each detector's beam on the sky at
    each instant in time.
    We compute this location in two steps:
    1) we calculate the location of each bolometer relative
    to the center of the array and  
    2) we then determine the absolute coordinates of the 
    center of the array.

\begin{figure}
  \plotone{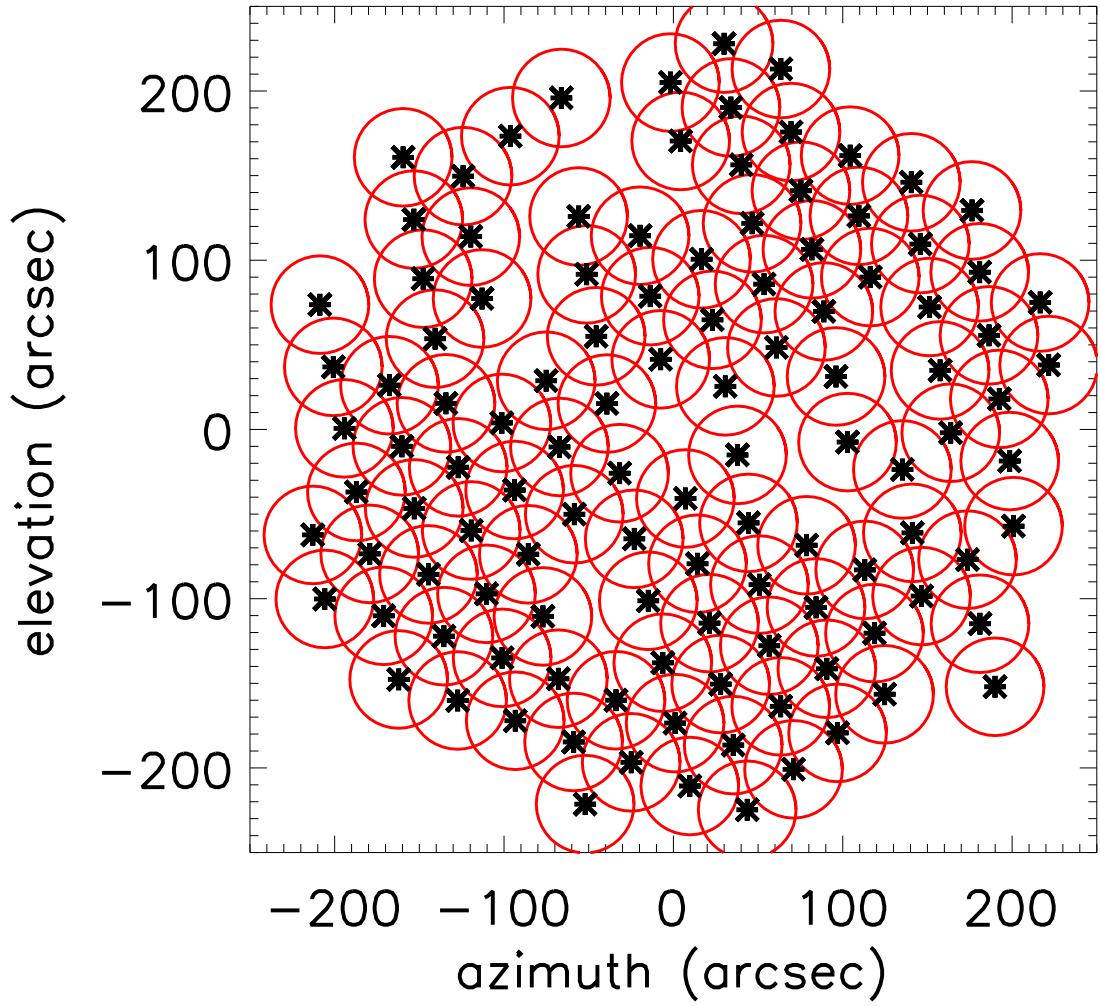}
  \caption{Location of the beam center of every detector
    relative to the center of the array.
    The red rings around each beam center represent the 
    approximate FWHM of the beam.}
  \label{fig:beam_locations}
\end{figure}

    To determine the relative locations of the bolometers, we 
    observed Uranus or Neptune for approximately fifteen minutes
    every other night.
    These planets are bright enough to appear at high
    signal-to-noise in a map made from a single bolometer, 
    so they can be used to determine the position of each detector
    relative to the array center.
    Since Bolocam was held at a fixed angle in the alt/az coordinate system
    for the entire observing run, each bolometer
    views the optics in the same way for the entire run
    and the coordinates on the sky in alt/az units remain fixed.
    Therefore, we combined the data from all the planet observations 
    to determine
    the average position of each beam on the sky. 
    See Figure~\ref{fig:beam_locations}.
    The uncertainties on these average positions were
    $\sim 1$~arcsecond,
    which is negligible when compared to
    the 60~arcsecond FWHM
    of a Bolocam beam.
    We found no evidence for a systematic difference in the 
    beam positions derived from any single observation to
    the average beam position found from all the observations.
    This indicates that the optical system was very stable
    over the entire observing run, including a wide range of
    telescope elevation angles.

    To determine the absolute location of the center of the
    array,  we observed a bright
    quasar with a known position
    near the science field for approximately ten minutes once every 
    two hours.
    Three different quasars were used for the SDS1 field
    (0106+013, 0113-118, and 0336-019), and two different
    quasars were used for the Lynx field
    (0804+499 and 0923+392).
    Each source was observed for five minutes while scanning parallel to
    RA, then for five minutes scanning parallel to dec (analogous to how the
    science fields were observed).
    We found no systematic offset based on scan direction;
    the maps made while scanning parallel to RA produce the same 
    source location as the maps made while scanning parallel to dec.
    The difference in the centroid location for these consecutive
    observations was then used to determine the 
    measurement uncertainty for the centroided location of each source.
    As expected, the uncertainty in the centroided location
    of the five sources is inversely 
    proportional to the flux of the source.
    Additionally, we found no evidence that the measurement
    uncertainty degrades or improves as a function of time 
    during the night for our typical observing times between
    20:00 and 07:00 local time.

\begin{figure}
  \plottwo{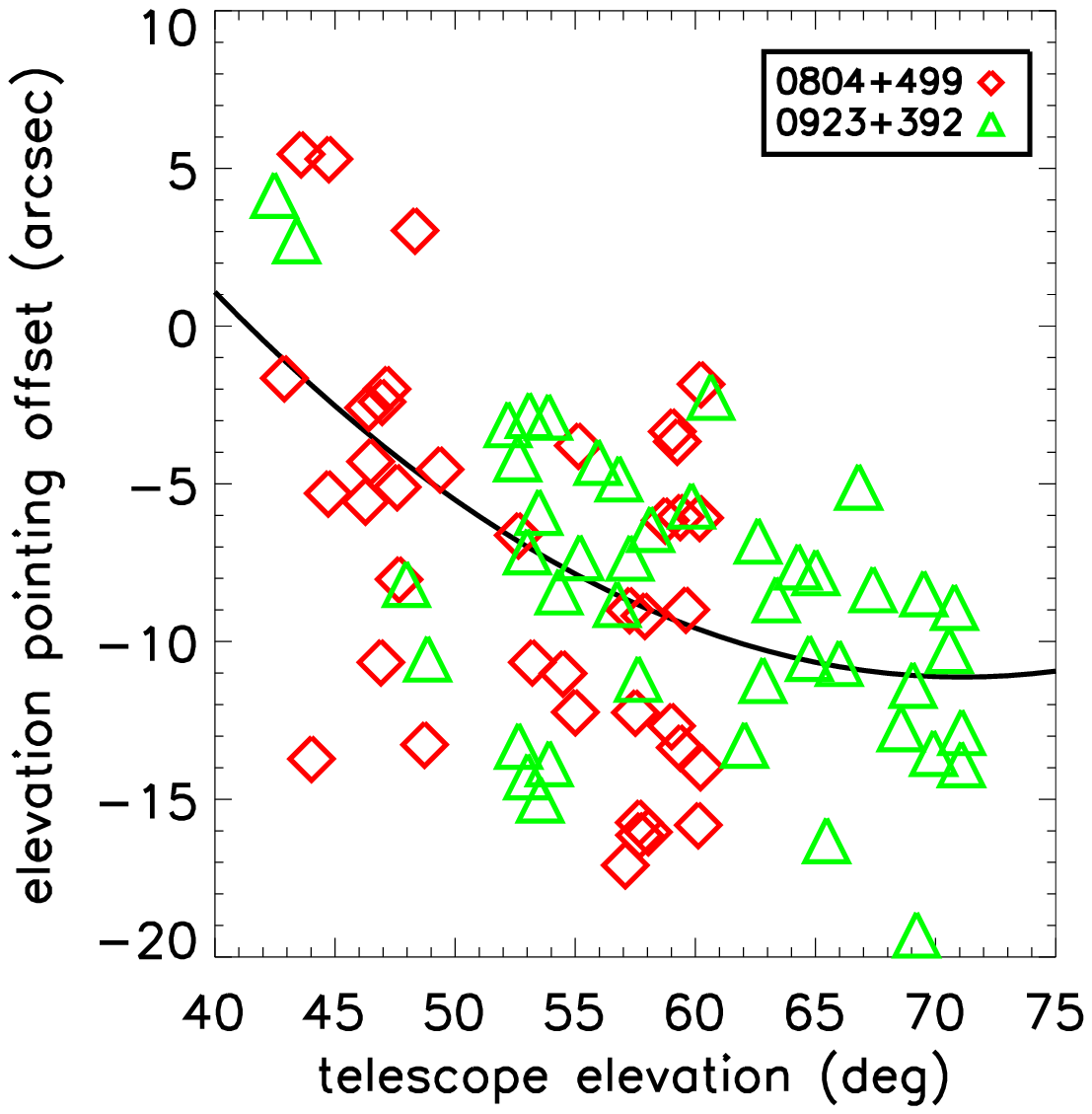}{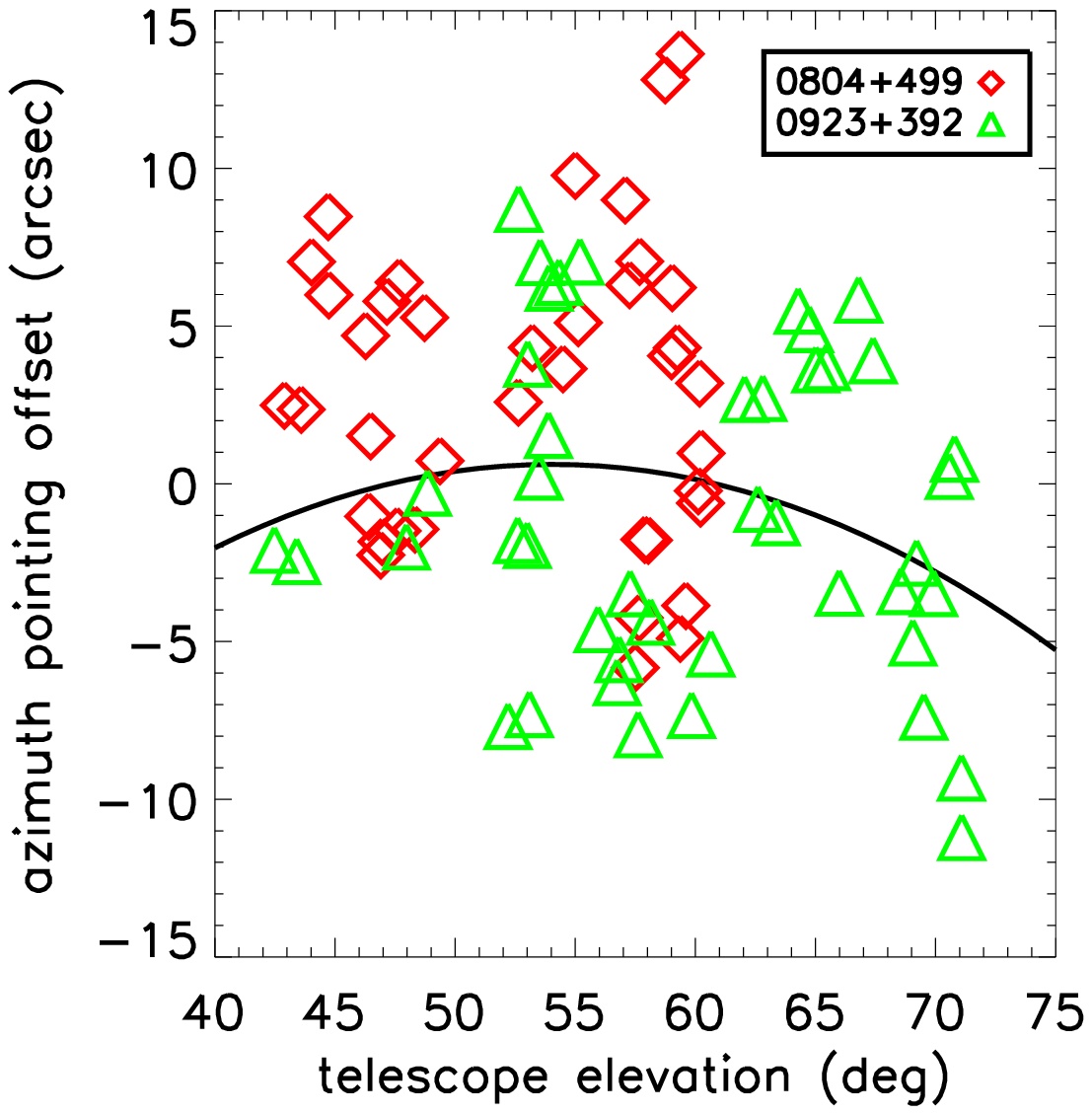}
  \caption{All of the
    raw pointing data for Lynx at azimuth angles between
    -90~and~90 degrees.  
    The pointing model (quadratic fit) is overlaid.
    Similar models were fit to the SDS1 pointing data and the
    Lynx data at azimuth angles between 270~and~360 degrees.}
  \label{fig:raw_pointing}
\end{figure}

    The pointing data were broken up into three distinct subsets
    corresponding to the azimuthal position of the telescope:
    SDS1 was observed between azimuth angles of 90 and 270 (in the south), 
    while Lynx was observed between azimuth angles 
    of -90 and 90 (in the north approaching from the east), and
    also between azimuth angles of 270 and 360
    (in the north approaching from the west).
    Most of the Lynx data were taken between an azimuth angle 
    of -90 to 90, 
    so the third subset of data is considerably smaller than the first
    two (about 1/5 the size).
    Note that the slewing limits of the telescope are roughly
    equal to azimuth angles of -90 to 360.
    There is a correlation between the elevation angle of the 
    telescope and the pointing offset for each of these subsets.
    We attempted to model 
    this correlation with several low-order polynomials, 
    but we found that a quadratic fit of pointing offset
    versus elevation was sufficient since higher-order fits
    did not significantly reduce the scatter of the data.
    We found no correlation between the telescope azimuth angle
    and the residual offset, other than the slight difference
    between the pointing models determined for the three subsets.
    Therefore, a simple quadratic fit of 
    pointing offsets versus telescope elevation angle
    served as our only pointing model. 
    Plots of this final model can be found in 
    Figure~\ref{fig:raw_pointing}.

    For each of the three subsets, we calculated the uncertainty
    in the pointing model by analyzing the residual offset
    of each centroid location from the model.
    Some residual scatter is expected due to the measurement
    uncertainty of each centroid,
    however the scatter we find is slightly larger.
    The difference between the actual scatter and the
    predicted scatter is consistent for all three subsets,
    and translates to an uncertainty in the pointing
    model of 4.9~arcseconds.
    This uncertainty is small compared to our beam size
    and thus made a negligible difference in the beam shape
    used in the final science analysis.

  \subsection{Flux Calibration}
    \label{sec:flux_cal}      

    Our flux calibration technique,
    summarized below,
    has been used previously with Bolocam
    to calibrate 1.1~mm data \citep{laurent05}.
    Since the amount of astronomical signal attenuation by the
    atmosphere is a function of opacity and airmass,
    the standard flux calibration technique for millimeter-wave
    instruments requires frequent observations of calibration
    sources that are close to the science field.
    However, we were able to use a more advanced technique
    with Bolocam because we continuously monitor the
    operating resistance of the bolometers using the 
    carrier amplitude measured by the bolometer voltage
    at the bias frequency.
    When the atmospheric transmission decreases, the optical
    loading from the atmosphere increases, which
    lowers the bolometer resistance.
    Additionally, the bolometer responsivity is a monotonically
    decreasing function of the bolometer resistance.
    Therefore, by fitting the flux calibration as a function
    of the bolometer operating resistance, we can simultaneously account
    for changes in the atmospheric transmission and 
    bolometer responsivity.

    Six different flux calibrations were needed for our data set.
    The base temperature 
    of the sub-Kelvin refrigerator
    was changed on November 4, 2003, and the
    bias voltage applied 
    to the bolometers was changed on November 5, 8 (twice),
    and 10, 2003.
    Each of the bias changes caused a change in 
    the responsivity of the bolometers, so a different flux calibration
    is needed after each change.
    Since the first five data sets are relatively short in duration,
    the observing conditions were relatively constant
    within each set.
    Therefore, a constant flux calibration, rather than a flux
    calibration that varies as a function of bolometer operating resistance,
    was adequate to describe the data for these five sets.
    However, a fit of the flux calibration as a function
    of bolometer operating resistance was required for the final data set.

    The relative calibration of the detectors was determined
    from the science field observations.
    Since these observations covered regions of the sky with negligible
    amounts of astronomical flux, 
    the fluctuations in thermal emission from the atmosphere are the
    dominant source of the signal recorded by each bolometer.
    Additionally, this signal should be the only one that is correlated
    among all the bolometers
    since the beams from 
    all bolometers overlap to a high degree when passing
    through the atmosphere.
    Therefore, this signal should be the same in each 
    bolometer, weighted by the responsivity of that bolometer. 
    So, by determining how correlated the data from each bolometer
    is with this common signal, it is possible to determine the relative 
    calibration of each bolometer.
    The uncertainties in the relative calibrations determined
    using this method are less than 1\%.

\begin{figure}
  \plotone{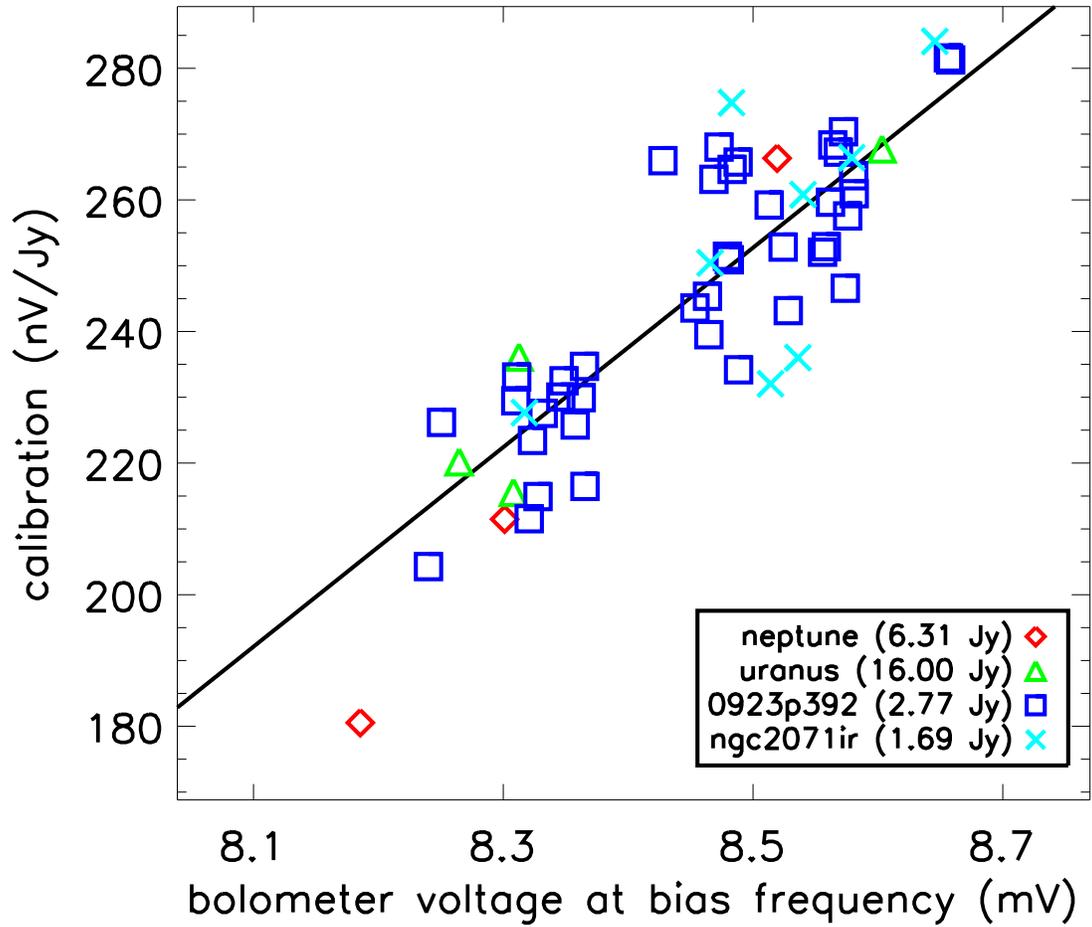}
  \caption{Flux calibration for one of the six calibration data
    sets, overlaid with
    a linear fit of calibration versus 
    bolometer operating resistance measured by the 
    bolometer voltage at the bias frequency.
    Note that the bolometer voltage at the bias frequency is 
    a monotonic function of the atmospheric opacity and
    the bolometer responsivity.}
  \label{fig:flux_slope}
\end{figure}

    The absolute flux calibration was determined from
    observations of Uranus, Neptune, 
    0923+392, and NGC2071IR.
    Since we did not have enough observations of Uranus and Neptune
    to adequately determine the shape of the calibration versus
    bolometer operating resistance, we used 0923+392 and NGC2071IR as
    secondary calibrators.
    These two sources are known to have minimal 
    variations in emitted flux as a function of time, so they are well
    suited to be used for determining the functional form
    of the flux calibration versus bolometer operating
    resistance relationship \citep{peng00,sandell94}.
    Note that we did not use any of the published fluxes
    for 0923+392 or NGC2071IR, rather the fluxes were left as
    free parameters and they were used to determine
    the shape of the calibration curve versus
    bolometer operating resistance.
    We used the peak signal and median bolometer operating resistance
    from each observation to determine the fit parameters in the 
    function
    \begin{displaymath}
      V_{j}(R_{bolo}) = F_{j} ( \alpha_1 + \alpha_2 R_{bolo} ), 
    \end{displaymath}
    where $V_{j}$ is the peak bolometer signal (in nV) recorded
    for the $j^{th}$ source, $R_{bolo}$ is the bolometer operating resistance,
    $F_j$ is 
    equal to the flux of the $j^{th}$ source (known for Uranus and
    Neptune, left as a free parameter for NGC2071IR and 0923+392), 
    and $\alpha_1$ and $\alpha_2$ free parameters. 
    The planet fluxes were determined from the
    temperature spectra 
    given in \citet{griffin93} or \citet{orton86},
    along with the planet solid angles
    calculated from the planet flux calculator at the
    James Clerk Maxwell Telescope website.\footnote{
      http://www.jach.hawaii.edu/jacbin/planetflux.pl.}
    For reference, 
    the absolute calibration ranges from approximately 180~nV/Jy up to
    280~nV/Jy over the range of 
    bolometer operating resistances recorded during 
    our observing run.
    See Figure~\ref{fig:flux_slope}.

    Our flux calibration uncertainty was determined as follows.
    First, the temperature profiles of Uranus and Neptune were derived
    by Griffin and Orton using Mars as an absolute
    calibrator \citep{griffin93}.
    To determine the surface brightness of Mars at millimeter
    wavelengths, Griffin and Orton used the model
    developed by Wright based on observations
    made at far-infrared wavelengths \citep{wright76},
    along with the logarithmic interpolation to 
    longer wavelengths described by \citet{griffin86}.
    The estimated uncertainty on this interpolated model is 
    approximately 5\% \citep{wright76}.\footnote{
      There is also a brightness model based on a
      physical model of the dielectric properties
      of the Martian surface that was developed 
      by Rudy \citep{rudy87, rudy87_2}.
      This model was constrained by measurements at 
      centimeter wavelengths, and also needs to be
      extrapolated to millimeter wavelengths.
      Griffin and Orton, along with Goldin, et al.,
      compared the results of these
      two models at millimeter wavelengths, and 
      found that they agree within their estimated
      uncertainties \citep{griffin93, goldin97}.
      Based on the comparison of these two models, 
      Griffin and Orton conclude the the 
      uncertainty in the Martian brightness based
      on the Wright model is 5\%.}
    Second, the uncertainties on the temperature profiles of Uranus and 
    Neptune are estimated to be less than 
    1.5\% relative to Mars \citep{griffin93}.\footnote{
      Griffin and Orton find that the uncertainty is 1.7~K
      for both their Uranus and Neptune models.
      Since the temperature of these planets in our band
      is approximately 115~K, this translates to an
      uncertainty of $\simeq 1.5$\%.}
    Additionally, the observations of 
    Uranus and Neptune were taken with a precipitable
    water vapor of $1.5 \pm 0.5$~mm, which results in a calibration
    uncertainty of $\sim 1.4$\%.
    Finally, the error inferred by the scatter of our measurements
    results in calibration uncertainties between 0.6\% and 3.0\% for
    each of the data sets.
    The end result is an overall flux calibration uncertainty
    of approximately 5.5\%, limited by the uncertainty
    in the temperature of Mars.

  \subsection{Beam Calibration}

    Since the astronomical signals in our maps are inherently
    smoothed based on the profile of the Bolocam
    beams, it is important to understand their shapes.
    Additionally, 
    our flux calibration is based on observations of point sources,
    so our maps have units of flux density.
    However, since the CMB or SZE signal we are looking for is a surface
    brightness or temperature, we need to know the area of our beam
    in solid angle to convert our maps to
    surface brightness units.
    Therefore, any error in our determination of the beam area will
    show up as a surface brightness or temperature calibration
    error.
    To determine the profile of our beam, we used the	
    observations of Uranus and Neptune.
    These planets are well suited for measuring our
    60~arcsecond FWHM beams;
    they have semi-diameters of $\simeq 1$~arcsecond,
    which means they are essentially point-like and thus
    will appear in our maps with shapes given
    by our beam profile.

\begin{figure}
  \plotone{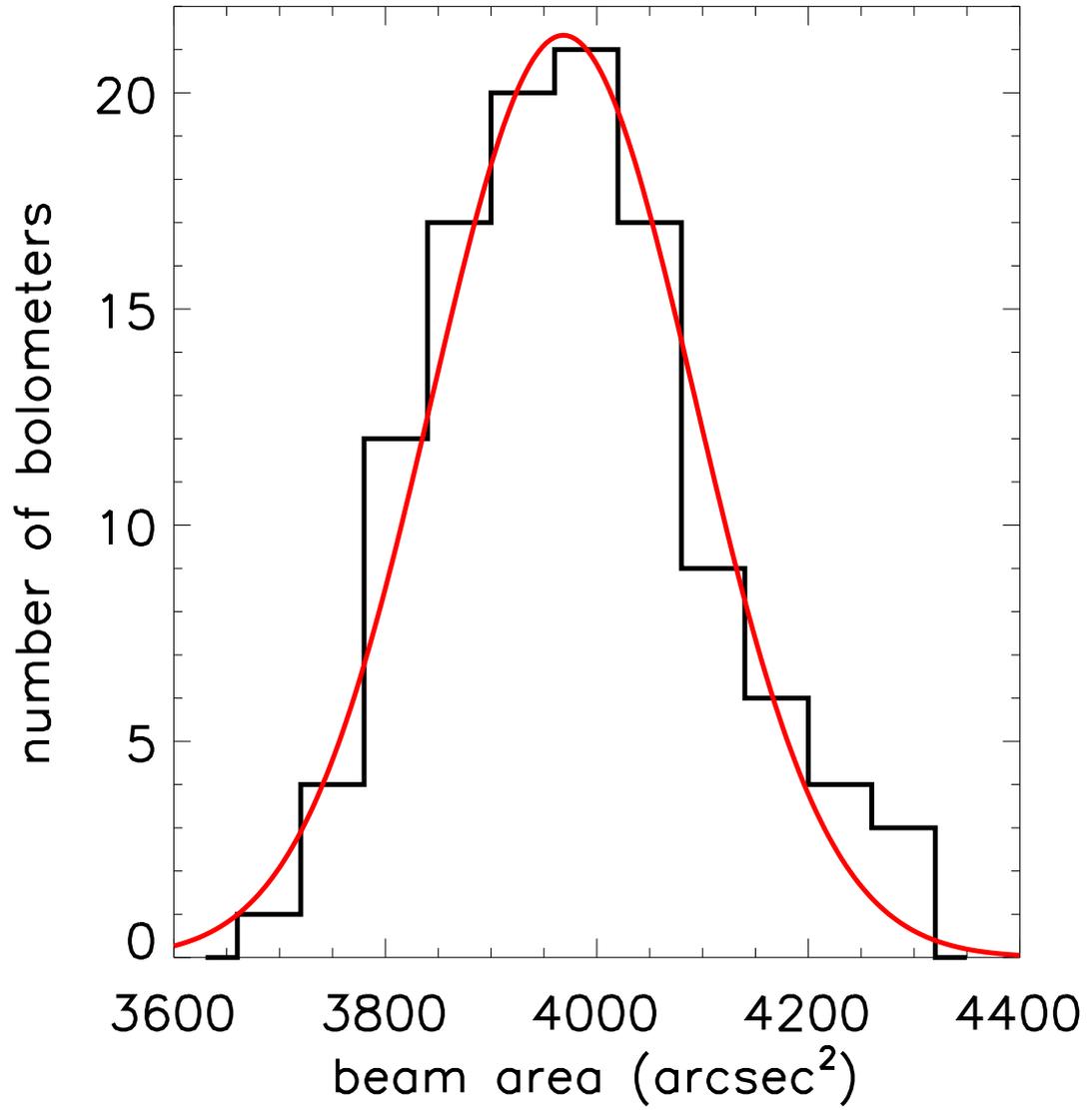}
  \caption{A histogram of 
    the beam area calculated 
    for each bolometer, with a Gaussian
    fit overlaid.}
  \label{fig:beam_area_variation}
\end{figure}

    Based on simulations, we expected all of the beams
    to have a similar profile.
    However, 
    we first calculated the beam for each bolometer 
    separately to validate this expectation.
    There was not enough data from a single planet observation
    to make a high signal-to-noise measurement of the beam
    for an individual bolometer, so we averaged the data
    for groups of four bolometers that are close to each
    other on the focal plane.
    Nearby bolometers have beams with similar paths through the
    optics, so they should also have similar profiles.
    Each bolometer was grouped into four distinct sets, each
    of which contained four nearby bolometers, and the 
    average profile from these four sets was determined.
    The measurement uncertainty on these profiles can be 
    quantified by the standard deviation
    of the peak-normalized
    areas of the beam profiles, which was approximately
    3.1\%.
    See Figure~\ref{fig:beam_area_variation}.
    Within our measurement uncertainty, all of the individual
    bolometer beam profiles were consistent, so a single
    beam profile can be used to describe every bolometer.
    To measure this single beam profile, we averaged
    the data from all of the planet observations
    for all of the bolometers.
    The peak-normalized area of this profile is 3970~arcseconds$^2$,
    which is the area of a Gaussian beam with a FWHM
    of 59.2~arcseconds.
    However, the beam profile is not exactly Gaussian, and
    the measured profile was used for all of our analysis.
    Since we cannot rule out systematic variations in the beam area
    from one bolometer to the next at the level
    of our single bolometer measurement uncertainty, 
    we have conservatively
    estimated the uncertainty in this beam area measurement to be 3.1\%.

\section{Map Making}

  \subsection{Least Squares Map Making Theory}

    The astronomical signals we seek can be 
    thought of as two-dimensional objects, which
    can be represented by a map with finite pixelization.
    For simplicity, this two-dimensional map can be thought
    of as a vector, $\vec{m}$.
    This map is stored in the bolometer time-streams,
    $\vec{d}$, according to
    \begin{equation}
      \vec{d} = \mathbf{p} \vec{m} + \vec{n},
      \label{eqn:TOD_map}
    \end{equation}
    where $\mathbf{p}$ is a matrix containing the pointing
    information and $\vec{n}$ is noise.
    Note that we represent matrices with a bold symbol, and
    vectors with an arrow.
    Since $\vec{m}$ is what we are fundamentally interested
    in obtaining, we need to find a solution to 
    Equation~\ref{eqn:TOD_map} that yields the optimum unbiased
    estimate of $\vec{m}$ given $\vec{d}$.
    There are several methods that can be used to estimate
    $\vec{m}$, including the commonly used least
    squares method described below \citep{tegmark97, wright96}.

    Solving the least squares problem for Equation~\ref{eqn:TOD_map}
    requires minimizing
    \begin{equation}
      \chi^2 = (\vec{d} - \mathbf{p}\vec{m})^T
      \mathbf{w} (\vec{d} - \mathbf{p}\vec{m}),
      \label{eqn:chi_map}
    \end{equation}
    where $\mathbf{w}$ is the inverse of the time-stream noise
    covariance matrix, $\left< \vec{n}\vec{n}^T \right>^{-1}$.
    The estimator for $\vec{m}$ derived from 
    Equation~\ref{eqn:chi_map} is
    \begin{equation}
      \vec{m}' = \mathbf{c}
      \mathbf{p}^T \mathbf{w} \vec{d},
      \label{eqn:map_est}
    \end{equation}
    where $\mathbf{c} = (\mathbf{p}^T \mathbf{w} \mathbf{p})^{-1}$
    is the map-space noise covariance matrix.
    If the time-stream noise, $\vec{n}$, has a white spectrum, then
    the various terms in Equation~\ref{eqn:map_est} are easy to
    understand because $\mathbf{w}$ and $\mathbf{c}$ are both
    diagonal.
    $\mathbf{w}$ is the inverse of the time-stream noise
    variance, and applies the appropriate weight to each sample
    in the time-stream.
    $\mathbf{p}^T$ then bins the data time-stream into a map,
    and $\mathbf{c}$ corrects for the fact that $\mathbf{p}^T$ sums
    all of the data in a single map bin instead of averaging it.
    The general idea is the same for non-white time-stream noise, 
    but $\mathbf{w}$
    will mix time samples and $\mathbf{c}$ will mix map pixels.

    If the time-stream noise is stationary, then
    the time-stream noise covariance matrix can be diagonalized
    by applying the Fourier transform operator, 
    $\mathbf{F}$.
    In this case, any element
    of the inverse time-stream noise covariance matrix can be described by
    \begin{displaymath}
      \mathbf{w}(t_1,t_2) = 
      \left< \vec{n(t_1)} \vec{n(t_2)}^T \right>^{-1} = 
      \mathbf{w}(\Delta t),
    \end{displaymath}
    where $t_1$ and $t_2$ are any two time samples separated 
    by $\Delta t$.
    The corresponding elements of the Fourier transform
    of the inverse covariance matrix, 
    $\mathbf{W} = \mathbf{F} \mathbf{w} \mathbf{F}^{-1}$,
    can be written as 
    \begin{displaymath}
      \mathbf{W}(f_1, f_2) = 
      \mathbf{W}(f_1) \delta_{f_1,f_2},
    \end{displaymath} 
    where $\delta_{f_1,f_2}$ represents a Kronecker delta and 
    $f$ is frequency in Hz.\footnote{
    Note that physical space values are denoted with a lower
    case letter, and the corresponding frequency space
    values are denoted with an upper case letter.}
    The diagonal elements of $\mathbf{W}$ are equal to 
    1/(PSD*$\Delta f$), where PSD is the noise power spectral density
    and $\Delta f$ is the frequency resolution of the time-stream.
    The Kronecker delta ensures that all of the off-diagonal
    elements are equal to zero.
    Returning to Equation~\ref{eqn:map_est}, 
    the estimate for $\vec{m}$ can be rewritten as
    \begin{equation}
      \vec{m}' = ((\mathbf{p}^T \mathbf{F}^{-1})
      (\mathbf{F} \mathbf{w} \mathbf{F}^{-1}) 
      (\mathbf{F} \mathbf{p}))^{-1}
      (\mathbf{p}^T \mathbf{F}^{-1}) 
      (\mathbf{F} \mathbf{w} \mathbf{F}^{-1})
      (\mathbf{F} \vec{d}),
      \label{eqn:map_est_fourier1}
    \end{equation}
    using the fact that $\mathbf{F}^{-1} \mathbf{F} = 1$.
    Finally, taking the Fourier transform of the various terms in 
    Equation~\ref{eqn:map_est_fourier1} yields
    \begin{displaymath}
      \vec{m}' = (\mathbf{P}^T \mathbf{W} \mathbf{P})^{-1}
      \mathbf{P}^T \mathbf{W} \vec{D}
    \end{displaymath}
    as an alternate expression to estimate the value of $\vec{m}$,
    where $\mathbf{P} = \mathbf{F}\mathbf{p}$, 
    ${\vec{D}} = \mathbf{F}{\vec{d}}$,
    and $\mathbf{W} = \mathbf{F}\mathbf{w}\mathbf{F}^{-1}$.
    Note that 
    $\mathbf{c} = (\mathbf{p}^T \mathbf{w} \mathbf{p})^{-1} = 
    (\mathbf{P}^T \mathbf{W} \mathbf{P})^{-1}$ does not
    in general simplify as a result of Fourier transforming.

  \subsection{The Bolocam Algorithm: Theory}
  \label{sec:bolo_map_theory}

    The science field maps produced by Bolocam each contain
    $n_p \simeq 20000$ pixels, and an extremely large matrix
    must be inverted to calculate $\mathbf{c}$ since
    $\mathbf{P}^T\mathbf{WP} = \mathbf{p}^T\mathbf{wp}$
    has dimensions of $n_p \times n_p$.
    Direct inversion of such a matrix is possible, but is not practical
    on a typical high-end desktop computer.
    The map could be determined on a desktop computer via
    a conjugate gradient solver, but determining the covariance
    matrix would require a significant amount of simulation
    power.
    Therefore, we developed an algorithm to approximate
    $\vec{m}'$ by exploiting the simplicity of our
    scan pattern, which involved raster scanning
    parallel to either the RA or dec axis.
    This approximation allows us to make maps in a relatively
    short amount of time using a standard desktop computer,
    which is extremely convenient.

    To illustrate this simplification, consider the map
    made from a single bolometer for a single scan
    within an observation.
    This scan will produce a one-dimensional map
    at a single dec value (for an RA scan) or 
    a single RA value (for a dec scan).
    Each data point in the time-stream is separated
    by 24~arcseconds in map-space 
    since our data are sampled at 10~Hz and
    the telescope scans at 240~arcseconds/sec.
    Therefore, our data is approximately Nyquist sampled
    for Bolocam's $\simeq 60$~arcsecond FWHM beams.
    The maps are binned with 20~arcsecond pixels 
    (1/3 of the beam FWHM, and slightly finer than
    Nyquist sampled),
    so $\mathbf{p}^T$ will map either one or zero time-stream
    samples to each map pixel.
    Note that $n_s$, the number of time-stream samples, 
    will be slightly less than $n_p$, the number of map-space pixels.
    Since $\mathbf{p}^T$ has dimensions of $n_s \times n_p$,
    the sum of each row in $\mathbf{p}^T$ is either one or zero and 
    the sum of each column is one.
    Consequently, we will make the approximation that 
    $\mathbf{p}^T = 1$.
    From Equation~\ref{eqn:map_est}, 
    this means that 
    \begin{equation}
      \mathbf{c} = \mathbf{w}^{-1}, 
      \label{eqn:map_corr_single_scan}
    \end{equation}
    and therefore $\vec{m} = \vec{d}$ for a single scan of
    time-stream data.
    If we Fourier transform Equation~\ref{eqn:map_corr_single_scan},
    then we find that
    \begin{displaymath}
      \mathbf{C} = \mathbf{W}^{-1}.
    \end{displaymath}
    Since $\mathbf{W}$ is diagonal, the inversion is trivial,
    and the result is that 
    the Fourier transform of the map-space noise covariance
    matrix is diagonal with elements equal to the time-stream
    PSD*$\Delta f$.
	
    The next step is to consider a map made from a
    single bolometer for a full
    observation, which contains twenty scans.
    We move the telescope in the orthogonal direction to
    the scan between scans by more than the size of 
    a single map pixel,
    so we can still approximate $\mathbf{p}^T \approx 1$.
    There are almost no correlations between scans
    because the atmospheric-noise subtraction coefficients
    are calculated scan-by-scan along with subtraction
    of the mean signal level.
    The covariance of maps made for a single
    observation from alternate scans is negligible, supporting
    this assumption that individual scans are uncorrelated.
    Therefore, the time-stream data and map-space data
    for different scans are essentially independent.\footnote{
      To verify that the data from different scans are independent,
      we created maps for each observation from all the 
      odd-numbered (right-going) scans and from all the
      even-numbered (left-going) scans.
      The cross PSDs of the right-going maps with the left-going
      maps were consistent with noise, indicating that the
      data from separate scans are independent.}
    Consequently, the noise in map-space will be
    stationary, which means that the noise covariance matrix
    can be diagonalized by Fourier transforming it.
    The Fourier transform of the full-map noise covariance
    matrix, $\mathbf{C}$, can be visualized by noting that each
    diagonal element corresponds to a single Fourier-space map
    pixel (or equivalently, a single Fourier-space time-stream sample).
    So, this visualization of $\mathbf{C}$ will be
    equal to the single scan time-stream PSD*$\Delta f$ for rows of 
    map-space pixels that are parallel to the scan
    direction, and will have a white spectrum
    for columns of map-space pixels that are perpendicular
    to the scan direction.
    Alternatively, since there is a one-to-one correspondence
    between time-stream samples and map-space pixels,
    this visualization of the diagonal elements of $\mathbf{C}$ is
    equal to the full map-space PSD*$\Delta f_{\Omega}$,
    where $\Delta f_{\Omega}$ is the angular frequency
    resolution of the map.

    At this point, we need to add together all of the 
    individual observations to make a single map.
    Since we have shown that the map-space data are 
    equivalent to the time-stream data for a single
    observation, the easiest way to co-add data from
    separate observations is to use the single
    observation maps.
    Since the noise in separate observations is uncorrelated,
    the maps can be co-added according to
    \begin{equation}
      \vec{m} = \left( \sum_i \mathbf{c}^{-1}_i \right)^{-1}
      \sum_j \mathbf{c}^{-1}_j \vec{m}_j,
      \label{eqn:map_obs_coadd}
    \end{equation}
    where the subscripts $i$ and $j$ refer to observation number.
    The easiest way to evaluate Equation~\ref{eqn:map_obs_coadd}
    is to Fourier transform it so that the noise
    covariance matrices are all diagonal.
    The result is
    \begin{equation}
      \vec{M} = \left( \sum_i \mathbf{C}^{-1}_i \right)^{-1}
      \sum_j \mathbf{C}^{-1}_j \vec{M}_j,
      \label{eqn:map_obs_coadd2}
    \end{equation}
    where $\vec{M}$ is the Fourier transform of the map
    and $\mathbf{C}$ is the Fourier transform of the noise covariance
    matrix, with diagonal elements equal to the PSD*$\Delta f_{\Omega}$
    of the map.
    Since all of the $\mathbf{C}$s are diagonal, we can simplify
    Equation~\ref{eqn:map_obs_coadd2} to
    \begin{equation}
      M = \left( \sum_i \frac{1}{\mathcal{P}_i} \right)^{-1}
      \sum_j \frac{M_j}{\mathcal{P}_j},
      \label{eqn:map_obs_coadd3}
    \end{equation}
    where $M$ is the two-dimensional Fourier transform of the 
    map and $\mathcal{P}$ is the two-dimensional PSD
    of the noise in the map.
    At this point we have dropped the vector and matrix notation
    since $M_i$ and $P_i$ have the same dimensions.
    Note that $\Delta f_{\Omega}$ is the same for every map, so the constant
    factor of $\Delta f_{\Omega}$ from the first sum in 
    Equation~\ref{eqn:map_obs_coadd3} cancels the factor
    of 1/$\Delta f_{\Omega}$ from the second sum in 
    Equation~\ref{eqn:map_obs_coadd3}.

    Finally, to make a map using all of our data,
    we need to consider every bolometer, not just a single
    detector.
    To properly weight the data from each bolometer
    prior to co-adding, we
    calculate the expected variance, $(\sigma_{pf})^2_i$,
    in measuring
    the peak flux of a point-like source from a single
    scan through the center of the source for 
    bolometer $i$.
    This variance is calculated using the scan-averaged 
    time-stream PSD for each bolometer, 
    $PSD_i(f)$, and the Fourier transform
    of the expected signal shape of a point-like
    astronomical signal, $S(f)$,
    according to
    \begin{equation}
      (\sigma_{pf})^2_i = \left( \int df \frac{S(f)^2}{PSD_i(f)}
      \right)^{-1}
      \label{eqn:relsens}
    \end{equation}
    where $f$ is temporal frequency.
    Note that $S(f)$ is the beam profile, not a delta function.
    Then, the data from each bolometer is weighted
    by a factor proportional to $1/(\sigma^2_{pf})$
    prior to co-adding it with data from other bolometers.
    This is the optimal way to co-add the
    data for point-like signals; it is nearly
    optimal for signals of any shape if the
    PSDs have similar profiles for every bolometer,
    which is largely true for our bolometer signals since they
    are dominated by atmospheric noise.

    However, due to atmospheric noise, along with our noise
    removal algorithms, there are correlations between
    the bolometers.
    But, most of these correlations are instantaneous in time
    and constant over the observation.\footnote{
      We have been able to find a small amount of correlated
      atmospheric signal that is not time-instantaneous.
      However, the time lag of these correlations is generally
      much less than one time sample, which means they
      will also be less than one map pixel.}
    Additionally, the relative positions of the bolometers
    do not change during the observation, so the 
    map-space separation of the correlations does not change.
    Therefore, the correlations are stationary in time
    with separations that are fixed in map-space,
    so the correlations are an additional time-independent
    covariance between map pixels that are sampled at the 
    same time by different bolometers.
    This additional covariance is approximately stationary
    over the entire map, except where it breaks down near
    the edges because part of the focal plane is outside
    the map region.
    Since this additional covariance between map pixels is
    approximately stationary in space, its contribution to
    $\mathbf{W}^{-1}$ will be diagonal.
    Since $\mathbf{W}^{-1}$ is still diagonal, co-addition of the maps of 
    individual observations can proceed according to
    Equation~\ref{eqn:map_obs_coadd3}.
    Therefore, Equation~\ref{eqn:map_obs_coadd3} can
    be used as the algorithm to produce our final
    science field maps.

    Note that we were forced to make several simplifying 
    assumptions in order to develop Equation~\ref{eqn:map_obs_coadd3}.
    We have assumed that the pointing matrix, $\mathbf{p}^T$,
    is equal to one.
    We have also assumed that the noise in our time-streams
    is stationary for each eight-minute-long observation.
    Additionally, we have assumed that the PSD
    of the correlations between bolometers is white, and
    that all of the correlations are time-instantaneous.
    Finally, we have assumed that the map coverage
    (\emph{i.e.}, the number of time-stream samples that are
    binned in each map-space pixel) is uniform,
    so that the Fourier transform of the map is
    a valid description of the time-stream data.
    Deviations from these assumptions will alter the
    map estimate we compute from the optimal least squares
    map estimate.
    But, these deviations only affect how each time-stream
    sample is weighted before it is mapped.
    This means our final map will have more noise than
    an optimal map, but it will not be biased in any way.
    In other words, since the map-making operation is linear,
    the resulting map will be unbiased no matter what
    weightings are used to co-add the data, as 
    long as the weights are properly normalized.
    We have confirmed this lack of bias via simulation,
    as we discuss below.

  \subsection{The Bolocam Algorithm: Implementation}
  \label{sec:mapmaking}

    To start, we must first produce a map from the time-stream
    data for each eight-minute-long observation.
    As mentioned in Section~\ref{sec:bolo_map_theory},
    this is done by calculating the variance in measuring
    the peak flux of a point-like source under the assumption that 
    the profile of the time-stream PSD
    is similar for every bolometer.
    To determine this variance, we calculate the 
    PSD for each bolometer for each scan.
    These spectra are then averaged over all twenty scans
    for each bolometer, thereby making the assumption that the
    noise properties do not change over the course
    of the observation.
    Then, we determine the expected shape of a point-like
    source using our measured beam profile and scan speed.
    Finally, Equation~\ref{eqn:relsens} is used to determine
    the variance in measuring the peak flux
    of a point-like source for each bolometer,
    which is inversely proportional to the weighting factor 
    applied to the time-stream data
    for that bolometer.

\begin{figure}
  \plottwo{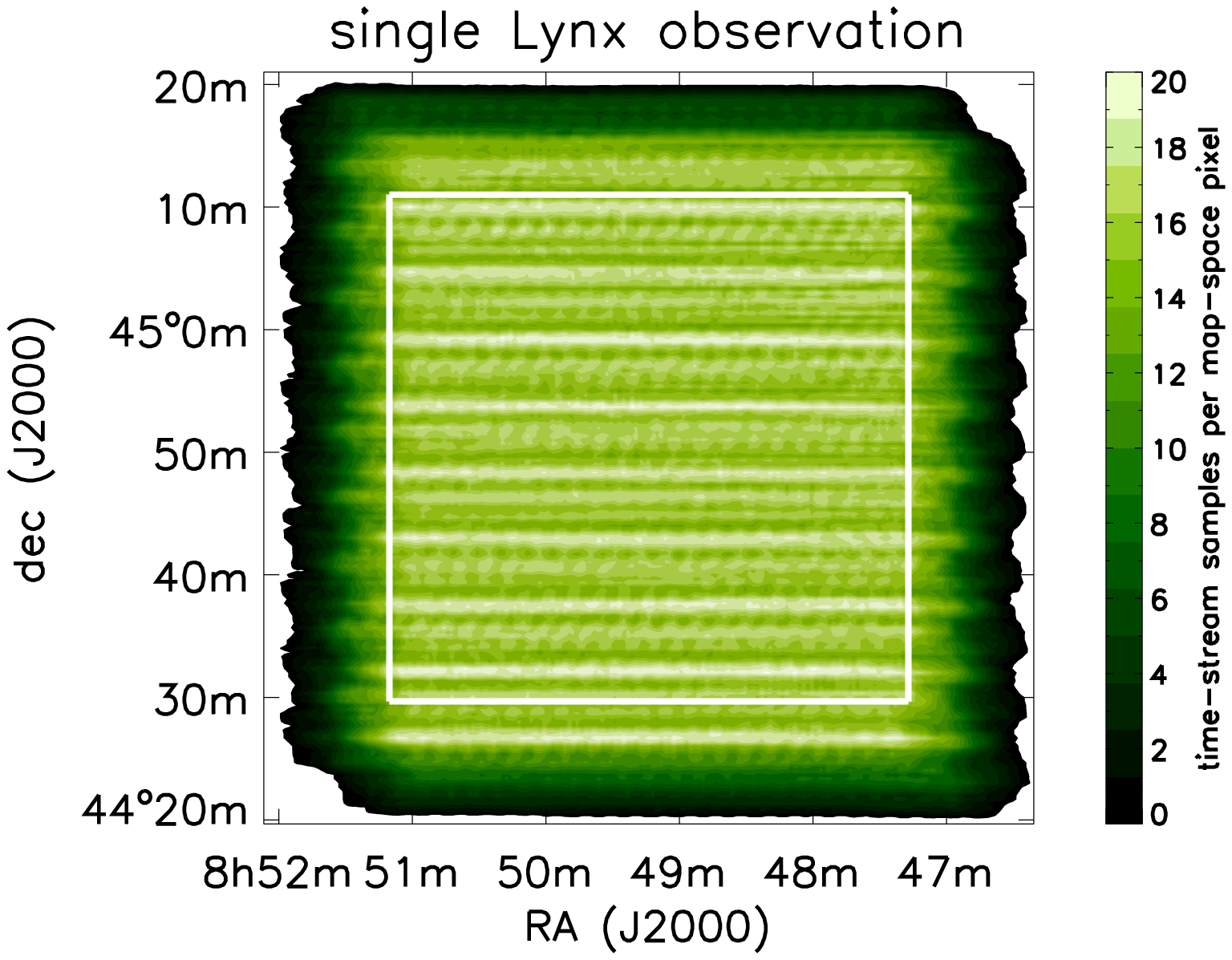}{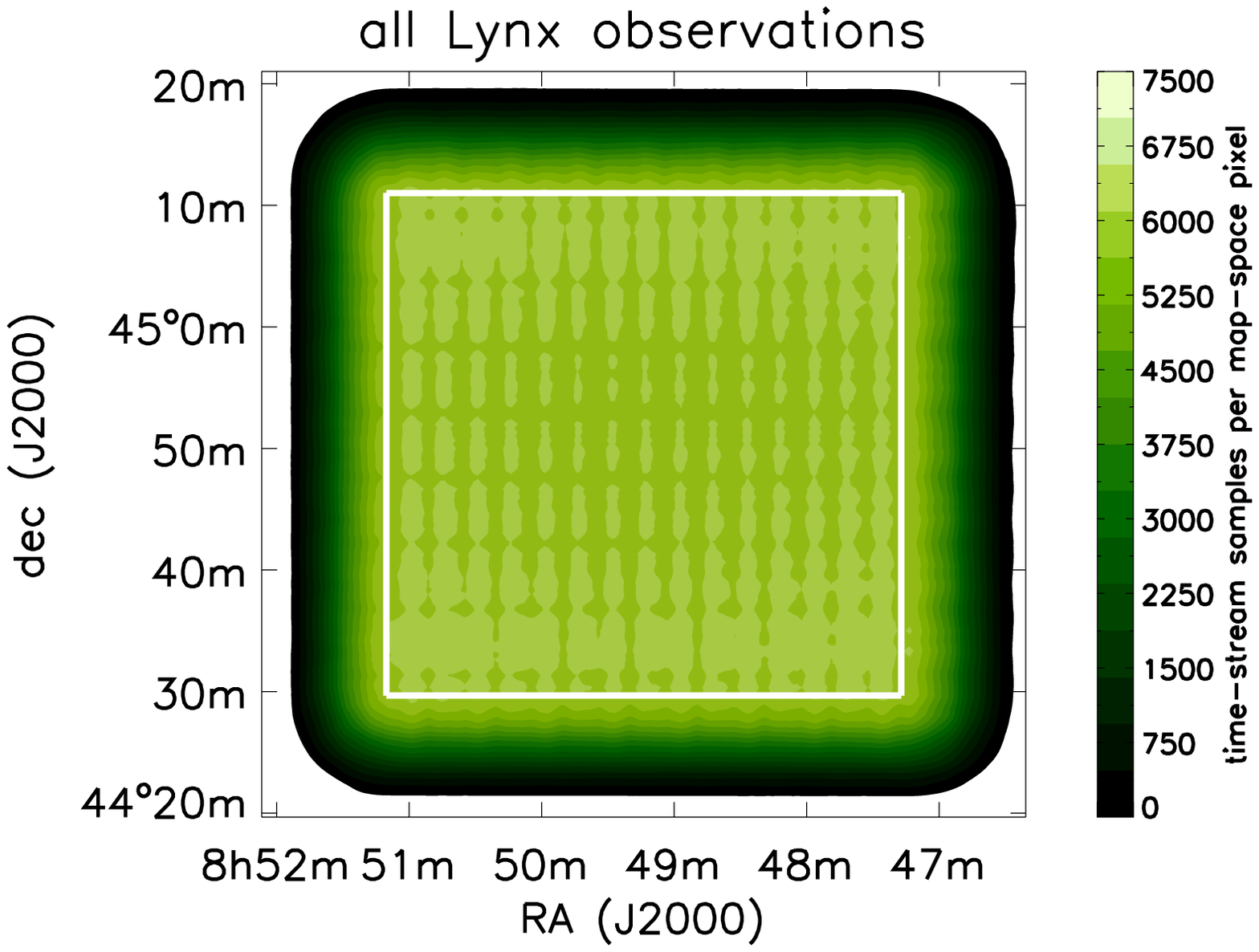}
  \caption{Map coverage, quantified by the number of time-stream
    samples that correspond to a particular map-space pixel
    for a single observation of the Lynx field and for
    the co-add of all observations of the Lynx field.
    The white square has sides of 
    approximately 42~arcminutes and 
    contains the region of the map
    defined to have uniform coverage.
    The RMS deviations in coverage within this region
    relative to the 
    average coverage within the region
    are approximately 8 -- 9\% for a single observation
    and around 1.5\% for the co-add of all observations.}
  \label{fig:mapcov_single_obs}
\end{figure}

    At this point, we have individual observation maps
    for every observation,
    and we can make a map from all of the data using
    Equation~\ref{eqn:map_obs_coadd3}.
    But, one of the main assumptions made in developing 
    Equation~\ref{eqn:map_obs_coadd3} was that the
    map coverage is uniform for each observation.
    If this assumption fails, then the Fourier transform of the
    map is not a good description of the time-stream data.
    Our scan strategy produced highly uniform 
    coverage in the central region of the map, and this 
    coverage falls rapidly to zero at the edges of the map.
    See Figure~\ref{fig:mapcov_single_obs}.
    To obtain sufficiently uniform coverage, we restrict our map
    to have 
    sides of 42~arcminutes; the fractional RMS variations in
    coverage within this region for a single eight-minute-long observation
    are only about 8 - 9\%.
    Since the coverage variations are minimal, we will assume
    that this square central region has uniform coverage,
    and therefore uniform noise properties.
    This assumption of uniform coverage allows us to directly
    compute the Fourier transform and noise properties
    of the map.
    We emphasize that, even if the assumption of uniform coverage fails
    and our algorithm is non-optimal, it is never biased
    because Equation~\ref{eqn:map_obs_coadd3} is linear
    in the map-space maps.

\begin{figure}
  \plottwo{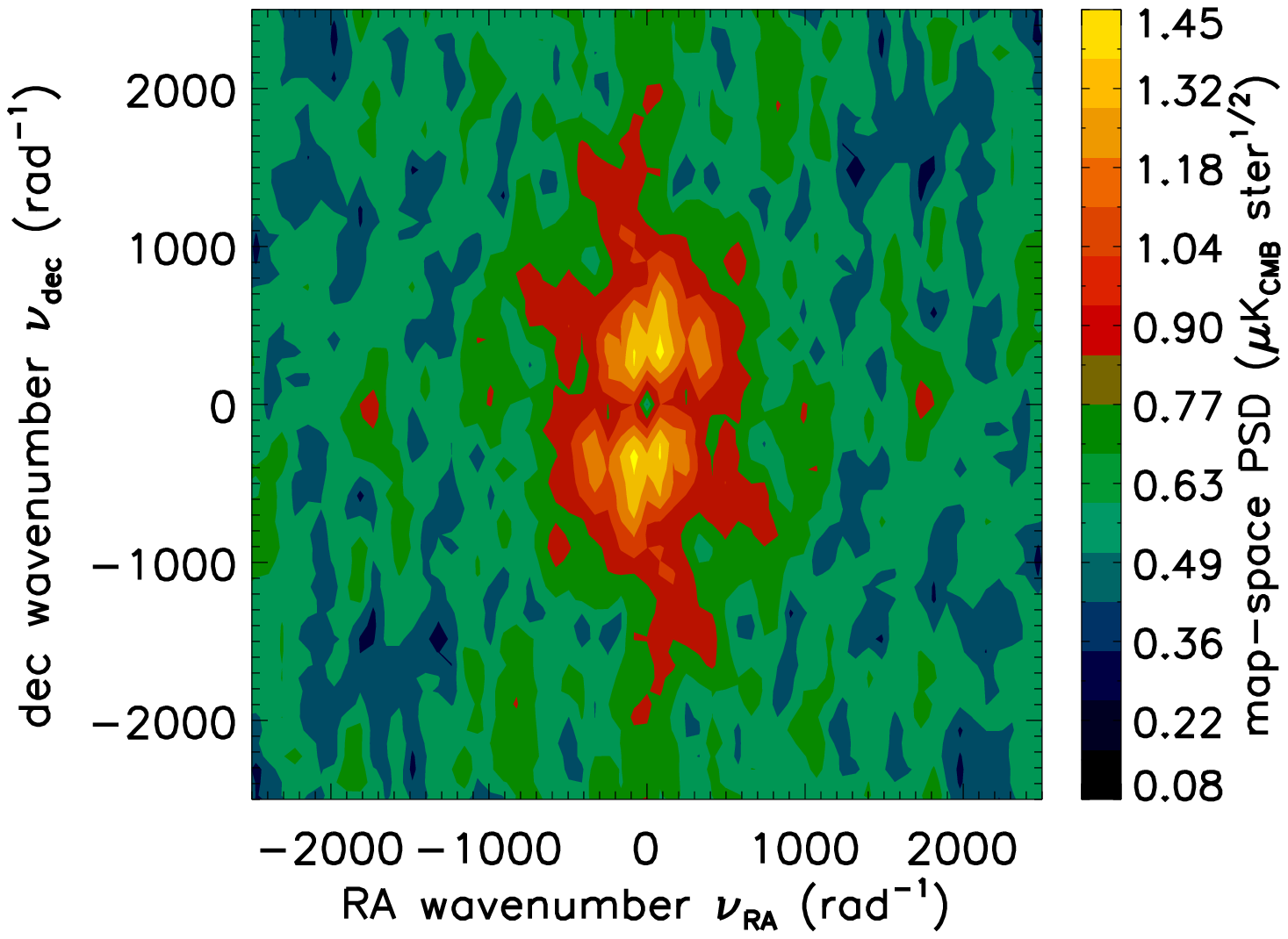}{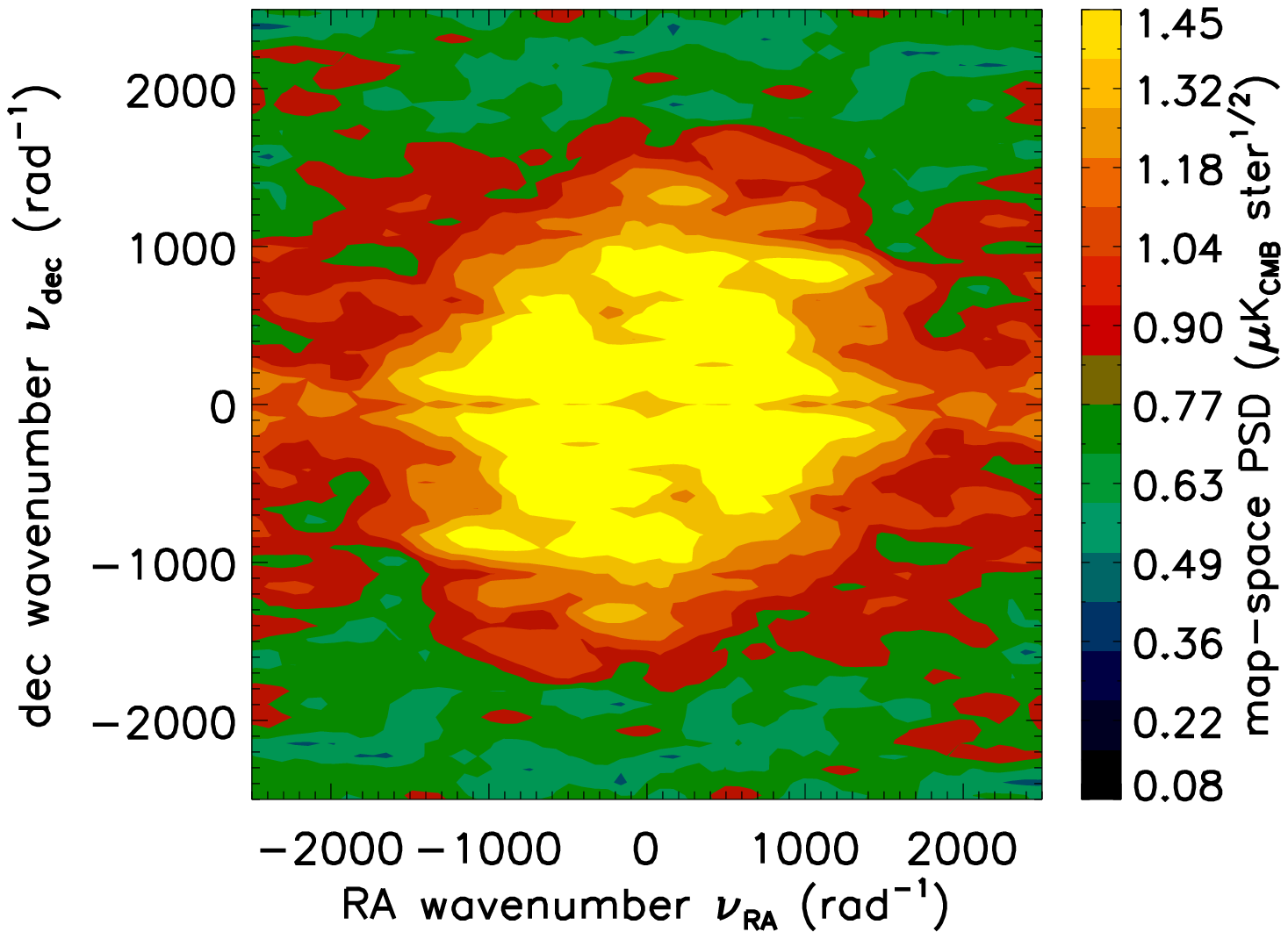}
  \caption{Map-space PSDs, $\mathcal{P}_{\vec{\nu}}$, for single observations.
    The plot on the left shows an observation made in relatively good weather
    while scanning in the RA direction, and the plot on the right
    shows an observation made in relatively poor weather
    while scanning in the dec direction.
    In each case, note that there is a stripe of increased
    noise at low frequency along the scan direction,
    due to time-stream noise with a $1/f$ spectrum.}
  \label{fig:map_psd_single_obs}
\end{figure}

    We now have a uniform coverage map for each observation,
    which can easily be Fourier transformed to produce
    the $M_i$s needed in Equation~\ref{eqn:map_obs_coadd3}.
    But, we still need to determine the two-dimensional 
    PSD of each single observation map.
    Due to residual correlations between bolometers, 
    we do not understand the noise properties of our
    data well enough to determine the map PSD from simulation,
    so we instead estimate the PSD by generating a large
    number of jackknifed maps from our real data.
    In each jackknifed map, a different subset of 
    the time-streams from half
    of the scans within each observation was multiplied by $-1$.
    Note that the data from all of the bolometers within a single
    scan are multiplied by $-1$, so the residual atmospheric noise
    that is correlated between bolometers is preserved.
    This multiplication leaves the noise properties of the
    map unchanged,\footnote{
      Each time-stream sample (and therefore each map-space pixel)
      can be expressed as the sum of two signals:
      1) an astronomical signal and 2) a random noise signal that
      is drawn from the underlying distribution of the noise in the
      Bolocam system.  The astronomical signal 
      corresponding to a particular map-space pixel will be the
      same for any scan, and will disappear in the jackknife
      realizations when
      time-stream data from half of the scans is multiplied 
      by $-1$.  But, if the underlying distribution of the noise
      is Gaussian, then the distribution of signals it
      will produce is symmetric about 0.
      Therefore, the statistical properties of the noise
      will be unchanged when half of the data are multiplied
      by -1.}
    while allowing us to produce a large
    number of noise realizations for each map.
    Note that the residual atmospheric noise correlations are
    time-instantaneous, so they remain in the
    jackknifed realizations.
    We then generate 100 realizations for each observation,
    and we set the true PSD for each observation
    equal to the average of the map-space PSD computed for
    each realization.
    See Equation~\ref{eqn:map_psd_est}.
    Examples of the PSDs we calculated are given in
    Figure~\ref{fig:map_psd_single_obs}.
    This method of determining the map-space PSDs assumes that
    the time-stream data for each scan is uncorrelated with the
    data from all other scans, which we argued in 
    Section~\ref{sec:bolo_map_theory}.
	
    To determine the validity of the map-space PSDs we estimated
    from the jackknifed map realizations,
    we examined the distribution of PSD values for each
    realization.
    If the noise properties of the data are Gaussian, as we have assumed,
    then the PSD measured at any given Fourier map-space pixel
    will be drawn from 
    \begin{equation}
      f(X_{i,\vec{\nu}}) = (1/\mathcal{P}_{\vec{\nu}})
      e^{(-X_{i,\vec{\nu}}/\mathcal{P}_{\vec{\nu}})},
      \label{eqn:PDF_map_PSD}
    \end{equation}
    where $X_{i,\vec{\nu}}$ is the measured PSD for
    realization $i$ at pixel $\vec{\nu}$,
    $\mathcal{P}_{\vec{\nu}}$ is the true PSD for
    pixel $\vec{\nu}$, and $f(X_{i,\vec{\nu}})$ is the probability
    density function of $X_{i,\vec{\nu}}$.
    Note that $\vec{\nu}$ has units of spatial frequency
    (i.e., radians$^{-1}$), and describes
    a pixel in the spatial Fourier transform of the map.
    See Appendix~\ref{sec:map_var} for a derivation of $f(X_{i,\vec{\nu}})$.
    The true PSD is estimated from
    \begin{equation}
      \widehat{\mathcal{P}_{\vec{\nu}}} = 
      \frac{1}{N_r} \sum_{i=1}^{i=N_r} X_{i,\vec{\nu}},
      \label{eqn:map_psd_est}
    \end{equation}
    where $N_r = 100$ is the number of realizations.
    To compare our measured PSDs to the probability density
    function (PDF) given in Equation~\ref{eqn:PDF_map_PSD},
    we created the dimensionless value
    \begin{equation}
      Y_{i,\vec{\nu}} = 
      \frac{X_{i,\vec{\nu}}}{\widehat{\mathcal{P}_{\vec{\nu}}}},
      \label{eqn:dimless_PSD}
    \end{equation}
    with associated PDF
    \begin{equation}
      f'(Y_{i,\vec{\nu}}) = e^{-Y_{i,\vec{\nu}}}.
      \label{eqn:PDF_map_PSD2}
    \end{equation}
    Then, we compared our measured values of $Y_{i,\vec{\nu}}$ to
    the PDF in Equation~\ref{eqn:PDF_map_PSD2}.
    In general, we found that
    our measured $Y_{i,\vec{\nu}}$ follow a
    distribution extremely close to $f'(Y_{i,\vec{\nu}})$,
    except that the number of $Y_{i,\vec{\nu}}$ 
    with values near zero is slightly less than expected.
    Therefore, the map-space PSDs estimated from the jackknife
    realizations should be a good estimate of the true map-space
    PSDs.

    Let us consider the possible effects of imperfect signal removal in  
    the single-observation jackknife maps.  
    Because we only use the single-observation noise estimates as weights
    for coadding,
    the result of 
    residual signal in the jackknife maps will be
    coaddition weights that are non-optimal.
    This non-optimality may degrade the noise of the final maps,
    but will not cause them to be biased.
    We may estimate the size of the signal leakage 
    to determine how large the deviation
    from optimality could be. 
    Our final flat band power anisotropy upper limit is  
    approximately 1000~$\mu$K$_{CMB}^2$.  
    With our effective $\Delta \ln(\ell)$ of 0.63  
    (derived in Section~\ref{sec:xfer}), 
    this upper limit corresponds to an excess  
    variance in our final maps of 
    $\simeq 500$~$\mu$K$_{CMB}^2$.
    These final maps have a variance of 
    10000~$\mu$K$_{CMB}^2$ (see Figure~\ref{fig:final_maps}).
   Given that $\simeq 500$ observations contribute to each map,
   the single-observation map variance is 
   $\simeq 5 \times 10^6$~$\mu$K$_{CMB}^2$, 
   or approximately 10000 times larger than our upper limit 
   on the astronomical signal contribution. 
   Even if we did not remove the astronomical signal using 
   the jackknifing procedure, 
   it would affect the single-observation PSDs,
   and therefore the weights, 
   at only the 0.01\% level. 
   Using jackknife-generated PSDs reduces the 
   effect of signal contamination further. 
   Therefore, the effect of signal leakage 
   into the single-observation jackknife maps is negligible.

\section{Transfer Functions}
\label{sec:xfer}

  The transfer function describes the fraction 
  of the astronomical signal that 
  remains after processing as a function of 
  map-space Fourier mode.
  In order to determine the transfer function of our
  data processing algorithms, we first generate
  a simulated map of the expected astronomical signal.
  This map is then reverse-mapped into
  a time-stream using the pointing
  information in a real observation.
  Next, this simulated time-stream is added to the
  real bolometer time-streams from the observation,
  and then processed and mapped in the standard way.
  A map made from data that did not have a simulated
  signal added to it is then subtracted from this map,
  producing a map with the simulated signal after
  processing.
  Finally, the PSD of this map is divided by the PSD
  of the original simulated signal map to determine
  how much of the signal remains.
  Note that we are computing the transfer function
  for a PSD because we are interested in measuring an
  excess noise and not a specific signal shape,
  which means we do not need the phase of the transfer function.

\begin{figure}
  \plottwo{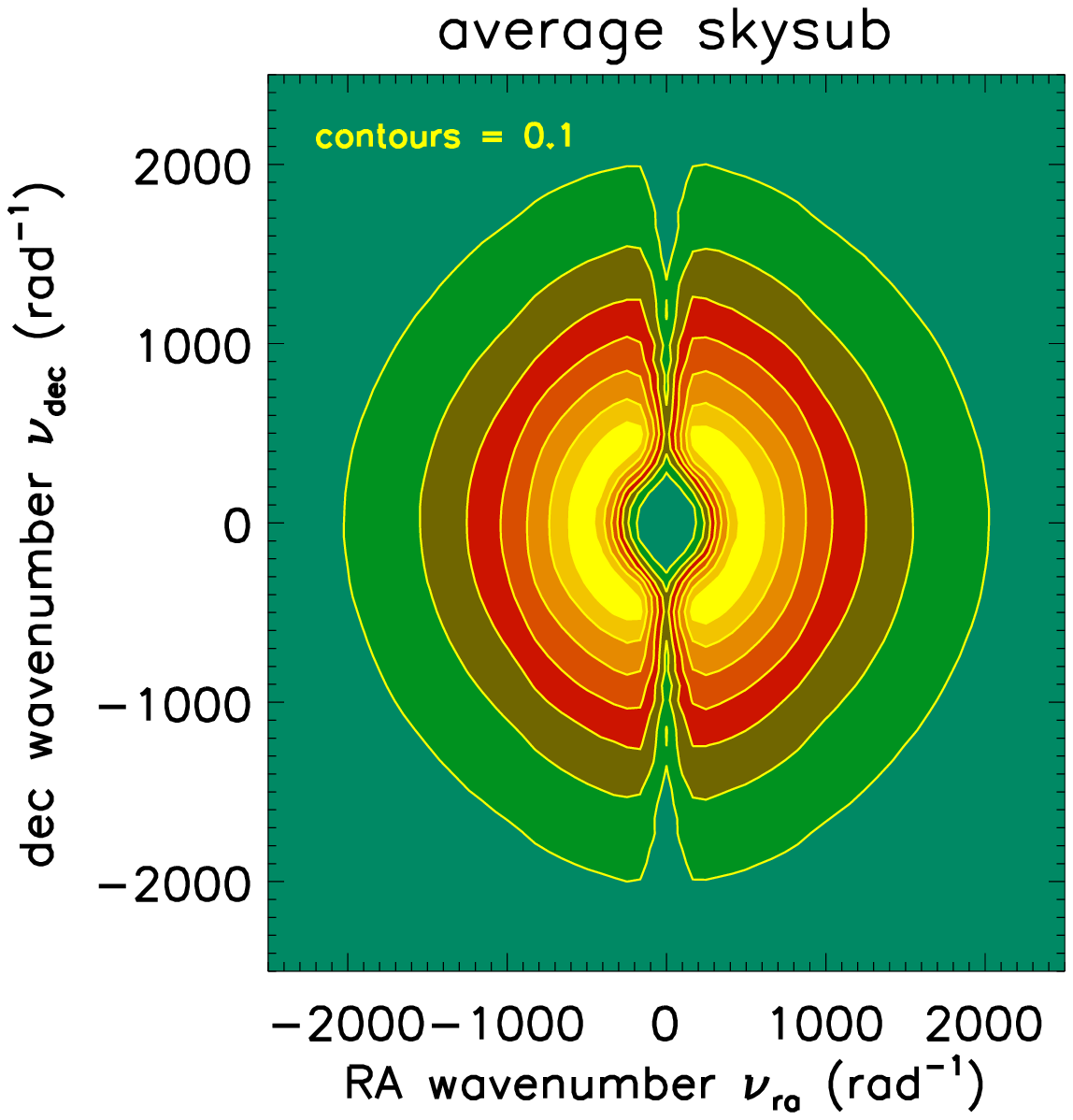}{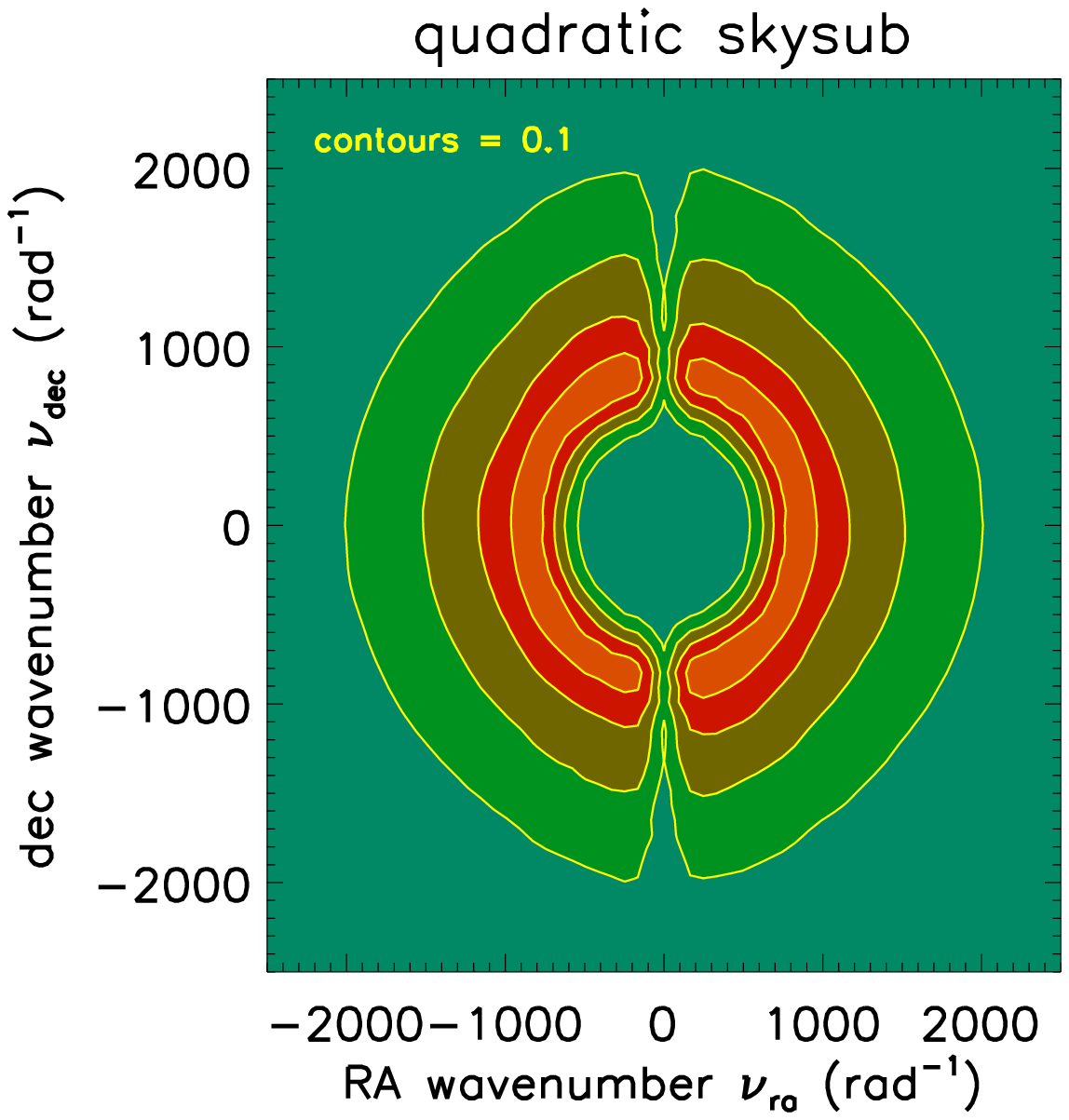}
  \caption{Contour plots of the
    transfer function, $T_{\vec{\nu}} B^2_{\vec{\nu}}$, 
    for observations made while scanning
    parallel to RA for average subtraction and
    quadratic subtraction.
    Note the large amount of attenuation at low frequencies along
    the scan direction and at scales larger than the focal plane
    size of approximately 500~radians$^{-1}$.}
  \label{fig:RA_xfer}
\end{figure}

\begin{figure}
  \plottwo{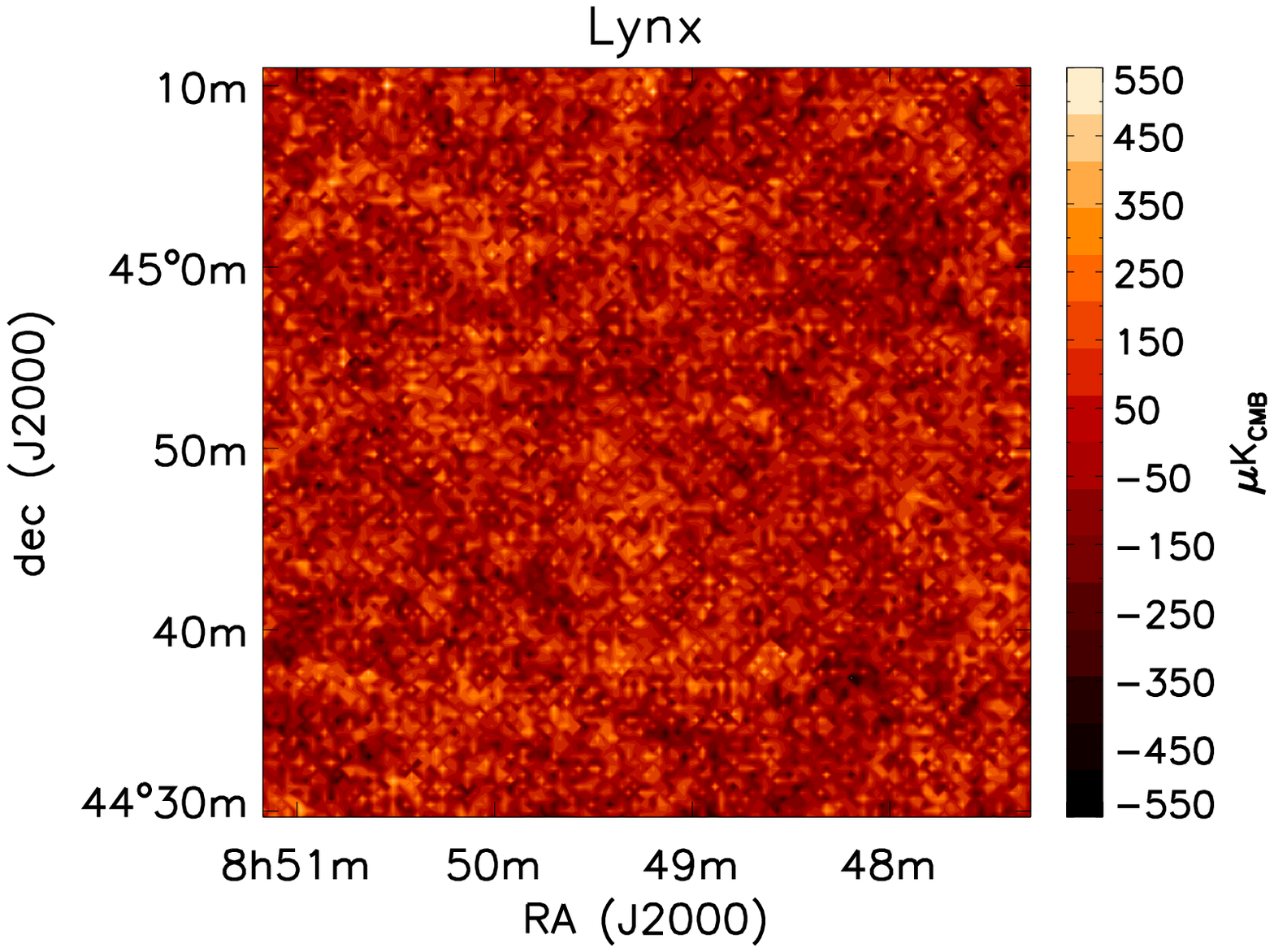}{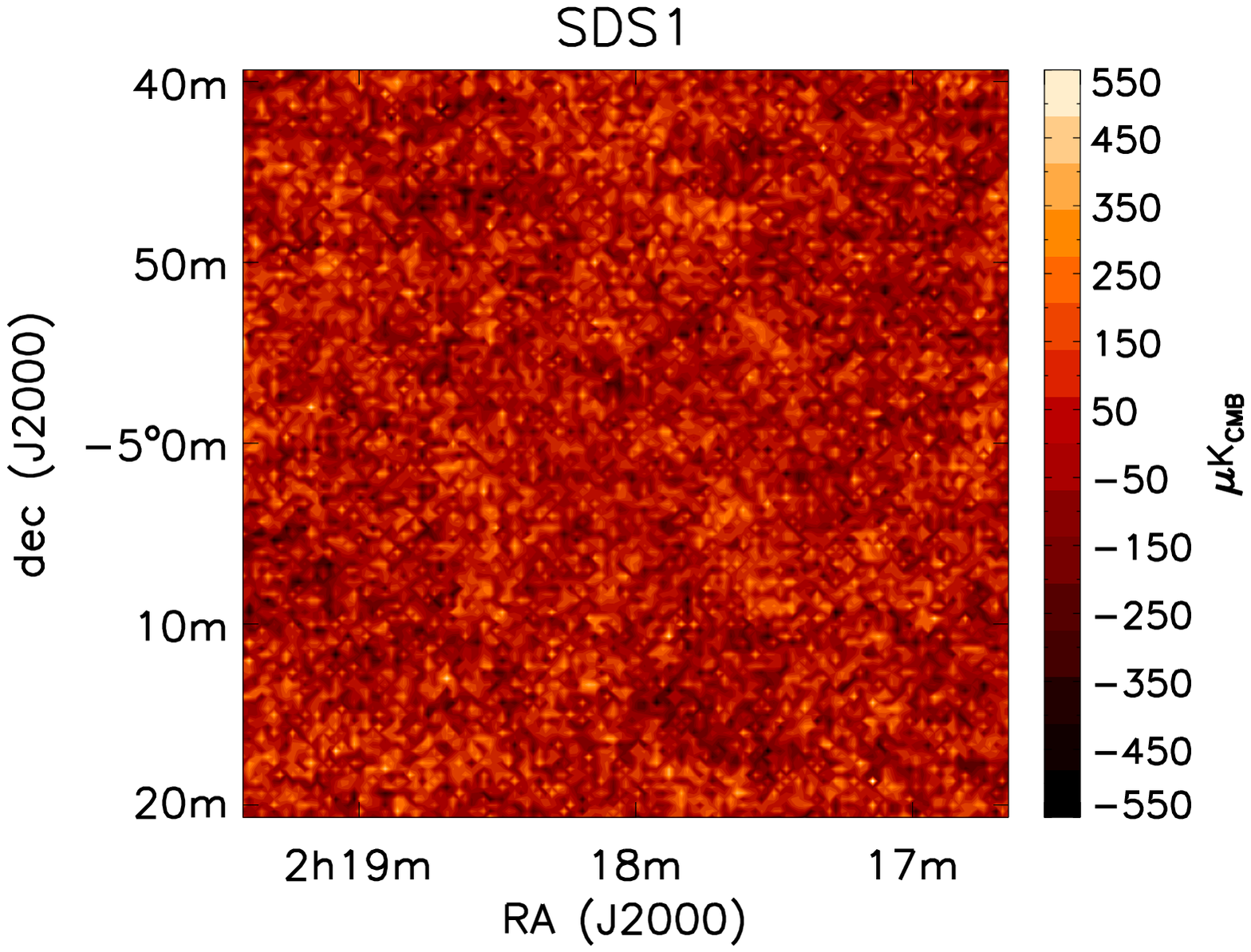}
  \caption{Maps of the science fields.
    Note that the astronomical signal in
    each map has been convolved with the
    transfer functions of the data processing
    and the beam, $\sqrt{TB^2}$, but the noise
    has not been filtered in any way.
    The RMS of these unfiltered maps
    is approximately 90~$\mu$K$_{CMB}$ per beam,
    and the RMS after optimally filtering
    for point sources is $\simeq 70$~$\mu$K$_{CMB}$
    per beam.}
  \label{fig:final_maps}
\end{figure}

  This transfer function was computed for twenty randomly selected
  observations, ten taken while scanning parallel to
  RA and ten taken while scanning parallel to dec.
  Realizations of the expected flat-band power anisotropy signal were
  used as the simulated signal.
  These realizations were generated in Fourier map-space
  assuming Gaussian fluctuations and a flat 
  band power in $\mathcal{C}_{\ell} = C_{\ell} \ell (\ell+1) / 2\pi$
  of 50~$\mu$K$_{CMB}^2$.
  Note that although a flat band power of 50~$\mu$K$_{CMB}^2$ was used
  for the simulated signal maps, we found that the
  transfer function is independent of 
  the amplitude of the flat-band power anisotropy signal.
  For each observation, we averaged the transfer 
  function obtained from 100 different signal realizations
  to determine the average transfer function.
  We then compared the average transfer function for each of the ten
  observations taken with a similar scan pattern.
  The result is that
  the transfer functions were the same within our measurement
  uncertainty for all of the observations.
  Therefore, we averaged the transfer function from all
  ten observations to produce a high signal-to-noise
  measurement for each atmospheric-noise removal method:
  average, planar, and quadratic.
  See Figure~\ref{fig:RA_xfer}.

  Since all of the data processing is performed on the time-streams,
  the attenuation caused by the processing has a preferred 
  orientation based on the scan strategy.
  The result is a transfer function that is not azimuthally 
  symmetric because of the large amount of attenuation
  at low frequencies parallel to the scan direction
  due to atmospheric noise removal.
  Additionally, there is massive attenuation on scales
  larger than the Bolocam focal plane ($\simeq 500$~radians$^{-1}$)
  because of the atmospheric noise removal algorithms.
  This occurs because these algorithms are designed to remove
  all time-instantaneous signals at each data sample,
  which is equivalent to subtracting any signals
  that vary slowly compared to the size of the focal plane.

  In addition to the signal attenuation caused by the 
  data processing, the Bolocam system also attenuates
  some of the astronomical signal.
  By scanning across the sky, we are effectively convolving
  any signal with the profile of a Bolocam beam;
  since the beams have a non-zero width, this convolution
  will act like a low-pass filter on all of the
  astronomical signals.
  This filter will be approximately symmetric because
  the Bolocam beam profiles have a high degree
  of rotational symmetry.
  Additionally, since the beams are nearly Gaussian,
  the filter will be approximately Gaussian
  with a HWHM in variance of about 1000~radians$^{-1}$
  (which is equivalent to a HWHM$_{\ell} \simeq 6000$
  in angular multipole space).
  See Figure~\ref{fig:RA_xfer}.

  In order to quantify the amount of signal	
  attenuation by each atmospheric noise removal algorithm,
  it is useful to determine the effective bandwidth
  of the transfer function.
  The effective bandwidth describes the range of angular
  multipoles to which we are sensitive,
  as quantified by the transfer function, 
  and can be used to convert an angular
  power, $C_{\ell}$, to a map-space variance in $\mu$K$^2_{CMB}$.
  In general, the effective bandwidth is calculated by
  integrating the transfer function over all
  angular multipoles.
  However, since the expected SZE power spectrum is
  approximately flat in $\mathcal{C}_{\ell}$,
  which results in a spectrum in $C_{\ell}$
  that falls like $1/\ell(\ell+1)$,
  it is more useful to weight the transfer function
  by a factor of $1/\ell(\ell+1)$.
  This weighting will produce an effective logarithmic,
  rather than linear, bandwidth, and can
  be used to convert an angular power
  in $\mathcal{C}_{\ell}$ to a map-space variance.
  This effective logarithmic bandwidth,
  ${\rm BW}_{eff}$, is defined as
  \begin{displaymath}
    {\rm BW}_{eff} = \int_{\vec{\nu}} d\vec{\nu} S_{\vec{\nu}}^2 
    T_{\vec{\nu}} B_{\vec{\nu}}^2,
  \end{displaymath}
  where $\vec{\nu}$ is the two-dimensional spatial frequency,
  $S_{\vec{\nu}}$ is the expected signal spectrum,
  $T_{\vec{\nu}}$ is the transfer function of the
  data processing in squared units, and $B_{\vec{\nu}}$ is the
  profile of the Bolocam beam.
  Since the expected anisotropy signal has a flat band power
  in $\mathcal{C}_{\ell}$, 
  \begin{displaymath}
    S_{\vec{\nu}}^2 \propto \frac{1}{\ell(\ell + 1)}
  \end{displaymath}
  for $\ell = 2 \pi |\vec{\nu}|$.\footnote{
  We have used the 
  small-scale flat sky approximation, $\ell = 2\pi |\vec{\nu}|$.}
  Assuming this spectrum for $S_{\vec{\nu}}^2$,
  a top-hat window between $\ell = \ell_{min}$ and
  $\ell = \ell_{max}$ will produce a bandwidth approximately 
  equal to
  \begin{displaymath}
    {\rm BW}_{eff} \propto \ln(\ell_{max}) - \ln(\ell_{min}) = 
    \Delta \ln(\ell).
  \end{displaymath}
  Although the Bolocam transfer functions are 
  not azimuthally symmetric, it is still useful to determine
  the effective $\Delta \ln(\ell)$ for each
  of the atmospheric noise removal algorithms,
  with $\Delta \ln(\ell) = 0.98$, 0.58, and 0.37
  for average, planar, and quadratic subtraction.
  Note that our final map, which consists of observations
  processed with different atmospheric noise removal
  algorithms as described in Section~\ref{sec:opt_skysub},
  has a bandwidth of $\Delta \ln(\ell) \simeq 0.63$.

\section{Optimal Atmospheric Noise Subtraction}
\label{sec:opt_skysub}

  Each of the science field observations were processed
  with average, planar, and quadratic sky subtraction,
  creating three separate files for each observation.
  Quadratic subtraction removes the most atmospheric noise,
  while average subtraction retains the most astronomical
  signal, so there is an optimal sky subtraction
  algorithm for each observation
  based on the type of astronomical signal we are 
  looking for.
  To determine which algorithm is optimal, we computed
  a figure of merit, FOM, for each subtraction method.
  Since the anisotropy signal appears as a variance in the map,
  the variance on the amplitude of the
  anisotropy signal will be proportional
  to the square of the map PSD divided by the 
  transfer function of the experiment.
  This can be seen
  in Equations~\ref{eqn:est_a}~and~\ref{eqn:est_var_a}.
  Therefore, the FOM is defined as the inverse
  of this variance on the anisotropy signal summed over
  all angular scales according to
  \begin{equation}
    {\rm FOM} = \sum_{\vec{\nu}} 
    \frac{(S_{\vec{\nu}}^2)^2 T_{\vec{\nu}}^2 (B_{\vec{\nu}}^2)^2}
    {\mathcal{P}_{\vec{\nu}}^2},
    \label{eqn:fom}
  \end{equation}
  where $\vec{\nu}$ is a two-dimensional spatial frequency with
  units of radians$^{-1}$, $S_{\vec{\nu}}^2$ is the expected
  anisotropy power spectrum, $T_{\vec{\nu}}$ is the transfer
  function of the data processing in squared units, $B_{\vec{\nu}}$
  is the profile of the Bolocam beam, and
  $\mathcal{P}_{\vec{\nu}}$ is the PSD of the noise
  in the map in squared units.
  Note that we have included the $\simeq 5$~arcsecond
  uncertainty in our pointing model in $B_{\vec{\nu}}$,
  and this pointing uncertainty effectively broadens
  the beam.
  To be precise, 
  \begin{displaymath}
    B_{\vec{\nu}} = \mathsf{B}_{\vec{\nu}} 
    e^{-|\vec{\nu}|^2 / 2 \sigma_{\nu}^2},
  \end{displaymath}
  where $\mathsf{B}_{\vec{\nu}}$ is the measured beam profile, 
  and $\sigma_{\nu} = 1/ 2 \pi \sigma_p$ for 
  a pointing uncertainty of $\sigma_p$.
  For the anisotropy spectrum, we assumed a flat band power in
  $\mathcal{C}_{\ell}$, so
  \begin{displaymath}
    S_{\vec{\nu}}^2 = \frac{1}{\ell(\ell+1)}
  \end{displaymath}
  for $\ell = 2 \pi |\vec{\nu}|$.
  The figure of merit is inversely proportional to the
  variance on an estimate of the anisotropy amplitude
  (in $\mu$K$_{CMB}^2$),
  so it characterizes the signal-to-noise ratio of the map.

  In the end, average subtraction was the optimal method for just over
  50\% of the observations, planar subtraction was the 
  optimal method for just over 40\% of the observations,
  and quadratic subtraction was the optimal method for
  just under 10\% of the observations.
  We can calculate how much the observations optimally cleaned
  by each method contribute to our final S/N from
  \begin{displaymath}
    \textrm{S/N} = \sqrt{\frac{\sum_{i \, \epsilon \, T} FOM_i^{-2}}
    {\sum_{i} FOM_i^{-2}}},
  \end{displaymath}
  where $T$ denotes the set of observations optimally cleaned
  by a given method
  and $FOM_i$ is the figure of merit from Equation~\ref{eqn:fom}.
  The S/N contributed by the average/planar/quadratic
  observations is 70/29/1\%.
  These ratios are different from the number of observations
  optimally cleaned by each method
  because the amount of atmospheric noise in the data
  generally determines which subtraction algorithm is optimal,
  and the observations optimally cleaned with average
  subtraction were made in the best observing conditions.
  Note that quadratic subtraction is the optimal method
  only when the weather conditions are extremely poor.
  This is because the anisotropy power
  spectrum falls quickly at high frequency,
  and the quadratic subtraction algorithm attenuates a large
  amount of signal at low frequency.
  For point-like sources, whose spectra are flatter, quadratic
  subtraction is the optimal processing method slightly more often.

\section{Final Map Properties}

      Once the FOM is determined for each subtraction method
      for each observation, we can then
      produce a map of all of the data using the optimally
      processed map for each observation.
      To produce this final map, we need to make a slight
      modification to Equation~\ref{eqn:map_obs_coadd3} to 
      account for the transfer function of the data processing
      and the Bolocam beam.
      We need to account for these effects because the transfer
      function depends on the scan direction and optimal 
      sky subtraction algorithm for each observation.
      Therefore, the amount of astronomical signal in the map
      is in general different for each observation.
      To account for the amount of signal attenuation in each 
      observation, the map PSD
      needs to be divided by the transfer function and
      the Fourier transform of the map needs to be divided
      by the square root of the transfer function.
      After making these modifications to Equation~\ref{eqn:map_obs_coadd3},
      we have
      \begin{equation}
	\mathcal{M} = \frac{\sum_i \left( \frac{M_i}{\sqrt{T_i B_i^2}} \right) 
	  \left( \frac{T_i B_i^2}{\mathcal{P}_i} \right) }
	{\sum_j \left( \frac{T_j B_j^2}{\mathcal{P}_j} \right) }
	\label{eqn:opt_sig_map}
      \end{equation}
      as the Fourier transform of the optimal map estimate, $\mathcal{M}$.
      $T_i$ is the transfer function of the data processing for 
      observation $i$ in squared units, $B_i$ is the Bolocam beam profile
      for observation $i$,
      $M_i$ is the Fourier transform of the map from observation $i$,
      and $\mathcal{P}_i$ is the noise PSD for observation $i$
      in squared units.
      Note that the astronomical signal in $\mathcal{M}$ will be 
      equal to the true astronomical signal, because we have 
      divided the Fourier transform of each single observation
      map, $M_i$, by the appropriate attenuation factor,
      $\sqrt{T_i B_i^2}$.\footnote{
        We have not included any phase information in the factor
	$\sqrt{T_i B_i^2}$ because both the signal and the noise
	PSD contain only noise; the phase is irrelevant.}
      However, for some pixels in Fourier space, 
      $T_i$ and/or $B_i$ take on extremely small values,
      which means that some pixels in both the
      numerator and denominator of $\mathcal{M}$ 
      have extremely small values.
      Therefore, before taking the ratio of the numerator
      and denominator in Equation~\ref{eqn:opt_sig_map}
      we apply a regularizing factor, so that
      \begin{equation}
	M' = \sqrt{\mathcal{R}} \mathcal{M} = 
	\frac{\frac{1}{\sqrt{\mathcal{R}}}
	      \sum_i \left( \frac{M_i}{\sqrt{T_i B_i^2}} \right) 
	  \left( \frac{T_i B_i^2}{\mathcal{P}_i} \right) }
	{\frac{1}{\mathcal{R}} 
	  \sum_j \left( \frac{T_j B_j^2}{\mathcal{P}_j} \right) },
	\label{eqn:regularizing}
      \end{equation}
      for
      \begin{equation}
	\sqrt{\mathcal{R}} = \frac{\sum_i \left( \sqrt{T_i B_i^2} \right)
	\left( \frac{T_i B_i^2}{\mathcal{P}_i} \right)}
        {\sum_j 
	\left( \frac{T_j B_j^2}{\mathcal{P}_j} \right)}.
	\label{eqn:regularizing2}
      \end{equation}
      Although $M'$ will be biased (\emph{i.e.}, it is not the 
      Fourier transform of the true map of the sky),
      this bias is accounted for by the final transfer
      function we calculate in Section~\ref{sec:sig_att}.\footnote{
	In Equations~\ref{eqn:regularizing} and 
	\ref{eqn:regularizing2}, $T_iB_i^2/\mathcal{P}_i$ 
	acts as a weighting factor for each observation.
	Therefore, $\mathcal{M}$ represents the
	weighted mean of the Fourier transform of 
	each single observation map divided by the square
	root of the transfer function for that map,
	$\overline{(M/\sqrt{TB^2})}$.
	Similarly, $\sqrt{R}$ represents the weighted
	mean of the square root of the transfer function
	for each observation, $\overline{\sqrt{TB^2}}$.
        So, $M' = 
        (\overline{\sqrt{TB^2}}) (\overline{(M/\sqrt{TB^2})})$,
        which reduces to the weighted mean
	of all the single observation map Fourier
	transforms, $M' \simeq \overline{M}$, in the limit
        that all of the single observation transfer
	functions, $T_iB_i^2$, are the same.}
      Note that $M'$ can be Fourier transformed back to
      map-space to produce a map $m'$,
      although $m'$ will be biased.
      The maps, $m'$, for each science field are given in 
      Figure~\ref{fig:final_maps}.

      \subsection{Noise PSDs}

\begin{figure}
  \plottwo{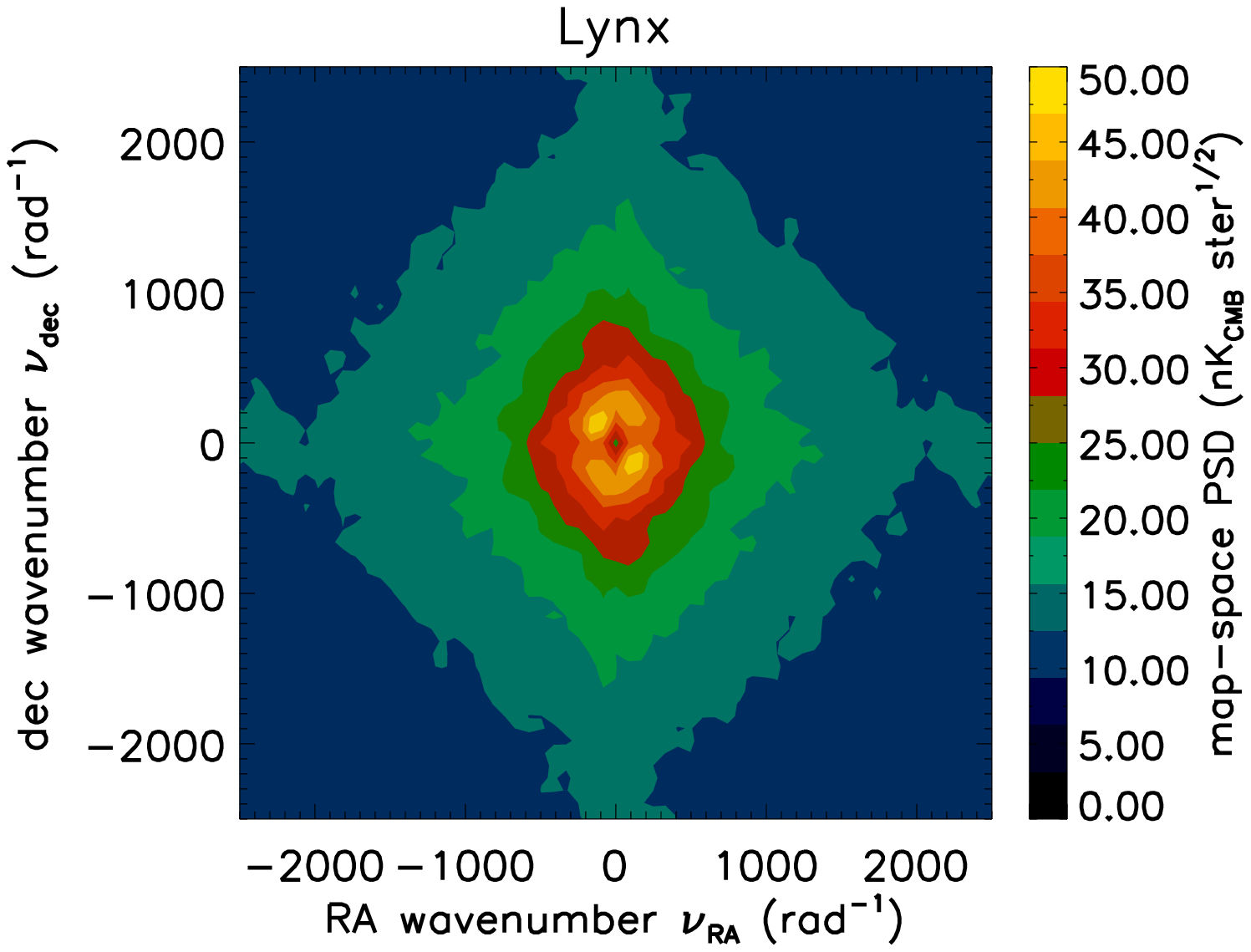}{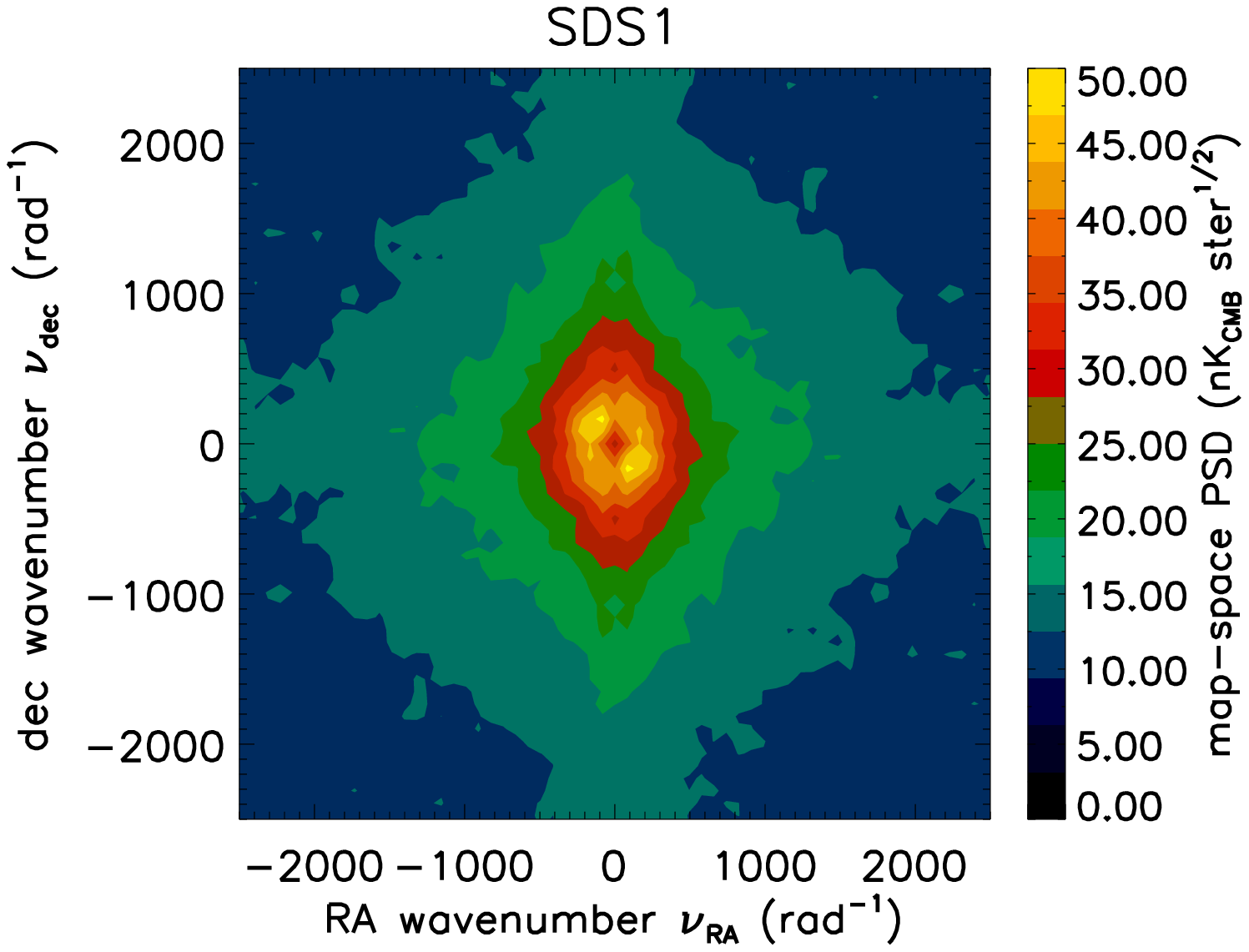}
  \caption{The map-space PSDs, $P_{\vec{\nu}}$,
     of the maps made from co-adding
    all observations for a each science field.
    Note that relative to these PSDs,
    the power spectra of
    any astronomical signals will have been multiplied by
    the transfer functions of the data processing and
    the beam, $T_{\vec{\nu}} B^2_{\vec{\nu}}$.}
  \label{fig:final_psd}
\end{figure}

      Analogous to the case of a single observation, we used
      jackknifed realizations of our data to estimate the 
      noise PSD of $m'$.
      In this case, each realization is 
      generated by multiplying a randomly
      selected set of half the observations in $m'$ by
      $-1$.
      The map-space PSD from
      1000 realizations were averaged to
      determine the best estimate of the
      noise PSD for 
      each science field,
      with the results shown in Figure~\ref{fig:final_psd}.
      In this section we establish that the noise PSD 
      estimated in this way is statistically well-behaved 
      (Gaussian) and unbiased. 
      These characteristics are critical to the remainder of our analysis.

      We analyzed the distribution of individual realization
      PSDs to determine if the underlying probability distribution
      describing the noise is Gaussian.
      As in the single observation case, we computed a dimensionless
      PSD value according to Equation~\ref{eqn:dimless_PSD},
      and compared the distribution of these values to the PDF
      given in Equation~\ref{eqn:PDF_map_PSD}.
      In general, the agreement is good, indicating
      the underlying noise distribution is well approximated
      by a Gaussian.
      See Figure~\ref{fig:final_PSD_PDF}.
      The Gaussianity of the noise PSDs of the jackknife maps is 
      important because it justifies the form of the likelihood 
      function we use, Equation~\ref{eqn:log_l} 
      (presented in Section~\ref{sec:overview_analysis}). 
      That form assumes the Fourier coefficients of the final map 
      are Gaussian-distributed random variables with variance given 
      by the noise PSD estimated from the jackknife maps.


      Next, we show that the noise PSD estimated from the jackknifes 
      is unbiased under two assumptions: 
      1) the covariance of any pair of distinct observations 
      vanishes on average; and 
      2) negligible signal leaks into the jackknife maps. 
      The first assumption is equivalent to the statement that there is 
      no scan-synchronous or fixed-pattern noise in the maps.
      We have checked this assumption empirically and find that the 
      average fractional covariance of distinct observations is 
      $\sim 2 \times 10^{-5}$, which is consistent
      with noise.
      We will discuss below how our non-detection of signal in the final
      map further justifies this assumption. With these assumptions, we can 
      prove lack of bias of the noise estimate in a straightforward fashion 
      by a simulation that obeys the assumptions. 
      We generate 
      a set of $N_{obs}=515$ single-observation maps 
      using the single-observation noise PSDs; these obviously have no 
      signal and are uncorrelated with one another. 
      Then, 
      we construct $N_{jack}=1000$ 
      simulated jackknife final maps and calculate 
      the noise PSD of these maps, which also obviously	   
      have no signal. 
      Next, we average these noise PSD estimates
      over all of the jackknife final
      maps, and divide by the input noise spectrum to determine how
      accurately we have recovered that input spectrum.
      The average (over all Fourier space pixels) of this normalized PSD
      is $0.9997 \pm 0.0006$, showing that we indeed recover the
      input noise spectrum within the measurement uncertainty of 
      our simulation.
      This exercise thus shows that the simulated final map noise PSD 
      estimated by jackknife maps is an unbiased estimate of the 
      simulated final map noise PSD. 
      Note that we do not claim that the 
      noise PSD generated in this fashion is the noise PSD of our true 
      final map; the simulation is
      intended only to show that the jackknife noise 
      PSD estimate method is unbiased.

      Let us now justify the assumption that negligible signal leaks
      into the jackknife final maps.
      In generating jackknife final maps, negative signs were applied to
      exactly one half of the observations.
      There are no fluctuations allowed in the number of negative signs,  
      only in which observations have them applied.
      Therefore, 
      residual signal can arise in the jackknife final maps in only two ways:  
      1) if the relative  
      calibrations of the different observations are imperfectly known or   
      2) if the weights of the 
      different observations are unequal.

      In the first case, consider a single-observation fractional relative  
      calibration error of $\psi$, but assume all the component observations  
      would otherwise be weighted equally 
      (\emph{i.e.}, no variation in noise between  
      observations).  
      This fluctuating relative calibration error does not  
      cause a bias; 
      in an ensemble of experiments, the final map has an  
      expected signal 
      value equal to the 
      signal value of the true final map and the jackknife final  
      map has an expected signal value of exactly zero.  
      But, the calibration fluctuations
      will cause an imperfect coaddition or cancellation
      of the signal
      in any given final map or jackknife final map realization,
      which will produce a fractional spread in the signal level
      of $\psi/\sqrt{N_{obs}}$ relative to the true signal.
      Given that $\psi \lesssim 3$\%\footnote{
	In Section~\ref{sec:flux_cal} we calculated our flux
	calibration uncertainty to be approximately 5.5\%.
	However, most of this uncertainty is due to systematics
	that will not change from one observation to the next.
	The uncertainty caused by fluctuations in the atmospheric
	opacity and the fit of our model are $\simeq 2-3$\%.}
      and $N_{obs} \simeq 500$ observations per map, this error  
      in both the final map and jackknife final maps is very small compared  
      to the signal.  Since we are not attempting a high signal-to-noise  
      measurement, the vanishingly small size of the error relative to  
      the signal is thus not a concern in the final map.  The error  
      affects the jackknifes in a more subtle way because it effectively  
      adds noise to the jackknife final maps,
      which means our noise estimate is slightly higher than the
      true noise level of our final map.
      However, this bias is negligible: the magnitude of the error 
      is of order the signal times $\psi/\sqrt{N_{obs}}$. 
      We know $\psi \lesssim 3$\%, $N_{obs} \simeq 500$, 
      and the signal is less than  2\% of the noise in the final maps 
      (in RMS units, c.f. Section~\ref{sec:mapmaking}), 
      so this bias is $< 0.003$\% 
      of the final map noise level in RMS units, or 
      $< 0.006$\% in variance units. The small size of the effect is 
      not surprising: it is proportional to the signal size, and we have 
      no detection of signal. Moreover, even if the effect were not 
      negligible, it would result in an overestimate of the noise PSD and 
      thus would result in an overly conservative upper limit. 
      This kind of effect would only be problematic if signal were 
      visible at high significance.

      In the second case, the argument is very similar, but now what  
      matters is the fractional variation in observation weight.  
      This fractional variation is large, $\lesssim 2$,
      due to the significant
      differences in atmospheric noise between observations.
      Here, the large number of  
      observations and the fact that the spread is proportional to the  
      signal size render the effect negligible.  
      The RMS spread of the  
      residual signal in the jackknifes will be 
      $\lesssim 2/\sqrt{N_{obs}} = 0.09$ times the  
      signal size.  
      Again, because of the small size of the error relative  
      to the signal and the lack of detected signal, 
      the error due to this  
      effect is insignificant.

      If the noise estimation approach has underestimated the noise 
      (for example, by failing to account for non-stationarity 
      or correlations, or in any other manner) then there  
      will be more noise in the final map than expected from the  
      jackknifes.  
      However, because we find in Section~\ref{sec:CMB_results} that
      our 90\% CL interval on the amplitude of astronomical anisotropy,
      $\hat{A}$, includes $\hat{A} = 0$, we do not see
      any significant excess of noise in the true map above what is 
      expected from the noise PSD estimate.
       This explicitly rules out scan-synchronous or 
      fixed-pattern 
      noise that would be averaged away in jackknife maps but would  
      remain in the coadded map at the level of interest for this  
      analysis.  
      Had there been an excess above the expectation from the noise 
      estimate, we would have had to show more explicitly that the 
      noise estimate was correct in order to claim a detection.

      Alternatively, we consider the effect of overestimating the noise. 
      The resulting final map PSD would be too low to be consistent
      with the noise PSD estimate. In our analysis 
      (see Section~\ref{sec:sci_anal}),
      this would yield a best fit value for the astronomical anisotropy 
      $\hat{A}$ of zero. We do not find 
      this to be true: since the best-fit value of $\hat{A}$ must 
      lie inside the confidence interval of any value,
      and our 68\% CL interval on $\hat{A}$ does 
      not include $\hat{A} = 0$ 
      (see Table~\ref{tab:SZE_result}), the best-fit value of
      $\hat{A}$ must therefore differ from zero.



      \subsection{Astronomical Signal Attenuation}
      \label{sec:sig_att}
      
\begin{figure}
  \plottwo{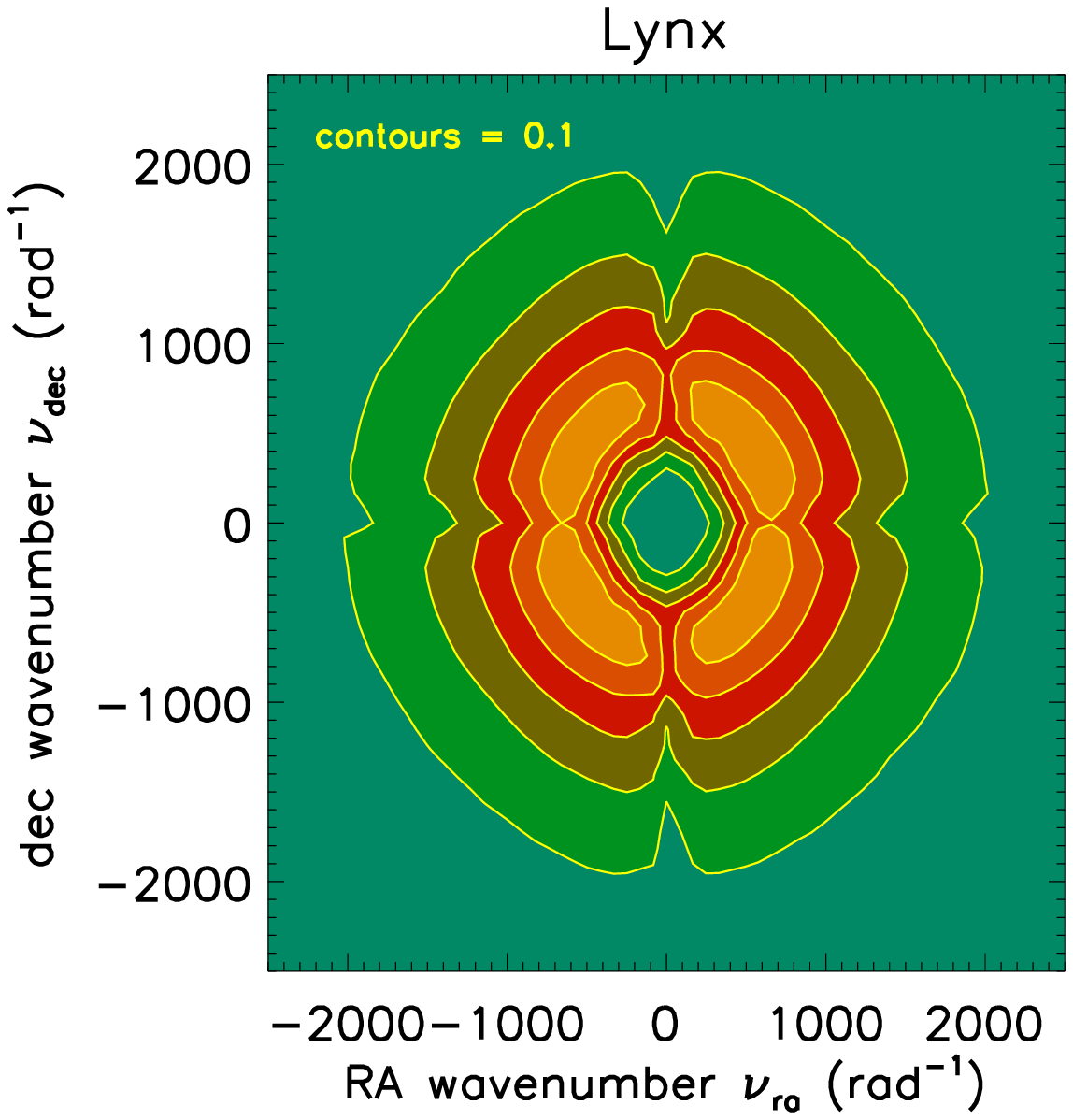}{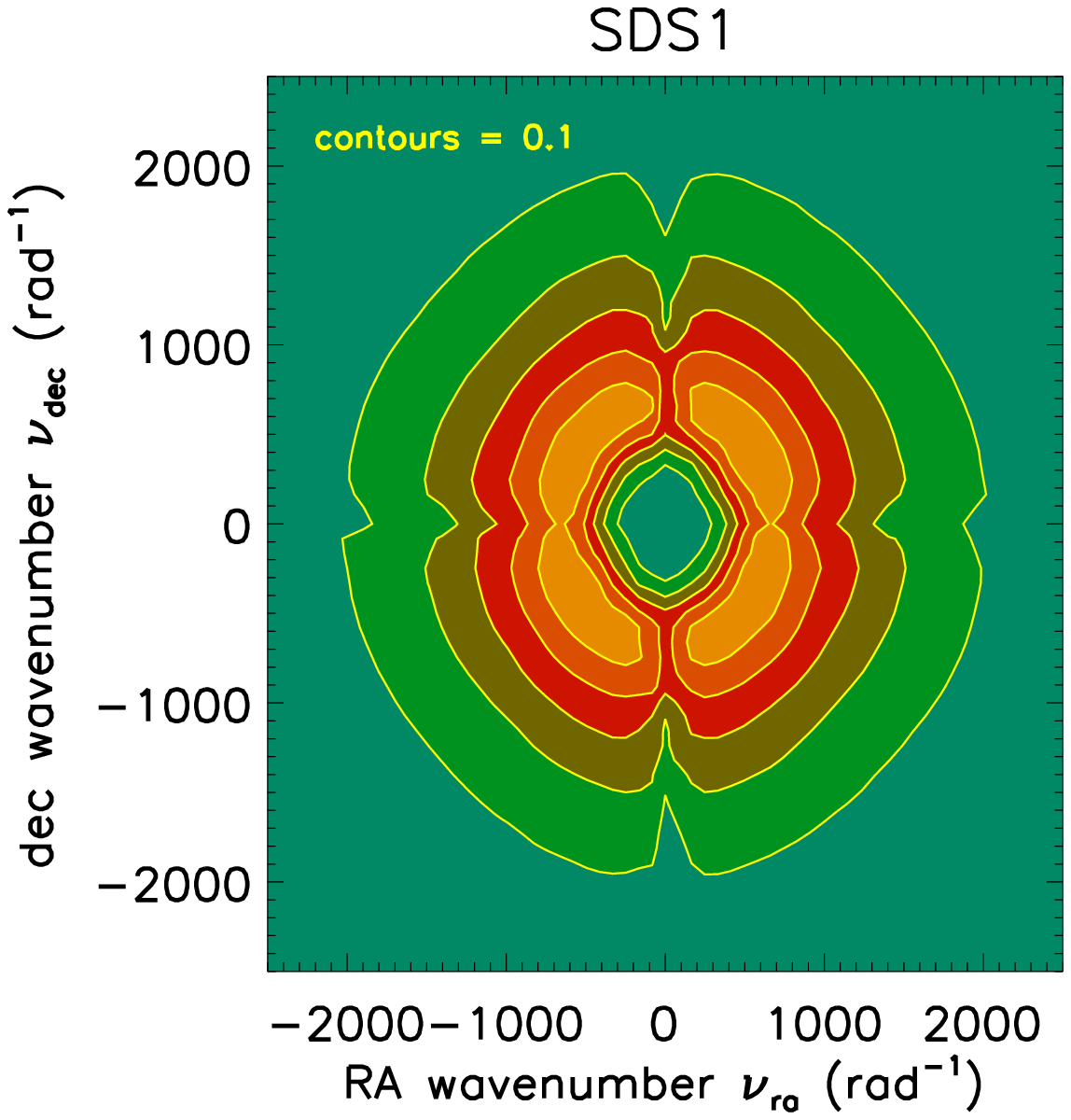}
  \caption{Contour plots showing the 
    transfer functions, $T_{\vec{\nu}} B^2_{\vec{\nu}}$,
    for the maps made from all observations
    of each science field.
    There is slightly more attenuation along the
    $\nu_{RA}$ axis compared to the $\nu_{dec}$ axis in the 
    maps because more observations were taken while scanning
    parallel to RA compared to scanning parallel to dec.}
  \label{fig:final_xfer}
\end{figure}

      Now that the noise properties of the maps are well
      described, we need to determine the amount of astronomical
      signal attenuation due to data processing, the Bolocam
      beam, and the regularizing factor in Equation~\ref{eqn:regularizing}.
      The method for calculating the transfer
      function of the data processing and regularizing factor
      is analogous to the method described in Section~\ref{sec:xfer}
      for single observations.
      Contour plots of the total astronomical signal attenuation
      are given in Figure~\ref{fig:final_xfer}.
      Compared to a single observation, the transfer functions
      for the final maps are much closer to being 
      rotationally symmetric.
      The difference 
      in the transfer functions 
      is at low spatial frequencies parallel to either
      RA or dec, and is caused by adding observations made while
      scanning in perpendicular directions.
      This is because the modes in single observation
      maps, where there is 
      a large amount of astronomical signal
      attenuation (\emph{i.e.}, 
      at low frequency parallel to the scan direction),
      do not contribute much to the final map.
      Therefore, most of the signal at low frequency along the 
      RA direction is obtained from maps made while scanning parallel
      to dec, and vice versa.
      This effect can be seen by comparing the plots in
      Figure~\ref{fig:RA_xfer} with
      the plots in Figure~\ref{fig:final_xfer}.

      \section{Noise from Astronomical Sources}

      \label{sec:astr_noise}

      Since the noise PSD of the final map is estimated from
      jackknifed realizations of the data, all of the 
      astronomical signal will be absent from the noise PSD.
      This is fine for the anisotropy signal we are looking for, 
      because we want to understand the noise of our system
      in the absence of our signal of interest.
      However, we need to estimate 
      the amount of noise produced by sources other than the
      SZE-induced CMB anisotropies,
      including galactic dust emission, radio point-source
      emission, emission from dusty submillimeter galaxies,
      and primary CMB anisotropies.

      The amount of galactic dust emission can be estimated 
      from maps of our science fields taken from
      the full-sky 100~$\mu$m DIRBE/IRAS dust 
      map \citep{dirbe_website, schlegel98}.
      To extrapolate the 100~$\mu$m data to our band at
      143~GHz~$\simeq$~2.1~mm, we have used the
      ``model 8'' extrapolation given in 
      \citet{finkbeiner99}.
      At 100~$\mu$m, the typical surface brightness of the dust
      emission in our science fields
      is just over 1~MJy/ster, which corresponds to 
      a surface brightness of around 5 -- 15~nK$_{CMB}$ for Bolocam.
      Using the maps that have been converted to a 
      thermodynamic temperature at
      143~GHz, we determined the map-space PSD of the 
      dust emission, which corresponds to
      a $\mathcal{C}_{\ell}$ less than $10^{-6}$~$\mu$K$_{CMB}^2$
      for $\ell \gtrsim 1000$.\footnote{
	Note that the resolution of the DIRBE/IRAS dust map
        is 6.1~arcminutes, which corresponds to
	HWHM in $\ell$-space of $\lesssim 2000$.
	Therefore, we have no direct knowledge of the power
	spectrum on scales smaller than $\simeq 6$~arcminutes,
	which are the angular scales Bolocam is most
	sensitive to.
	However, the power spectrum of the dust falls rapidly
	at small angular scales, so the estimate at $\ell < 2000$
	should provide a reasonable upper limit.}
      Since this is well below the expected SZE-induced CMB anisotropy
      we are looking for, it is safe to conclude that the signal 
      from the dust emission in our maps is negligible.

      Emission from radio point sources will also contribute to
      the astronomical signal in our maps.
      The power spectrum from these sources can be calculated
      from
      \begin{equation}
	C_{\ell} = \int_0^{S_{cut}} S^2 N(S) dS + w_{\ell} I^2,
	\label{eqn:ptsrc_ps}
      \end{equation}
      where $S$ is the flux of the source, $N(S)$ is the differential
      number of sources at a given flux in a given solid angle, 
      $S_{cut}$ is an estimate of the source-detection
      threshold in the map (\emph{i.e.}, the level at which
      sources may be detected and removed),
      $C_{\ell}$ is the angular power spectrum,
      $w_{\ell}$ is the Legendre transform of the two-point correlation
      function of the sources, and 
      \begin{displaymath}
	I = \int_0^{S_{cut}} S N(S) dS
      \end{displaymath}
      is the background contributed by the 
      sources \citep{white04, scott99}.
      We will assume $w_{\ell} = 0$, since there is a large amount of
      uncertainty in the clustering of these sources.\footnote{
      Note that the total number of sources and total integrated
      power in $\ell$-space will not change if $w_{\ell}$ is
      non-zero; the clustering modeled by $w_{\ell}$
      will only shift power from high-$\ell$ to low-$\ell$.}
      Differential number counts have been determined from measurements
      at 1.4, 5, and 8.44~GHz \citep{toffolatti98, danese87}, with
      \begin{equation}
	N(S)_{5 {\rm GHz}} = 150 \textrm{ } S^{-2.5} \textrm{  } 
	{\rm Jy}^{-1} {\rm ster}^{-1}.
	\label{eqn:ns_cm}
      \end{equation}
      Since the spectrum of the sources is nearly flat
      (i.e., $S_{\nu} \propto \nu^{\beta}$ with $\beta = 0$), 
      this equation is valid over a 
      wide range of frequencies.
      Additionally, the WMAP K, Ka, and Q bands have been used to
      determine the differential number counts at 22, 30, 
      and 40~GHz \citep{bennett03}.
      $N(S)$ is similar for all three WMAP bands, and is 
      $\lesssim 70$\% of the value of the model in Equation~\ref{eqn:ns_cm}.
      The differential number counts at 40~GHz are described by
      \begin{displaymath}
	N(S)_{40{\rm GHz}} = 32 \textrm{ } S^{-2.7} \textrm{  }
	{\rm Jy}^{-1} {\rm ster}^{-1}.
      \end{displaymath}
      To extrapolate this equation to the Bolocam band center at
      143~GHz, we will use the method described in \citet{white04}.
      Since there is evidence of the power law for $N(S)$ flattening
      out at higher frequencies, they describe the differential
      number counts according to\footnote{
	There is some uncertainty in the spectrum of $S_{\nu}$
	for these radio sources between 40~GHz and 143~GHz.
	White and Majumdar quote two spectra, one with
	$\beta = 0$, and one with $\beta = -0.3$.
	This uncertainty in the spectrum of the radio point sources
	results in a finite range for the normalization of the
	number counts after extrapolating to 143~GHz.}
      \begin{equation}
	N(S)_{143 {\rm GHz}} = (20-32) \textrm{ } S^{-2.3} \textrm{  }
	{\rm Jy}^{-1} {\rm ster}^{-1}.
	\label{eqn:ns_150}
      \end{equation}

      We also need to estimate $S_{cut}$ in order to evaluate the 
      power spectrum in Equation~\ref{eqn:ptsrc_ps}.
      This cutoff flux will necessarily be somewhat arbitrary, but, since
      $C_{\ell}$ is only weakly dependent on $S_{cut}$, it will not
      significantly alter our result.
      We have chosen $S_{cut} = 10$~mJy, which is approximately four
      times the RMS fluctuations per beam in maps made from
      data that have been optimally filtered for point sources.\footnote{
	From Equation~\ref{eqn:ns_150}, we only expect $1-2$ sources
	brighter than 10~mJy in our entire survey of 1 square
	degree, which is why we have not attempted to subtract
	out any sources prior to our anisotropy analysis.
        Additionally, the largest excursions in our maps are
        $\simeq 10$~mJy, further justifying our choice to set
        $S_{cut} = 10$~mJy.}
      Inserting this value of $S_{cut}$ into Equation~\ref{eqn:ptsrc_ps},
      along with Equation~\ref{eqn:ns_150}, yields
      $C_{\ell} \simeq 1.1-1.9$~Jy$^2$~ster$^{-1}$, or
      $C_{\ell} \simeq 7-12 \times 10^{-6}$~$\mu$K$_{CMB}^2$.
      To compare this angular power spectrum to the expected SZE-induced
      CMB anisotropies, we determine the amplitude
      of a flat band power, $\mathcal{C}^{\rm eff}_{\ell}$, 
      that is required
      to cause the same temperature fluctuation as 
      $C_{\ell}$ given our transfer function, $T_{\ell} B_{\ell}^2$,
      according to
      \begin{equation}
	\mathcal{C}_{\ell}^{\rm eff} = \frac{\sum_{\ell}
	  C_{\ell}
	  \frac{2\ell + 1}{4 \pi}
	  T_{\ell} B_{\ell}^2}{
	  \sum_{\ell} \frac{2\pi}{\ell(\ell+1)} \frac{2\ell + 1}{4 \pi} 
	  T_{\ell}B_{\ell}^2}.
	  \label{eqn:c_l_eff}
      \end{equation}
      For the radio point sources with
      $C_{\ell} = 7-12 \times 10^{-6}$~$\mu$K$_{CMB}^2$,
      the effective $\mathcal{C}_{\ell}$ given the Bolocam
      transfer function is
      $\mathcal{C}_{\ell}^{\rm eff} \simeq 35-60$~$\mu$K$_{CMB}^2$,
      which is comparable to the expected signal from the
      SZE-induced CMB anisotropies.

      Additionally, emission from dusty submillimeter galaxies will
      be present in our maps.
      The same method used to determine the power spectrum from
      radio point sources can also be used to
      estimate the power spectrum of these sources.
      We used the number counts distribution
      determined by \citet{aguirre08}, with
      \begin{displaymath}
	N(S)_{268 {\rm GHz}} = 1619 \textrm{ } S^{-2.26} e^{-303 S} 
	\textrm{  } 
	{\rm Jy}^{-1} {\rm ster}^{-1}.
      \end{displaymath}
      The spectrum of these objects can be described by 
      $S_{\nu} \propto \nu^{\beta}$, where 
      $2.5 \lesssim \beta \lesssim 3.5$ 
      \citep{borys03},
      which gives a differential number count at 143~GHz of 
      \begin{displaymath}
	N(S)_{143 {\rm GHz}} = (100-220) \textrm{ } S^{-2.26} 
	e^{-(2730-1460) S} \textrm{  } 
	{\rm Jy}^{-1} {\rm ster}^{-1}.
      \end{displaymath}
      Inserting the above formula into Equation~\ref{eqn:ptsrc_ps}
      gives $C_{\ell} = 0.4-1.2$~Jy$^2$~ster$^{-1}$, or 
      $C_{\ell} = 3 - 9 \times 10^{-6}$~$\mu$K$_{CMB}^2$.
      Equation~\ref{eqn:c_l_eff} can again be used to convert
      this to an effective constant $\mathcal{C}_{\ell}$ for
      our transfer function, giving
      $\mathcal{C}_{\ell}^{\rm eff} \simeq 15-45$~$\mu$K$_{CMB}^2$.
      Alternatively, we can compute a power spectrum
      using the differential number
      counts derived from SHADES data at 350~GHz \citep{coppin06},
      which is described by
      \begin{displaymath}
	N(S)_{350 {\rm GHz}} = 2.2 \times 10^4 \textrm{ } 
	\left[S^2 + (5.9 \times 10^7) S^{5.8} \right]^{-1} \textrm{  } 
	{\rm Jy}^{-1} {\rm ster}^{-1}.
      \end{displaymath}
      Converting this $N(S)$ to a differential number count at 143~GHz
      using the average spectrum of $\nu^3$ 
      yields a similar
      power spectrum,
      with $C_{\ell} = 1.0$~Jy$^2$~ster$^{-1}$, or 
      $C_{\ell} = 8 \times 10^{-6}$~$\mu$K$_{CMB}^2$,
      which is consistent with the result from the number
      counts given by \citet{aguirre08}.


      Finally, there will also be a signal in our map due to
      the primary CMB anisotropies,
      which are distinct from the SZE-induced anisotropies
      we are searching for.
      The power spectrum of the primary CMB anisotropies
      has been measured to high precision by
      WMAP at $\ell \lesssim 800$ \citep{nolta08},
      and by ACBAR at $500 \lesssim \ell \lesssim 2500$
      \citep{reichardt08}.
      This measured power spectrum is well fit by theory, with
      only a small number of free parameters.
      Therefore, we have generated a template of the primary 
      CMB power spectrum using the theoretical prediction generated
      by CMBFAST \citep{seljak96, zaldarriaga98,
      zaldarriaga00},
      with the best fit values to the free parameters from
      the WMAP 5-year data \citep{dunkley08}.
      Since the CMBFAST routine only computes the power spectrum
      up to $\ell = 3000$, we fit a decaying exponential
      to the $\mathcal{C}_{\ell}$ versus $\ell$ to extrapolate
      the primary CMB power spectrum to higher $\ell$.
      We can again use Equation~\ref{eqn:c_l_eff} to convert
      this power spectrum to an effective constant $\mathcal{C}_{\ell}$
      given our transfer function,
      with $\mathcal{C}_{\ell}^{\rm eff} \simeq 45$~$\mu$K$_{CMB}^2$.
      This band power is similar to what is expected
      from the SZE-induced CMB anisotropies.
      A summary of the expected signal from the various
      astronomical sources is given in Figure~\ref{fig:astr_noise}.

  \section{Science Analysis}
\label{sec:sci_anal}


\subsection{Overview of Analyses}
\label{sec:overview_analysis}

In addition to instrumental noise from the 
bolometers, electronics, etc.,
our maps will contain an excess noise from astronomical sources,
including
anisotropies due to primary CMB fluctuations, fluctuations due to the
SZE, and fluctuations due to unresolved astronomical point sources.
It is our goal to constrain the level of
these 
astronomically sourced noises, which we will
specify as the amplitude of flat band power anisotropy power spectrum
contributions
in $\mathcal{C}_\ell$.  To obtain such a constraint, we must calculate
the difference between the observed and expected power spectra of our
maps and obtain a best estimate of the excess noise, goodness-of-fit
of the data to the model, including any possible excess noise, and
confidence intervals for the amount of excess noise.  This section
describes how we obtain the estimate and intervals.

The first analysis we perform will simply constrain the total
astronomical anisotropy in the maps, without any interpretation of the
source, assuming only that the astronomical noise has a spectral shape
flat in $\mathcal{C}_\ell$.

The second analysis will statistically subtract the primary CMB
anisotropy power spectrum by using the precise constraints 
placed on it by a
variety of measurements \citep{reichardt08, nolta08}.  
The result will be a constraint
on the non-primary-CMB contributions to anisotropy, and will be mildly
more sensitive because of the subtraction.  
 We will do this by
adding the expected ``noise'' from the primary CMB to
our model of the instrumental noise.
This expectation will fully take
into account cosmic variance on the primary CMB anisotropy in a manner
that we will explain below.  

In the end, this analysis will yield an upper limit
on the astronomical noise.
Because it yields an upper limit, it is conservative to
immediately interpret the constraint as a limit on SZE anisotropy: if
there are point source contributions, as we expect there are, then the SZE
contribution 
will be smaller than the upper limit we 
obtain by the assumption that the point source contributions are
negligible.  The situation would of course be different, and that
assumption would not be conservative, were we claiming a detection of
excess non-primary-CMB anisotropy.

One could extend this methodology to statistical subtraction of the
non-negligible submillimeter and radio point source contributions, but the
large uncertainties in those contributions as well as the possibly
unknown systematic uncertainties lead us to conclude that the
improvement in sensitivity will be negligible and somewhat
untrustworthy.

\subsection{Deficiencies of a Bayesian Analysis}
\label{sec:bayesian_dificiencies}

      We have chosen to model astronomical anisotropies using a flat
      band power in $\mathcal{C}_\ell$, which corresponds to $C_{\ell}
      = A S_{\vec{\nu}}^2$, for $\ell = 2 \pi |\vec{\nu}|$ and
      $S_{\vec{\nu}}^2 = 2 \pi / \ell(\ell+1)$.
      With these definitions,
      and assuming the noise PSD, $\mathcal{P}_{\vec{\nu}}$, fully
      describes the noise properties of the data for the reasons we
      have explained in Section~\ref{sec:mapmaking}, the best fit
      amplitude for an astronomical anisotropy signal is determined by
      maximizing Equation~\ref{eqn:log_l},
      \begin{displaymath}
        {\rm log}(\mathcal{L}) = \sum_{\vec{\nu} \epsilon V} \left(
	- {\rm log}
	({\mathcal{P}_{\vec{\nu}} + A S_{\vec{\nu}}^2 
	  B_{\vec{\nu}}^2 T_{\vec{\nu}}})
	- \frac{X_{\vec{\nu}}}{
	  {\mathcal{P}_{\vec{\nu}} + A S_{\vec{\nu}}^2 
	   B_{\vec{\nu}}^2 T_{\vec{\nu}}}} 
	\right),
      \end{displaymath}
      with respect to $A$, where $X_{\vec{\nu}}$ is the measured PSD
      of the science field map in squared units, 
      $T_{\vec{\nu}}$ is the transfer
      function of our data processing in squared units, 
      and $B_{\vec{\nu}}$ is the
      profile of our beam.\footnote{ 
        Note that we are calculating the anisotropy amplitude for
        a single bin in $\ell$-space.
	However, the technique can be applied to multiple bins
	in $\ell$-space by windowing the appropriate terms
	in Equation~\ref{eqn:log_l}
        (\emph{i.e.}, if an $\ell$-space bin is described by the
        transfer function $\mathcal{T}_{\vec{\nu}}$, then
        $\mathcal{P}_{\vec{\nu}} \rightarrow \mathcal{T}_{\vec{\nu}}
	  \mathcal{P}_{\vec{\nu}}$, 
	$S_{\vec{\nu}}^2 \rightarrow \mathcal{T}_{\vec{\nu}} 
	S^2_{\vec{\nu}}$, 
	and $X_{\vec{\nu}} 
	\rightarrow \mathcal{T}_{\vec{\nu}} X_{\vec{\nu}}$.)}
      Since our maps are real,
      $X_{\vec{\nu}} = X_{-\vec{\nu}}$, $\mathcal{P}_{\vec{\nu}} =
      \mathcal{P}_{-\vec{\nu}}$, etc., so the sum only
      includes half the 
      $\vec{\nu}$-space pixels, denoted by the set $V$.
      For reference, a detailed
      derivation of the above equation is given in
      Appendix~\ref{sec:map_var}.  
      Note that Equation~\ref{eqn:log_l}
      allows for $A < 0$.  Although such values are not physical,
      fluctuations in the noise can cause the most likely value of $A$
      to be less than zero when the expected value of $A$ is small
      compared to the non-astronomical noise.

      The above expression is incorrect at some level because the
      $\vec{\nu}$-space pixels are slightly correlated,
      approximately
      1 - 4\% for nearest-neighbor pairs of pixels and less than 1\%
      for all other pairs of pixels, while Equation~\ref{eqn:log_l}
      treats all Fourier modes as independent.  
      Note that the correlation 
      function, $c_{\vec{\nu},\vec{\nu}'}$, is largely translation
      invariant ($c_{\vec{\nu},\vec{\nu}'} \approx c(\vec{\nu} -
      \vec{\nu}')$).  
      This error due to pixel correlations raises
      three questions: 1) Does maximization of the likelihood given
      above result in an unbiased estimator of $A$?  2) Is this
      an approximately minimum variance estimator? and 3) Can we derive
      Bayesian credibility intervals on $A$ from it?  
      We have demonstrated using simulations that
      Equation~\ref{eqn:log_l} remains an unbiased and 
      approximately minimum variance
      estimator for $A$ in spite of these correlations, presumably
      because ignoring these fairly uniform correlations does not
      shift the peak of $\mathcal{L}$.
      See Table~\ref{tab:log_l}.
      However, the width of
      $\mathcal{L}$ is certainly dependent on these correlations: we
      are essentially over-counting the number of independent data
      points entering the likelihood and thus assuming more
      statistical power than we really have.

      We can make an approximate, unrigorous correction for the
      effective number of independent modes  
      by calculating
      \begin{displaymath}
      N_{\rm eff} = N_{\rm true} \left( 
        \frac{1}{2 N_{\rm true}} 
	\left( \sum_{\vec{\nu} \epsilon V} \sum_{\vec{\nu}' \epsilon V} 
	c_{\vec{\nu},\vec{\nu}'} + \sum_{\vec{\nu} \epsilon V} 
	c_{\vec{\nu},\vec{\nu}} \right)
	\right)^{-1},
      \end{displaymath}
      where $N_{\rm true}$ is the
      total number of $\vec{\nu}$-space pixels in $V$, $N_{\rm eff}$ is the
      effective number of $\vec{\nu}$-space pixels, and
      $c_{\vec{\nu},\vec{\nu}'}$ is the correlation between pixel
      $\vec{\nu}$ and pixel $\vec{\nu}'$.  
      The factor of 1/2 inside the
      parentheses arises from the fact that we have
      double counted the correlations with our
      sums over $\vec{\nu}$ and $\vec{\nu}'$;
      the factor of $1/N_{\rm true}$ is a normalization factor.
      $c_{\vec{\nu},\vec{\nu}'}$
      is calculated from the Fourier transform of the map, $M$,
      according to
      \begin{displaymath}
	c_{\vec{\nu},\vec{\nu}'} = \left|
	\frac{\left< M_{\vec{\nu}}^{*} M_{\vec{\nu}'} \right>}{
	  \left<|M_{\vec{\nu}}|\right> \left<|M_{\vec{\nu}'}|\right>} \right|,
      \end{displaymath}
      where the averages are taken over jackknife realizations.
      Equation~\ref{eqn:log_l} is then multiplied 
      by $N_{\rm eff}$/$N_{\rm true}$ 
      to account for these correlations when
      calculating the Bayesian likelihood, with $N_{\rm eff}/N_{\rm
      true} \simeq 0.43$ for our data.
      When
      $\ln(\mathcal{L})$ is exponentiated, this scaling factor will
      cause $\mathcal{L}$ to fall off less quickly than it would with
      $N_{\rm eff} = N_{\rm true}$, thereby increasing the width of
      $\mathcal{L}$.

      The lack of rigor behind the above correction implies
      that there will be statistical problems in placing
      constraints using the above likelihood function.  
      Were the above likelihood function correct, we could
      use it to set a $\alpha$\% Bayesian credibility interval
      on the parameter $A$ by finding an interval
      $[A_1,A_2]$, $A_1, A_2 \ge 0$, such that
      \begin{displaymath} 
      \frac{\alpha}{100} = 
      \frac{\int_{A_1}^{A_2} dA\, \mathcal{L}(X_{\vec{\nu}}|A)}
      {\int_0^{\infty} dA\, \mathcal{L}(X_{\vec{\nu}}|A)}
      \end{displaymath}
      where we have assumed a flat prior $A \ge 0$
      because
      of the non-physical nature of $A < 0$.  If the likelihood function's
      width in $A$ is not to be trusted, then such credibility
      intervals are not valid.  Not even a simulation permits
      one to set a credibility interval because $\mathcal{L}$ is
      simply not the correct likelihood, even if its distribution can
      be characterized by simulation.  

      Additionally,
      determining the goodness-of-fit for the best-fit value of $A$,
      $\hat{A}$, will require simulation.  That is, if
      Equation~\ref{eqn:log_l} was correct, then
      we should be able to determine an analytic expression
      for the distribution of $\ln(\mathcal{L})$ for $\hat{A}$ that would
      allow us to calculate the goodness-of-fit of the 
      data to the model.
      But, since the above likelihood 
      function is incorrect, 
      we must simulate an ensemble of measurements,
      with appropriate correlations in the Fourier modes, to determine
      the distribution of $\ln(\mathcal{L})$ for the value
      $\hat{A}$.
      Therefore, the Bayesian approach offers no simplifications
      or reductions in computing time relative to the 
      simulation-based frequentist technique we employ below.

      \subsection{Overview of Frequentist, Feldman-Cousins Analysis
      Technique}

      It is possible to deal with all of the above problems
      approximately correctly with a frequentist technique for
      establishing goodness-of-fit confidence levels and frequentist
      confidence (as opposed to Bayesian credibility) intervals on $A$
      that incorporate the prescriptions of Feldman and Cousins for
      dealing with a physical boundary~\citep{feldman98}.  The
      technique has two main features: \begin{enumerate}

      \item First, we use jackknife maps with signal added based on an
      input value $A_{\rm sim}$ in the physically allowed region
      $A_{\rm sim} \ge 0$ to determine the distribution of
      $\mathcal{L}(X_{\vec{\nu},i}(A_{\rm sim})|A_{\rm sim})$ as
      defined in Equation~\ref{eqn:log_l} for an ensemble of
      experiments with outcomes $X_{\vec{\nu},i}(A_{\rm sim})$ 
      With this distribution,
      we may determine whether $\mathcal{L}(X_{\vec{\nu}}|A_{\rm
      sim})$, the value of $\mathcal{L}$ for the true data and the
      value $A_{\rm sim}$, is among the $\alpha$\% most likely
      outcomes for that input value $A_{\rm sim}$, thereby determining
      a goodness-of-fit confidence level.  In doing this, we make the
      reasonable approximation that, although $\mathcal{L}$ is not a
      rigorously correct likelihood, it maps in a one-to-one,
      monotonic fashion to the true likelihood function
      $\mathcal{L}_{\rm true}$.  Specifically, if we consider two
      realizations $X_{\vec{\nu},1}$ and $X_{\vec{\nu},2}$, we assume
      that the sign of $\ln \mathcal{L}(X_{\vec{\nu},1}|A) - \ln
      \mathcal{L}(X_{\vec{\nu},2}|A)$ is the same as that of $\ln
      \mathcal{L_{\rm true}}(X_{\vec{\nu},1}|A) - \ln \mathcal{L_{\rm
      true}}(X_{\vec{\nu},2}|A)$.  This assumption is far looser than
      the assumption that rescaling $\ln \mathcal{L}$ by $N_{\rm
      eff}/N_{\rm true}$ is correct; we are only assuming that the
      ordering of realizations in $\mathcal{L}$ and $\mathcal{L}_{\rm
      true}$ are the same, even if the numerical values are not the
      same.

      \item Second, we want to define a confidence interval of
      confidence level $\alpha$\% on $A$.  Since we are taking a
      frequentist approach, these confidence intervals are defined to
      include the values of $A$ for which, if $A$ is the true value of
      the anisotropy amplitude, then the observed outcome
      $X_{\vec{\nu}}$ is within the $\alpha$\% most likely
      outcomes (as defined below, a definition that is different than
      the usual likelihood) for that value of $A$.  We use the same
      set of simulations with the following procedure based on the
      Neyman construction as modified by \citet{feldman98}.  
      We now calculate for each simulation
      realization $i$ for each input parameter value $A_{\rm sim}$ the
      ratio
      \begin{displaymath}
      R_i(A_{\rm sim}) = \frac{\mathcal{L}(X_{\vec{\nu},i}(A_{\rm
               sim})|A_{\rm sim})}
               {\mathcal{L}(X_{\vec{\nu},i}(A_{\rm sim})|\hat{A}_i)}
      \end{displaymath}
      where $A_{\rm sim}$ is the simulation input parameter value
      ($A_{\rm sim} \ge 0$) and $\hat{A}_i$ is the best-fit value of $A$
      {in the physically allowed region $A \ge 0$} for the given
      realization $i$.  We order the realizations in order of
      decreasing $R_i(A_{\rm sim})$ until $\alpha$\% of the realizations
      have been included; the value of $R_i(A_{\rm sim})$ defining this
      boundary is denoted by $R_\alpha(A_{\rm sim})$.  The input parameter
      value $A_{\rm sim}$ is then included in the $\alpha$\% confidence
      interval if the likelihood ratio for the real data, $R_{\rm
      data}(A_{\rm sim})$, is among the $\alpha$\% largest $R_i(A_{\rm sim})$
      values, $R_{\rm data}(A_{\rm sim}) \ge R_\alpha(A_{\rm sim})$.  The
      interpretation is that, for values $A_{\rm sim}$ belonging to the
      confidence interval of confidence level $\alpha$\%, the data is
      among the $\alpha$\% most likely outcomes, where ``likely'' is
      quantified by $R_{\rm data}(A_{\rm sim})$ {instead of}
      $\mathcal{L}(X_{\vec{\nu},i}|A_{\rm sim})$.

      The above procedure can be conveniently visualized as follows.
      The simulations indicate that there is a smooth relationship
      between $R_i(A_{\rm sim})$ and $\hat{A}_i$ at a given value of
      $A_{\rm sim}$.  This is generically true, not specific to this
      analysis.  Therefore, each simulation realization may be
      labeled by its value of $\hat{A}$ and we may write $R(A_{\rm
      sim},\hat{A})$ in place of $R_i(A_{\rm sim})$.  We may visualize
      $R(A_{\rm sim},\hat{A})$ as a function of $\hat{A}$ for a given
      value of $A_{\rm sim}$; the cutoff value $R_\alpha(A_{\rm sim})$
      is a horizontal line in this plot, and so points with $R(A_{\rm
      sim},\hat{A}) > R_\alpha(A_{\rm sim})$ map to a set of intervals
      in $\hat{A}$; in fact, in our case, there is a single interval
      for each $A_{\rm sim}$.  These intervals, called {confidence
      belts}, can be displayed as intervals $[\hat{A}_1, \hat{A}_2]$
      at the given value $A_{\rm sim}$ in a plot of $A_{\rm sim}$
      vs. $\hat{A}$, as illustrated in
      Figure~\ref{fig:full_conf_belts}.  Do not confuse
      confidence belts, which are intervals along the $\hat{A}$ axis,
      with confidence intervals, which are intervals along the $A_{\rm
      sim}$ axis as defined below.

      Then, to determine the confidence interval of confidence level
      $\alpha$\% on $A$ given a data set $X_{\vec{\nu}}$, one finds
      $\hat{A}_{\rm data}$, draws a vertical line on the plot of
      confidence belts at the value $\hat{A}_{\rm data}$ on the
      horizontal ($\hat{A}$) axis, and includes all values of
      $A_{\rm sim}$ for which the vertical line lies inside the confidence
      belt at that value of $A_{\rm sim}$.  This confidence belt
      construction is equivalent to the above description based on
      $R(A_{\rm sim},\hat{A})$ because the smooth relationship between
      $R(A_{\rm sim},\hat{A})$ and $\hat{A}$ ensures that, for a given
      $A_{\rm sim}$, if $\hat{A}_{\rm data}$ is inside the confidence belt
      at a given value $A_{\rm sim}$, then $R_{\rm data} \ge
      R_{\alpha}(A_{\rm sim})$ for that value $A_{\rm sim}$.

      \end{enumerate}
      
      We comment on two important aspects of this construction of the
      confidence intervals.  First is the ordering of the simulation
      realizations by $R$, not by $\mathcal{L}(X_{\vec{\nu},i}(A_{\rm
      sim})|A_{\rm sim})$.  Feldman and Cousins discuss both possible
      constructions (the latter originally proposed by~\citet{crow59}) 
      and argue that the latter has a
      serious deficiency in that it ties the confidence level of the
      confidence interval to the goodness-of-fit confidence level;
      essentially, it is possible for the confidence interval to not
      provide the advertised frequentist coverage if the
      goodness-of-fit is poor.  This typically happens when the
      experimental outcomes yield best-fit parameter values near or
      outside a physical boundary.  In our application, this can occur
      if the simulation realization has a bit less anisotropy than
      expected, which would yield a best-fit $\hat{A}$ that is
      negative.  In such a case, the (approximate) likelihood of the
      data set, $\mathcal{L}(X_{\vec{\nu}} | A_{\rm sim})$, will in
      general be small.  However, the approximate likelihood of that
      data set may not be small compared to the approximate
      likelihood, $\mathcal{L}(X_{\vec{\nu}} | 0)$, of the most
      probable physically allowed alternative hypothesis of $\hat{A}_i
      = 0$. Feldman and Cousins show that, with this ordering
      principle, the confidence intervals never contain unphysical
      values for the observable.  Additionally, there is a smooth
      transition from the case of an upper limit to a central
      confidence region, eliminating intervals that under-cover due to
      choosing between an upper limit and a central region based on
      the result.  There is not room here to reproduce their arguments
      in detail, we refer the reader to \citet{feldman98}.

      The second important aspect is that the construction is done
      entirely by simulation so that the only way in which we depend
      on $\mathcal{L}$, which we know to be deficient, is in the
      ordering it provides.  We have assumed above that, in spite of
      its inaccuracy, $\mathcal{L}$ provides the same ordering of
      points as $\mathcal{L}_{\rm true}$, and hence $\hat{A}$ and
      $R(A_{\rm sim},\hat{A})$ for a given (simulated or real) data
      realization and $R_\alpha(A_{\rm sim})$ for a given $A_{\rm
      sim}$ will be the same regardless of whether we use
      $\mathcal{L}$ or $\mathcal{L}_{\rm true}$.

      \subsection{Construction of Simulated Data Sets for
      Frequentist Technique}
      
      To apply this method to our data, we first create a simulated
      map of the astronomical anisotropy for a given value of the
      astronomical anisotropy amplitude, $A_{\rm sim}$, using our assumed
      profile $S_{\vec{\nu}}^2$.  This simulation is produced by drawing
      a value for each pixel, $\vec{\nu}$, from an underlying Gaussian
      distribution,\footnote{
        We have also determined confidence intervals using 
	non-Gaussian distributions for the SZE-induced anisotropy
	signal. 
	The results are described in Section~\ref{sec:SZE_results}.}
      then multiplying it by $A_{\rm sim} S_{\vec{\nu}}^2$.
      The PSD of this simulated map is multiplied by our full transfer
      function and added to the jackknifed realization of our data,
      $X_{i,\vec{\nu}}$.\footnote{ The reason we add the simulated
      astronomical anisotropy map to the jackknifed realization map
      instead of the time-streams is to reduce the amount of
      computational time required.  Since the transfer functions of
      the maps are well measured, there is no reason to 
      go all the way back to the time-streams to add the simulated
      signal.}
       Note that a
      different simulated map is created for each jackknifed
      realization of the data to allow for cosmic variance.  Then,
      we use Equation~\ref{eqn:log_l} to determine the most likely
      value of the astronomical anisotropy amplitude, $\hat{A}_{i}$,
      for realization $i$.  By using jackknifes of our actual data, we
      are including all of the correlations between pixels, and by
      simulating the astronomical anisotropy maps we are accounting
      for cosmic variance in the astronomical anisotropy.  For a given
      value of $A_{\rm sim}$, we repeat this process for each jackknifed
      realization of the data.

      The data sets are then ordered based on the ratio of their
      likelihood to the likelihood of the most probable physically
      allowed outcome, $R_i(A_{\rm sim})$, as defined above. The procedure
      outlined in the previous section for defining
      $R_\alpha(A_{\rm sim})$, finding confidence belts for each
      $A_{\rm sim}$, and then determining a confidence interval in
      $A_{\rm sim}$ is then employed as described.

      To determine the goodness-of-fit of our data to the
      model given by $S_{\vec{\nu}}^2$, we compare the 
      likelihood of the actual data at the best fit value
      of $\hat{A}_{data}$, $\mathcal{L}(X_{\vec{\nu}}|\hat{A}_{data})$,
      to the likelihoods of a set of jackknifed realizations
      of our data with simulated spectra added according to
      $S_{\vec{\nu}}^2$ with amplitude $\hat{A}_{data}$,
      $\mathcal{L}(X_{\vec{\nu},i}(\hat{A}_{data})|
      \hat{A}_{data})$.
      For the observations of the Lynx field 
      $\mathcal{L}(X_{\vec{\nu}}|\hat{A}_{data})$ is greater than
      $\mathcal{L}(X_{\vec{\nu},i}(\hat{A}_{data})|
      \hat{A}_{data})$ for 17\% of the realizations,
      and for the observations of the SDS1 field
      $\mathcal{L}(X_{\vec{\nu}}|\hat{A}_{data})$ is greater than
      $\mathcal{L}(X_{\vec{\nu},i}(\hat{A}_{data})|
      \hat{A}_{data})$ for 43\% of the realizations.
      Therefore, we can conclude that our model provides
      an adequate description of the data.

    \subsection{Total Anisotropy Amplitude Results}

    \label{sec:CMB_results}

      To determine the confidence intervals for the full data set,
      we make a joint estimate of $A$ using both the 
      Lynx and SDS1 data sets.
      A plot of the Bayesian likelihood, along with
      confidence belts computed using the Feldman and
      Cousins method are given in Figure~\ref{fig:full_conf_belts}.
      Uncertainties in our pointing model have already been included 
      in these calculations by an effective broadening of the
      Bolocam beam.
      Our upper limits on the total anisotropy amplitude are equal to
      590, 760, and 830~$\mu$K$^2_{CMB}$ at confidence levels
      of 68\%, 90\%, and 95\%.
      Note that the uncertainty on these limits due to the finite
      number of simulations we have run is $\simeq 10-15$~$\mu$K$_{CMB}^2$.
      
      To determine the effective angular scale of our
      anisotropy amplitude measurements we have computed
      our band power window function, $W^B_{\ell}/\ell$,\footnote{
	This band power window function is defined such that $
	   \left< \mathcal{C}_B \right> = \sum_{\ell}
	   (W^B_{\ell}/\ell) \mathcal{C}_{\ell}
	$, where $\left< \mathcal{C}_B \right>$ is the 
        experimental band power measurement for 
        the power spectrum, $\mathcal{C}_{\ell}$.
	Note that the transfer function of our data
	processing, $T_{\ell}$,
	is not the same as the band power window function,
	$W_{\ell}^B/\ell$.}
      using
      the method given by \citet{knox99}.
      A plot of the peak-normalized band power window function
      for the full data set is given in Figure~\ref{fig:knox}.
      From this band power window function we have calculated
      an effective angular multipole for our data
      set, $\ell_{eff}$, given by
      \begin{displaymath}
	\ell_{eff} = \frac{\sum_{\ell} \ell (W^B_{\ell}/\ell)}
	{\sum_{\ell} W^B_{\ell}/\ell},
      \end{displaymath}
      and equal to 5700.
      Additionally, the full-width half-maximum of the window
      function, FWHM$_{\ell}$, is equal to 2800.
      A plot comparing our result to other measurements of the
      CMB on similar scales is shown in Figure~\ref{fig:cmb_power}.

    \subsection{SZE-Induced CMB Anisotropy Results and Constraints on 
    $\sigma_8$}
    \label{sec:SZE_results}

      In order to determine the amplitude of the SZE-induced
      CMB power spectrum, we follow the same methods described 
      above
      to determine the total
      amplitude of the anisotropy power spectrum.
      However, we now have to statistically subtract the
      signal due to the primary CMB anisotropies by accounting
      for both the amplitude and fluctuations of its 
      expected power spectrum in the likelihood;
      these primary CMB anisotropies are effectively
      an additional noise in the map.
      The noise contributed to the map from the Bolocam
      system is given by $\mathcal{P}_{\vec{\nu}}$.
      Since the spectrum of the primary anisotropies in the CMB 
      is well understood,
      we can calculate the
      expected noise from the primary CMB anisotropies.
      To calculate this noise
      we first create a simulated map of the primary CMB, assuming
      that the underlying distribution of 
      $\vec{\nu}$-space pixel values is Gaussian.
      This simulation is produced by drawing a value for each
      pixel, $\vec{\nu}$, from an underlying Gaussian distribution,
      then multiplying it by the best fit primary CMB
      spectrum given in Section~\ref{sec:astr_noise}.
      The PSD of this map is then multiplied by $T_{\vec{\nu}} B_{\vec{\nu}}^2$
      and added to a jackknifed realization of our data, $X_{i,\vec{\nu}}$,
      to give $X_{i,\vec{\nu}}^{[SZE]}$.
      A different simulated map is generated for each
      jackknifed realization of the data to account for
      the cosmic variance in the CMB spectrum.
      These modified jackknifed realizations of the data
      are then be used to determine the expected PSD, 
      $\mathcal{P}_{\vec{\nu}}^{ \left[ SZE \right] }$,
      for the noise contributed by the Bolocam system
      and the primary CMB anisotropies.
      We note that such a simulation of the primary CMB
      contribution is more correct than simply adding the
      primary CMB power spectrum to the non-astronomical noise
      power spectrum because it correctly reproduces 
      pixelization and Fourier-mode correlation effects.

      Next, we select a model spectrum for the SZE anisotropies,
      $S_{\vec{\nu}}^{[SZE]}$.
      Using these new definitions, the 
      Bayesian likelihood function in Equation~\ref{eqn:log_l}
      can be written as
      \begin{displaymath}
        {\rm log}(\mathcal{L}) = \sum_{\vec{\nu}} \left(
	- {\rm log}
	({\mathcal{P}_{\vec{\nu}}^{[SZE]} + 
	  A^{[SZE]} (S_{\vec{\nu}}^{[SZE]})^2 B_{\vec{\nu}}^2 T_{\vec{\nu}}})
	- \frac{X_{\vec{\nu}}^{[SZE]}}{
	  {\mathcal{P}_{\vec{\nu}}^{[SZE]} + 
	    A^{[SZE]} 
	    (S_{\vec{\nu}}^{[SZE]})^2 B_{\vec{\nu}}^2 T_{\vec{\nu}}}} 
	\right),
      \end{displaymath}
      where $A^{[SZE]}$ is the amplitude of the SZE-induced
      CMB anisotropies, $B_{\vec{\nu}}^2$  is the profile of our beam,
      and $T_{\vec{\nu}}$
      is the transfer function of our data processing in squared units.
      As before, we create simulated SZE maps with an amplitude
      $A_{sim}^{[SZE]}$, add these to our jackknifed realizations
      after multiplying by $(S_{\vec{\nu}}^{[SZE]})^2 
      B^2_{\vec{\nu}} T_{\vec{\nu}}$,
      then use the ordering method developed by
      \citet{feldman98} to determine
      the width of the confidence belt at $A_{sim}^{[SZE]}$.
      By repeating this procedure for a range of physically
      allowed values of $A_{sim}^{[SZE]}$, we can construct 
      a full confidence belt that can be used
      to determine our confidence intervals.
      We emphasize that, while we use the Bayesian likelihood to
      construct our best estimator for $A$, a procedure that
      we have already demonstrated by simulation is unbiased and
      approximately 
      minimum variance, we in no way rely on the Bayesian likelihood
      to determine confidence intervals on $A$.
      The frequentist Feldman-Cousins method is used for the
      latter task.

      Additionally, we need to account for the flux calibration
      uncertainty.
      The uncertainty in the flux calibration model derived from
      point sources is 5.5\%, and the uncertainty in the area
      of our beam is 3.1\%.
      Therefore, the uncertainty in our surface brightness
      calibration is 6.3\%.
      To determine the effect of this flux calibration error
      on our confidence intervals, we multiplied each
      simulated primary and SZE-induced CMB  map by
      $\phi_i = 1 + y$, where $y$ is drawn from
      a Gaussian distribution with a standard deviation
      equal to our flux uncertainty of 0.063.
      A different $\phi_i$ was generated for each simulated
      CMB map.
      This means that each simulated map has a different
      flux calibration, distributed 
      according to our uncertainty in the calibration.
      New confidence belts were then calculated using 
      the same procedure described above.
      We have also determined the confidence intervals
      assuming that there is no uncertainty in the 
      known flux of Uranus and Neptune
      (\emph{i.e.}, the only flux calibration uncertainties are due
      to our measurement errors and observational
      techniques).
      In this case, the flux calibration uncertainty is
      3.5\% instead of 6.3\%.
      
      These flux calibration uncertainties produce
      non-negligible changes to the confidence intervals we determine
      for the anisotropy amplitude since it is a variance 
      (\emph{i.e.}, it depends
      quadratically on the flux calibration).
      Therefore, for a simulated amplitude $A_{sim}$,
      a fractional flux calibration uncertainty of
      $\sigma_f$ will increase/decrease the upper/lower
      bounds of the 68\% CL confidence belt by an amount
      roughly proportional to $A_{sim} (1 + \sigma_f)^2$.
      The resulting fractional change to the 
      confidence interval limits
      will in general be non-trivial, but should be
      approximately equal to $(1 + \sigma_f)^2$ for
      68\% CL limits.
      So, for a 3.5\% flux calibration uncertainty
      we expect the 68/90/95\% CL upper limits to increase
      by approximately 7/12/14\% compared to the
      case of no flux calibration uncertainty.
      Similarly, for a 6.3\% flux calibration
      uncertainty we expect the 68/90/95\% CL upper
      limits to increase by approximately
      13/22/27\% compared to the case of no
      flux calibration uncertainty.
      After fully simulating the effect of the flux calibration
      uncertainty on our upper limits, we find results 
      that are comparable to the predictions given above.
      See Table~\ref{tab:SZE_result}.

      We have computed confidence intervals for two different
      SZE spectra: a flat
      spectrum, $(S_{\vec{\nu}}^{[SZE]})^2 = 2\pi/\ell(\ell+1)$
      for $\ell = 2\pi|\vec{\nu}|$ and
      the analytic spectrum calculated by \citet{komatsu02}.
      The results for both of these spectra are similar,
      which is reasonable since the analytic spectrum
      is nearly flat at the scales to which we are most 
      sensitive ($4000 \lesssim \ell \lesssim 7000$).
      See Table~\ref{tab:SZE_result}.
      In addition to the analytic spectrum calculated
      by Komatsu and Seljak, several SZE power
      spectra have been determined via hydrodynamic simulations using
      either MMH (moving-mesh 
      hydrodynamic) 
      or SPH
      (smoothed-particle hydrodynamic) algorithms.
      Examples of MMH simulations can be found in 
      \citet{zhang02}, \citet{seljak01},
      \citet{refregier00}, and \citet{refregier00_2}.
      Examples of SPH simulations can be found in
      \citet{dasilva01} and \citet{springel01}.
      Since most of the simulated  SZE spectra 
      are approximately flat at the 
      angular scales we are most sensitive to, we have
      not determined confidence levels using any
      of these spectra.
      See Figure 1 in \citet{komatsu02}.

    Komatsu and Seljak determined that the amplitude
    of the SZE-induced CMB anisotropies scales
    according to $\sigma_8^7 (\Omega_b h)^2$
    and is relatively insensitive to all other
    cosmological parameters \citep{komatsu02}.
    Using the results from the WMAP 5-year data,
    the best fit values for $\sigma_8$,
    $\Omega_b$, and $h$ are
    0.796, 0.0440, and 0.719 \citep{dunkley08}.
    These values produce a maximum SZE anisotropy
    amplitude of less than 10~$\mu$K$_{CMB}^2$
    at our band center of 143~GHz for the 
    analytic Komatsu and Seljak spectrum.
    For comparison, the 90\% confidence level
    upper limit on the average value of the analytic spectrum 
    weighted by the Bolocam transfer function 
    is 950~$\mu$K$_{CMB}^2$,
    including our flux calibration error.
    See Table~\ref{tab:SZE_result}.
    Based on this upper limit,
    assuming the scaling relation given
    by Komatsu and Seljak and holding
    all other parameters fixed, the corresponding
    90\% confidence level upper limit on the 
    three cosmological parameters is
    $\sigma_8^7 (\Omega_b h)^2 < 2.13$.
    Individually, the best constraint can be placed
    on $\sigma_8$ since the amplitude depends most
    strongly on this parameter, with
    $\sigma_8 < 1.55$ at a confidence level
    of 90\%.
    
    However, this upper limit has been derived by assuming
    the SZE-induced anisotropy signal is Gaussian, which
    is a poor assumption.
    To account for the non-Gaussianity of the signal, we
    have used a method similar to the one described by
    \citet{goldstein03} to analyze data collected with
    ACBAR.
    Based on the results of numerical simulations by 
    \citet{white02} and \citet{zhang02}, along with
    calculations of the trispectrum term from
    \citet{cooray01} and \citet{komatsu02},
    they determined that the sample variance of the 
    SZE-induced anisotropy signal should be a 
    factor of three larger than the Gaussian
    equivalent for the $\ell$-range that ACBAR
    is most sensitive to.
    For our data, at $\ell \simeq 6000$, the sample
    variance is approximately four times
    larger than the expectation for a Gaussian.
    When we account for this increased sample variance
    our 68\%, 90\%, and 95\% confidence level upper limits for the average 
    amplitude of the Komatsu and Seljak spectrum 
    are 790, 1060, and 1080~$\mu$K$_{CMB}^2$, which are
    approximately 10\% higher than the upper limits obtained
    from assuming the fluctuations in the SZE anisotropy signal
    are Gaussian.
    The changes in the upper limits we determine are relatively
    minor because our uncertainty is dominated by 
    Gaussian instrument noise rather than sample variance
    on the anisotropy signal.
    When we convert these upper limits on the anisotropy signal
    to an upper limit on $\sigma_8$, we find $\sigma_8 < 1.57$
    at a 90\% confidence level. 

    \section{Conclusions}

We have surveyed two science fields totaling one square degree with
Bolocam at 2.1 mm to search for secondary CMB anisotropies caused by
the Sunyaev-Zel'dovich effect.  The fields are in the Lynx and
Subaru/XMM SDS1 fields.  Our survey is sensitive to angular scales
with an effective angular multipole of $\ell_{eff} = 5700$ with
FWHM$_{\ell} = 2800$ and has an angular resolution of 60 arcseconds
FWHM.  Our data provide 
no evidence for anisotropy.  We are able to constrain the level of
total astronomical anisotropy, modeled as a flat band power in
$\mathcal{C}_\ell$, with frequentist 68\%, 90\%, and 95\% CL upper
limits of 560, 760, and 830 $\mu K_{CMB}^2$.  We statistically
subtract the known contribution from primary CMB anisotropy, including
cosmic variance, to obtain constraints on the SZE anisotropy
contribution.  Now including flux calibration uncertainty, our
frequentist 68\%, 90\% and 95\% CL upper limits on a flat band power in
$\mathcal{C}_\ell$ are 690, 960, and 1000 $\mu K_{CMB}^2$.  When we
instead employ the analytic spectrum suggested by \citet{komatsu02},
and account for the non-Gaussianity of the SZE anisotropy signal,
we obtain upper limits on the average amplitude of their spectrum
weighted by our transfer function of 790, 1060, and 1080 $\mu
K_{CMB}^2$.  We obtain a 90\% CL upper limit on $\sigma_8$, which
normalizes the power spectrum of density fluctuations, of 1.57.  These
are the first constraints on anisotropy and $\sigma_8$ from survey
data at these angular scales at frequencies near 150~GHz.
 
To calibrate the observations, beam maps were obtained using Uranus
and Neptune.  Pointing reconstruction was performed using frequent
pointing observations of bright sources near our science fields.  The
data were flux-calibrated using techniques similar to those developed
to analyze earlier Bolocam survey data collected at 1.1
mm~\citep{laurent05}, using Uranus and Neptune as absolute calibrators
and a number of other sources as transfer calibrators.  Internal
uncertainty on the pointing and flux calibration contributes
negligible uncertainty to the final result; calibration uncertainty in
the final result is dominated by uncertainty in models for the
absolute brightness temperatures of Mars, Uranus, and Neptune.
  
Our time-streams are dominated by fluctuations in atmospheric thermal
emission (sky noise) and we developed several algorithms to subtract
this noise from our data.  We made use of our simple yet cross-linked
scan strategy to develop a pseudo least-squares map-maker that can be
run in moderate amounts of time on a single desktop computer.  We used
simulations to calibrate the transfer function of our data-taking and
analysis pipeline and map-maker.  We determined the expected noise
properties of our final maps using jackknife realizations of the data
obtained by randomly signed combinations of the $\sim$500 independent
observations contributing to each science field map.  Our final
confidence intervals on anisotropy level are determined using these
jackknife realizations combined with the measured transfer function
for anisotropies.

\section{Acknowledgements}

We acknowledge the assistance of: Minhee Yun and Anthony D. Turner of
NASA's Jet Propulsion Laboratory, who fabricated the Bolocam science
array; Toshiro Hatake of the JPL electronic packaging group, who
wirebonded the array; Marty Gould of Zen Machine and Ricardo Paniagua
and the Caltech PMA/GPS Instrument Shop, who fabricated much of the
Bolocam hardware; Carole Tucker of Cardiff University, who tested
metal-mesh reflective filters used in Bolocam; Ben Knowles of
the University of Colorado, who contributed to the software pipeline,
the day crew and Hilo
staff of the Caltech Submillimeter Observatory, who provided
invaluable assistance during commissioning and data-taking for this
survey data set; and Kathy Deniston, who provided effective
administrative support at Caltech.  Bolocam was constructed and
commissioned using funds from NSF/AST-9618798, NSF/AST-0098737,
NSF/AST-9980846, NSF/AST-0229008, and NSF/AST-0206158.  JS and GL
were partially supported by
NASA Graduate Student Research Fellowships and
SG was partially supported by a R.~A.~Millikan Postdoctoral Fellowship
at Caltech.




{\it Facilities:} \facility{CSO}.



\appendix

\section{Appendix material}
  \label{sec:map_var}

    The goal of our analysis is to determine the
    amplitude of the power spectrum due to emission from astronomical
    sources by measuring
    an excess noise
    in the maps of the science fields.
    This excess noise is the difference between the
    actual noise of the map, and the
    expected noise of the map based on measurements
    of the noise in the Bolocam system and knowledge
    of the expected signal spectrum.
    Therefore, we need measurements of the following quantities:
    \begin{itemize}
      \item $X_{\vec{\nu}}$: The measured PSD of the science field map
	at pixel $\vec{\nu}$ in units of $\mu$K$_{CMB}^2$.
	$\vec{\nu}$ is a two-dimensional value,
	$\vec{\nu} = (\nu_{RA}, \nu_{dec})$, describing a location
	in the spatial Fourier transform of the map,
	and has units of 1/radians.
      \item $\mathcal{P}_{\vec{\nu}}$: 
        The predicted PSD of the science field map
	at pixel $\vec{\nu}$ in the absence of
	the desired astronomical signal.
	$\mathcal{P}_{\vec{\nu}}$ is estimated from jackknife
	realizations, along with the PSDs of unwanted astronomical
	sources in the map (\emph{i.e.}, primary CMB anisotropies in our case).
      \item $\mathcal{S}_{\vec{\nu}}^2$: The spatial power spectral profile
	of the expected astronomical signal.  For a flat band power
	$\mathcal{S}_{\vec{\nu}}^2 = 2 \pi / (\ell(\ell+1))$,
	where the angular multipole $\ell$ is 
	described by $\ell = 2 \pi |\vec{\nu}|$.
      \item $B_{\vec{\nu}}^2$: The peak-normalized square of the
	$\vec{\nu}$-space
	Bolocam beam profile.  
	Since astronomical signals are attenuated by the beam,
	$B_{\vec{\nu}}^2$ acts like a transfer function or filter.  
	Note that the broadening of the beam in 
	map-space due to our pointing uncertainty is included
	in $B_{\vec{\nu}}^2$.
      \item $T_{\vec{\nu}}$: The effective transfer function, or window
	function, of the data processing applied to the
	time-stream data.  Analogous to $B_{\vec{\nu}}^2$, 
	$T_{\vec{\nu}}$ describes how much astronomical signal
	is attenuated.

    \end{itemize}
    With this convention, the expected PSD of the map can
    be described by 
    \begin{equation}
      \left< X_{\vec{\nu}} \right> = 
      \mathcal{P}_{\vec{\nu}} + A S^2_{\vec{\nu}} B^2_{\vec{\nu}} 
        T_{\vec{\nu}},
      \label{eqn:exp_map_var}
    \end{equation}
    where $A$ is the amplitude of the excess anisotropy power, 
    in $\mu$K$_{CMB}^2$.

\begin{figure}
  \plotone{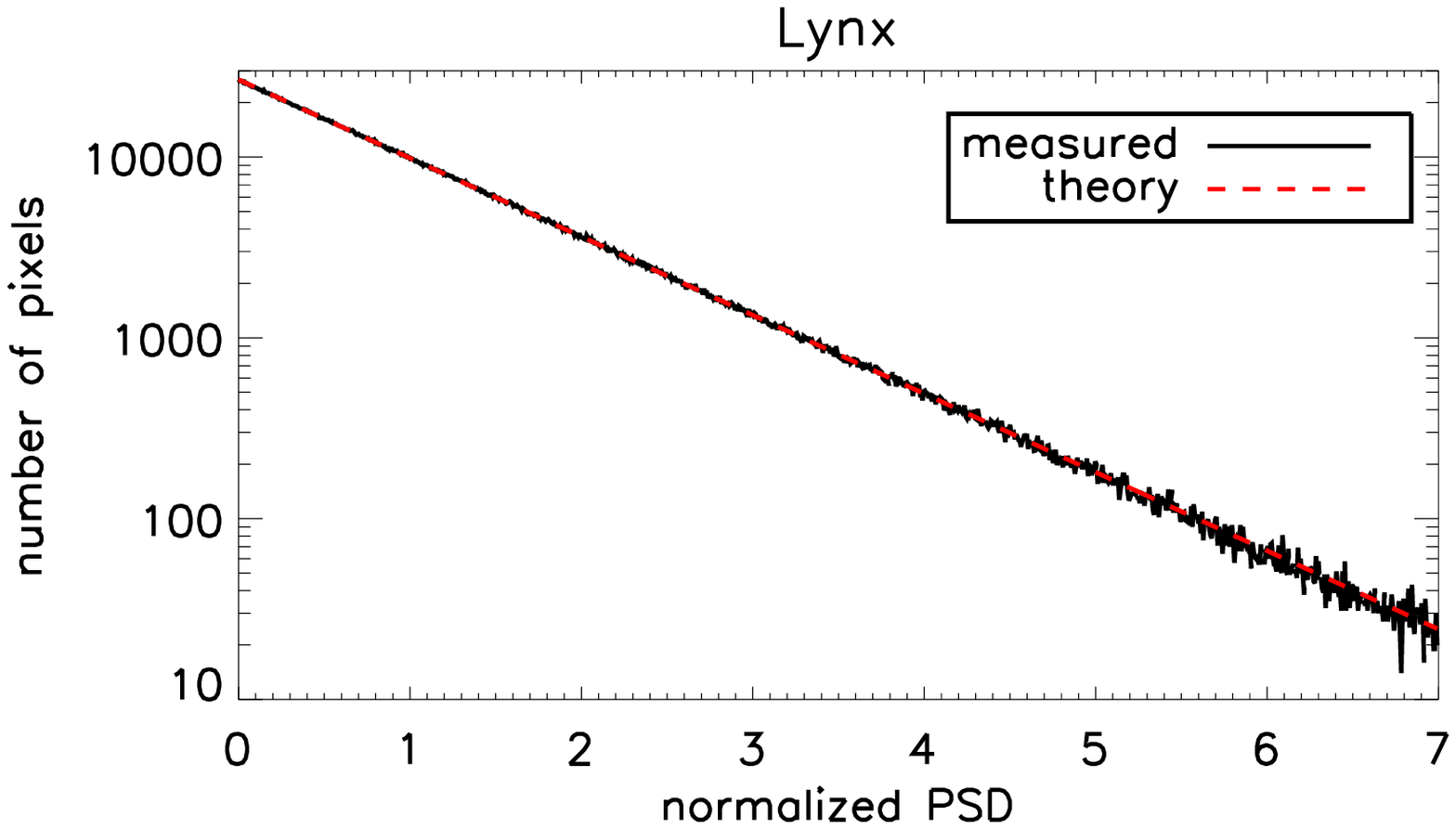}
  \plotone{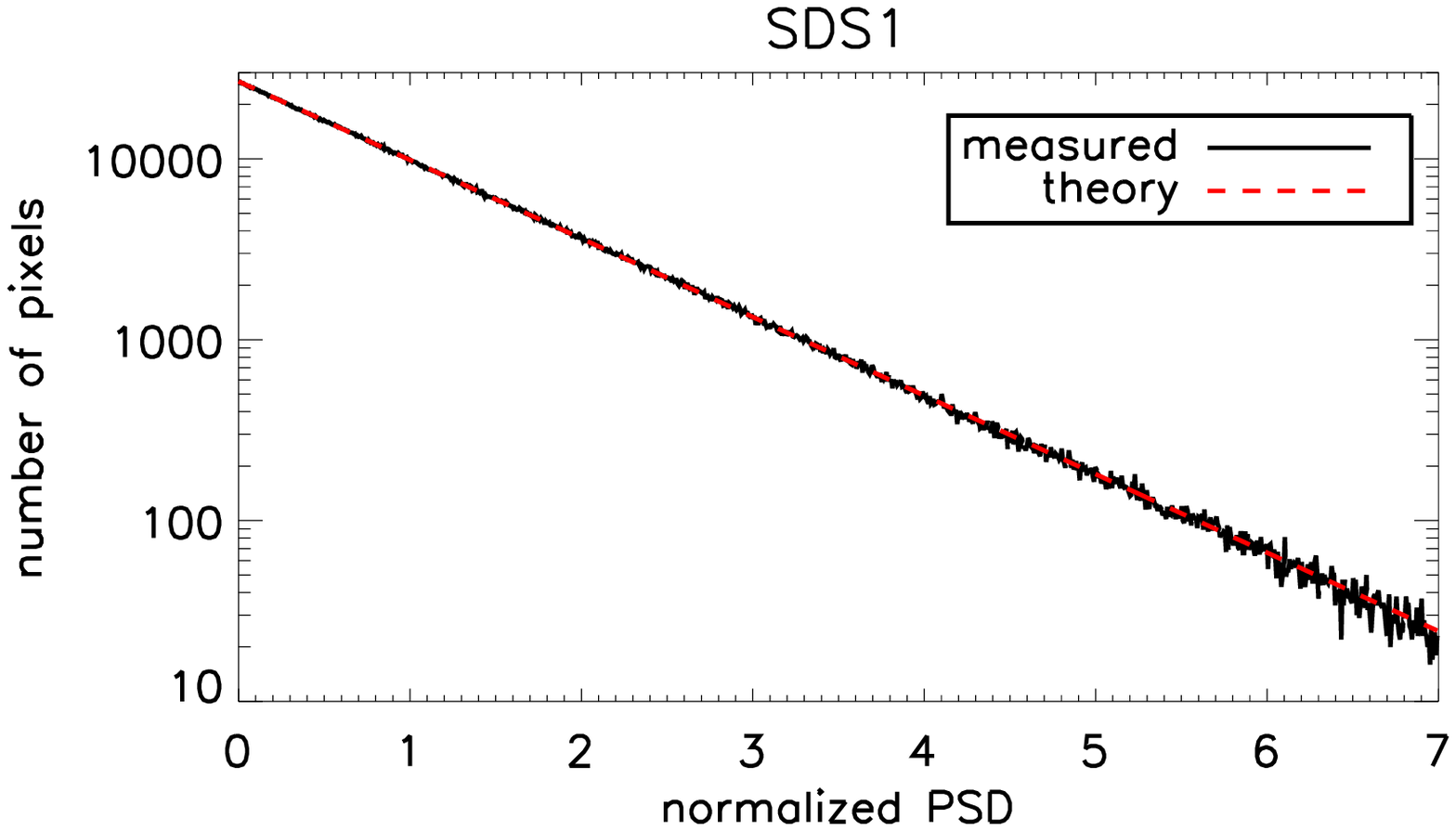}
  \caption{A comparison between the distribution of PSD values
    from the jackknifed realizations to 
    a Gaussian PDF for the data
    co-added over all observations for each
    Science field.
    See Equation~\ref{eqn:PDF_map_PSD}.
    The agreement is good, indicating
    that the underlying noise distribution is 
    approximately Gaussian.}
  \label{fig:final_PSD_PDF}
\end{figure}

\begin{figure}
  \plotone{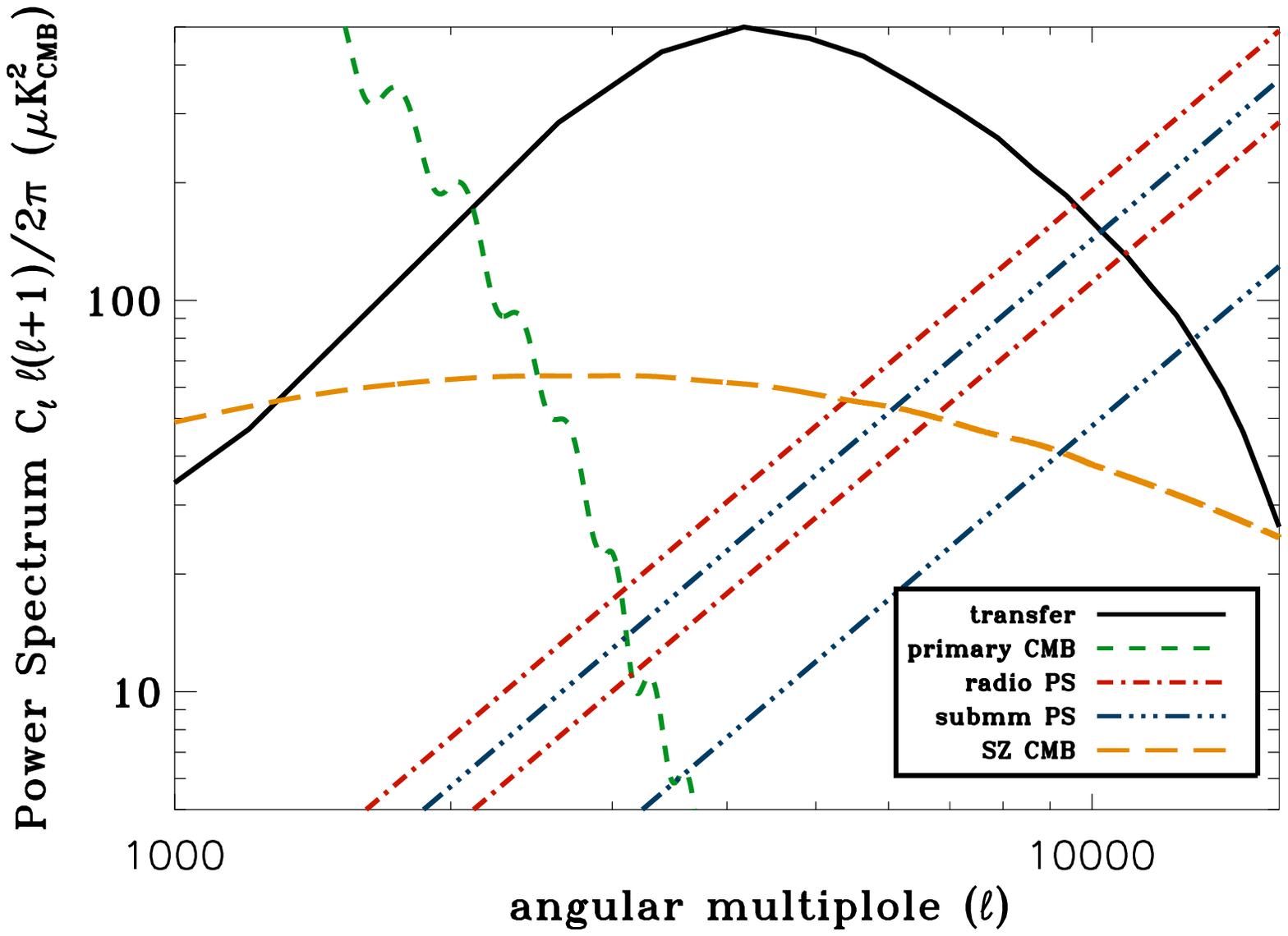}
  \caption{The power spectra from the primary CMB anisotropies 
    (short green dashes),
    high and low estimates for
    radio point sources (red dash-dot), 
    high and low estimates for submillimeter
    point sources (blue dot-dot-dot-dash), and the 
    analytically predicted
    SZE-induced CMB anisotropies from \citet{komatsu02} using the best fit
    value of $\sigma_8$ from \citet{dawson06}
    (long orange dashes).
    Note that the point-source power spectra assume 
    unclustered distributions.
    Also included as a solid black line
    is the transfer function of the final
    map of the Lynx field
    with arbitrary normalization.}
  \label{fig:astr_noise}
\end{figure}

    The anisotropy amplitude can be estimated by determining what value of $A$
    maximizes the likelihood of the measured map PSD, $X_{\vec{\nu}}$.
    Therefore, we need to determine the probability density function (PDF)
    describing $X_{\vec{\nu}}$, given $A$.
    First, note that 
    \begin{displaymath}
      {X}_{\vec{\nu}} = \left| \alpha + i \beta \right|^2,
    \end{displaymath}
    where $\alpha$ is the real part of the Fourier transform of the
    science field map and $\beta$ is the imaginary part of the 
    Fourier transform of the science field map.
    If we assume that the noise properties of the map are 
    Gaussian,\footnote{
      This is an extremely good assumption.  
      See Figure~\ref{fig:final_PSD_PDF}.
      Although the anisotropy signal may not follow a Gaussian
      distribution,
      $A S^2_{\vec{\nu}} B^2_{\vec{\nu}} T_{\vec{\nu}} \ll
      P_{\vec{\nu}}$ for a single $\nu$-space pixel, 
      so the underlying distribution function for
      $X_{\vec{\nu}}$ will still be well approximated by a Gaussian.}
    then the PDFs for $\alpha$ and $\beta$ are the same and are given by
    \begin{eqnarray}
      f(\alpha) = \frac{1}{\sqrt{2 \pi \sigma^2}} e^{-\alpha^2/2\sigma^2} 
      & {\rm and} &
      f(\beta) = \frac{1}{\sqrt{2 \pi \sigma^2}} e^{-\beta^2/2\sigma^2}, 
      \label{eqn:PDF_real_mapft}
    \end{eqnarray}
    where $\sigma^2 = \left< X_{\vec{\nu}} \right>/2$.
    Next, after a change of variables to
    $\alpha = r \cos(\theta)$ and $\beta = r \sin(\theta)$, the
    PDF in Equation~\ref{eqn:PDF_real_mapft} becomes 
    \begin{displaymath}
      f(r,\theta) = \frac{1}{2 \pi \sigma^2}
      r e^{-r^2/2\sigma^2}.
    \end{displaymath}
    Since the $\theta$ dependence of $f(r,\theta)$ is trivial, 
    we can reduce the above PDF to $f(r) = 2 \pi f(r,\theta)$,
    with
    \begin{displaymath}
      f(r) = \frac{r}{\sigma^2}e^{-r^2/2\sigma^2}.
    \end{displaymath}
    Finally, after one more change of variables
    using the relation $X_{\vec{\nu}} = r^2$, we find that the 
    PDF for $X_{\vec{\nu}}$ is equal to
    \begin{equation}
      f(X_{\vec{\nu}}) = 
      \frac{1}{2 \sigma^2} 
      e^{-X_{\vec{\nu}} / 2\sigma^2},
      \label{eqn:PDF_x}
    \end{equation}
    where the factor of $r$ has been replaced by $1/2$
    due to the change in the differential element.
    Equation~\ref{eqn:PDF_x} 
    can be written in terms of our measured parameters as
    \begin{equation}
      f(X_{\vec{\nu}}|A) = 
      \frac{1}{\mathcal{P}_{\vec{\nu}} + 
	A S^2_{\vec{\nu}} B^2_{\vec{\nu}} T_{\vec{\nu}}}
      {\rm exp} \left( \frac{-X_{\vec{\nu}}}
      {\mathcal{P}_{\vec{\nu}} + A S^2_{\vec{\nu}} B^2_{\vec{\nu}} 
        T_{\vec{\nu}}}
      \right)
      \label{eqn:PDF_X_a}
    \end{equation}
    using Equation~\ref{eqn:exp_map_var}.
    Note that we have made use of the fact that 
    $2\sigma^2 = \left< X_{\vec{\nu}} \right> = 
    \mathcal{P}_{\vec{\nu}} + A S^2_{\vec{\nu}} B^2_{\vec{\nu}} T_{\vec{\nu}}$
    to go from Equation~\ref{eqn:PDF_x} to Equation~\ref{eqn:PDF_X_a}. 

    The next step is to calculate a likelihood function,
    $\mathcal{L}$, from
    Equation~\ref{eqn:PDF_X_a} by multiplying $f(X_{\vec{\nu}}|A)$ 
    over all
    of the $\vec{\nu}$-space pixels.
    This product can be turned into a sum by taking the 
    logarithm of $\mathcal{L}$, with
    \begin{equation}
      {\rm log}(\mathcal{L}) = \sum_{\vec{\nu}} \left(
	- {\rm log}
	({\mathcal{P}_{\vec{\nu}} + A S^2_{\vec{\nu}} B^2_{\vec{\nu}} 
	T_{\vec{\nu}}})
	- \frac{X_{\vec{\nu}}}{
	  {\mathcal{P}_{\vec{\nu}} + A S^2_{\vec{\nu}} B^2_{\vec{\nu}} 
	  T_{\vec{\nu}}}} 
	\right).
	\label{eqn:log_l}
    \end{equation}
    Note that half of the $\nu$-space is discarded from the sum
    in Equation~\ref{eqn:log_l} since our maps are real 
    ($X_{\vec{\nu}} = X_{-\vec{\nu}}$).

    Then, the most probable value of the anisotropy amplitude
    for our measured map PSD can be determined 
    by maximizing ${\rm log}(\mathcal{L})$ with respect to $A$. 
    In practice, we maximize Equation~\ref{eqn:log_l} by evaluating
    ${\rm log}(\mathcal{L})$ at a range of values for $A$.
    Since the number of $\vec{\nu}$-space pixels is $\lesssim 10000$,
    the computational time required to evaluate ${\rm log}(\mathcal{L})$
    at each value of $A$ is minimal, which means that we
    can determine the best fit value of $A$ to almost 
    any desired precision using this numerical method.
    
    However, it is also instructive to analytically approximate the value
    of $A$ that maximizes Equation~\ref{eqn:log_l}.
    To start, we take the derivative of ${\rm log}(\mathcal{L})$ with
    respect to $A$, yielding
    \begin{equation}
      \left. \frac{\partial {\rm log}(\mathcal{L})}{\partial A} 
      \right|_{A = \hat{A}} = \left.
      \sum_{\vec{\nu}} \frac{\Theta_{\vec{\nu}}}
	  {(\mathcal{P}_{\vec{\nu}} + A \Theta_{\vec{\nu}})^2}
	  \left(X_{\vec{\nu}} - \mathcal{P}_{\vec{\nu}} - A \Theta_{\vec{\nu}}
	  \right) \right|_{A = \hat{A}} = 0,
	  \label{eqn:partial_l}
    \end{equation}
    where $\Theta_{\vec{\nu}} = S^2_{\vec{\nu}} B^2_{\vec{\nu}} 
    T_{\vec{\nu}}$ and
    $\hat{A}$ is the best fit value of $A$.
    For any given $\vec{\nu}$-space pixel, $\mathcal{P}_{\vec{\nu}} \gg
    A \Theta_{\vec{\nu}}$ for any physically reasonable value of $A$.
    Therefore, we can simplify
    Equation~\ref{eqn:partial_l} to
    \begin{displaymath} \left.
      \sum_{\vec{\nu}}
      \frac{\Theta_{\vec{\nu}}}{\mathcal{P}_{\vec{\nu}}^2}
      \left(1 - \frac{2 A \Theta_{\vec{\nu}}}{\mathcal{P}_{\vec{\nu}}}
      + \mathcal{O}
      \left( \frac{A^2 \Theta_{\vec{\nu}}^2}{\mathcal{P}_{\vec{\nu}}^2}
      \right) \right)
	  \left(X_{\vec{\nu}} - \mathcal{P}_{\vec{\nu}} - A \Theta_{\vec{\nu}}
	  \right) \right|_{A = \hat{A}} \simeq 0.
    \end{displaymath}
    If we rearrange some terms, and again keep only the lowest
    order terms in $A \Theta_{\vec{\nu}} / \mathcal{P}_{\vec{\nu}}$,
    then we find
    \begin{displaymath}
      \hat{A} \simeq
      \frac{\sum_{\vec{\nu}} 
	\frac{\Theta_{\vec{\nu}}^2}{\mathcal{P}^2}
	\left( \frac{X_{\vec{\nu}} - \mathcal{P}_{\vec{\nu}}}{\Theta_{\vec{\nu}}}
	\right)}
      {\sum_{\vec{\nu}}
	\frac{\Theta_{\vec{\nu}}^2}{\mathcal{P}^2}
	\left( \frac{ 2 X_{\vec{\nu}} - \mathcal{P}_{\vec{\nu}}}
	{\mathcal{P}_{\vec{\nu}}} \right) }.
    \end{displaymath}
    Finally, because
    $A \Theta_{\vec{\nu}} \ll \mathcal{P}_{\vec{\nu}}$,
    we can make the approximation that $\left< X_{\vec{\nu}} \right> \simeq 
    \mathcal{P}_{\vec{\nu}}$,
    which means that  $\left< 2 X_{\vec{\nu}} - \mathcal{P}_{\vec{\nu}} \right> 
    \simeq \mathcal{P}_{\vec{\nu}}$.
    With this approximation we find
    \begin{equation}
      \hat{A} \simeq
      \frac{\sum_{\vec{\nu}} 
	\frac{\Theta_{\vec{\nu}}^2}{\mathcal{P}^2}
	\left( \frac{X_{\vec{\nu}} - \mathcal{P}_{\vec{\nu}}}{\Theta_{\vec{\nu}}}
	\right)}
      {\sum_{\vec{\nu}}
	\frac{\Theta_{\vec{\nu}}^2}{\mathcal{P}^2}}.
      \label{eqn:est_a}
    \end{equation}      
    To understand this result, consider that
    for a single $\vec{\nu}$-space pixel the best estimate of $A$
    is $(X_{\vec{\nu}} - \mathcal{P}_{\vec{\nu}})/ \Theta_{\vec{\nu}}$.
    Therefore, Equation~\ref{eqn:est_a} determines the
    weighted mean of $A$ over all pixels, assuming that the 
    uncertainty on the value of $A$ for
    each $\vec{\nu}$-space pixel is proportional to
    $\mathcal{P}_{\vec{\nu}}/\Theta_{\vec{\nu}}$,
    which is a reasonable assumption.
    This means that the variance on $\hat{A}$ implied by  
    Equation~\ref{eqn:est_a} is proportional to
    \begin{equation}
      \sigma^2_{\hat{A}} \propto \frac{1}{\sum_{\vec{\nu}}
	\frac{\Theta_{\vec{\nu}}^2}{\mathcal{P}^2}}.
      \label{eqn:est_var_a}
    \end{equation}
      



\clearpage
\begin{deluxetable}{ccccc} 
  \tablewidth{0pt}
  \tablecaption{SZE-induced CMB anisotropy results} 
  \tablehead{\colhead{spectrum} & \colhead{flux uncertainty} &
    \colhead{68\% CL interval} & \colhead{90\% CL interval} &
    \colhead{95\% CL interval}}
  \startdata
    flat-total & 0 & $100 - 590$ $\mu$K$^2_{CMB}$ & 
    $0 - 760$ $\mu$K$^2_{CMB}$ &
    $0 - 830$ $\mu$K$^2_{CMB}$ \\ 
    flat-SZE & 0 & $90 - 580$ $\mu$K$^2_{CMB}$ & 
    $0 - 750$ $\mu$K$^2_{CMB}$ &
    $0 - 830$ $\mu$K$^2_{CMB}$ \\
    flat-SZE & 3.5\% (meas) & $90 - 630$ 
    $\mu$K$^2_{CMB}$ & $0 - 790$ $\mu$K$^2_{CMB}$ &
    $0 - 880$ $\mu$K$^2_{CMB}$ \\
    flat-SZE & 6.3\% (total) & $80 - 690$ $\mu$K$^2_{CMB}$ & 
    $0 - 960$ $\mu$K$^2_{CMB}$ &
    $0 - 1000$ $\mu$K$^2_{CMB}$ \\
    KS-SZE & 0 & $80 - 540$ $\mu$K$^2_{CMB}$ & 
    $0 - 690$ $\mu$K$^2_{CMB}$ & 
    $0 - 770$ $\mu$K$^2_{CMB}$ \\
    KS-SZE & 3.5\% (meas) & $80 - 570$ $\mu$K$^2_{CMB}$ & 
    $0 - 740$ $\mu$K$^2_{CMB}$ & 
    $0 - 830$ $\mu$K$^2_{CMB}$ \\
    KS-SZE & 6.3\% (total) & $70 - 730$ $\mu$K$^2_{CMB}$ & 
    $0 - 950$ $\mu$K$^2_{CMB}$ & 
    $0 - 990$ $\mu$K$^2_{CMB}$ \\
    KS-SZE (nG) & 6.3\% (total) & $90 - 790$ $\mu$K$^2_{CMB}$ & 
    $0 - 1060$ $\mu$K$^2_{CMB}$ & 
    $0 - 1080$ $\mu$K$^2_{CMB}$ \\
  \enddata
  \tablecomments{Confidence intervals
    for our estimates of the total and
    SZE-induced CMB anisotropy
    amplitude for both a flat SZE band power in 
    $\mathcal{C}_{\ell}$ and the SZE spectrum given
    by the analytic model of Komatsu and Seljak \citep{komatsu02}.
    The limits for the analytic model refer to the 
    average amplitude of the SZE spectrum weighted by
    our transfer function.
    The three rows for each SZE spectrum give the upper
    limits for no uncertainty in our flux calibration,
    the 3.5\% uncertainty in our flux calibration
    due to measurement error, and the
    6.3\% uncertainty in our flux calibration
    due to the combination of measurement error
    and uncertainty in the surface brightness
    of Uranus and Neptune.
    The final row gives the confidence intervals when
    the non-Gaussianity of the SZE anisotropy signal
    is accounted for.}
    \label{tab:SZE_result}
\end{deluxetable}

\clearpage
\begin{deluxetable}{ccccc} 
  \tablewidth{0pt}
  \tablecaption{Bias and efficiency of Equation~\ref{eqn:log_l}}
  \tablehead{\colhead{input $A_{sim}$} & \colhead{average $\hat{A}$} &
    \colhead{$\sigma_{\hat{A}}$} & \colhead{$\sigma_{\hat{A}}/N_{real}$} &
    \colhead{min. var. $\sigma_{\hat{A}}$}}
  \startdata
    0 $\mu$K$_{CMB}^2$ & -2 $\mu$K$_{CMB}^2$ & 365 $\mu$K$_{CMB}^2$ &
    12 $\mu$K$_{CMB}^2$ & 270 $\mu$K$_{CMB}^2$ \\
    100 $\mu$K$_{CMB}^2$ & 96 $\mu$K$_{CMB}^2$ & 366 $\mu$K$_{CMB}^2$ &
    12 $\mu$K$_{CMB}^2$ & 270 $\mu$K$_{CMB}^2$ \\
    200 $\mu$K$_{CMB}^2$ & 194 $\mu$K$_{CMB}^2$ & 367 $\mu$K$_{CMB}^2$ &
    12 $\mu$K$_{CMB}^2$ & 270 $\mu$K$_{CMB}^2$ \\
    400 $\mu$K$_{CMB}^2$ & 389 $\mu$K$_{CMB}^2$ & 371 $\mu$K$_{CMB}^2$ &
    12 $\mu$K$_{CMB}^2$ & 270 $\mu$K$_{CMB}^2$ \\
    800 $\mu$K$_{CMB}^2$ & 806 $\mu$K$_{CMB}^2$ & 380 $\mu$K$_{CMB}^2$ &
    12 $\mu$K$_{CMB}^2$ & 270 $\mu$K$_{CMB}^2$ \\
  \enddata
  \tablecomments{A comparison between the amplitude
  of a simulated power spectrum added to a
  jackknifed realization of the 
  data, $A_{sim}$, and the most likely
  amplitude determined from Equation~\ref{eqn:log_l},
  $\hat{A}$.
  In each case 1000 jackknifed realizations of the Lynx data were
  used, 
  and the table lists the average value of 
  $\hat{A}$ for these realizations along with the
  standard deviation of the values of $\hat{A}$.
  In each case the average value of $\hat{A}$ is 
  consistent with $A_{sim}$, indicating that 
  Equation~\ref{eqn:log_l} is an unbiased estimator
  of $A$.
  Additionally, 
  to determine whether Equation~\ref{eqn:log_l} is an
  efficient (minimum variance) estimator for $A$,
  we calculate the standard deviation of the estimates
  $\hat{A}$ for each input $A_{sim}$ and compare them to
  the standard deviation one would estimate using the
  Bayesian likelihood, Equation~\ref{eqn:est_var_a}.
  The latter underestimates the minimum possible standard
  deviation because the Bayesian likelihood is incorrect
  for the reasons presented in Section~\ref{sec:bayesian_dificiencies}.
  Thus, the fact that the observed standard deviation is only
  40\% larger than the Equation~\ref{eqn:est_var_a}-based
  estimate gives us confidence that our estimator
  for $A$ is reasonably efficient.}
  \label{tab:log_l}
\end{deluxetable}

\clearpage

\clearpage

\clearpage

\clearpage

\clearpage
	  
\clearpage

\clearpage

\clearpage

\clearpage

\clearpage

\clearpage

\clearpage

\clearpage

\clearpage
\begin{figure}
  \plottwo{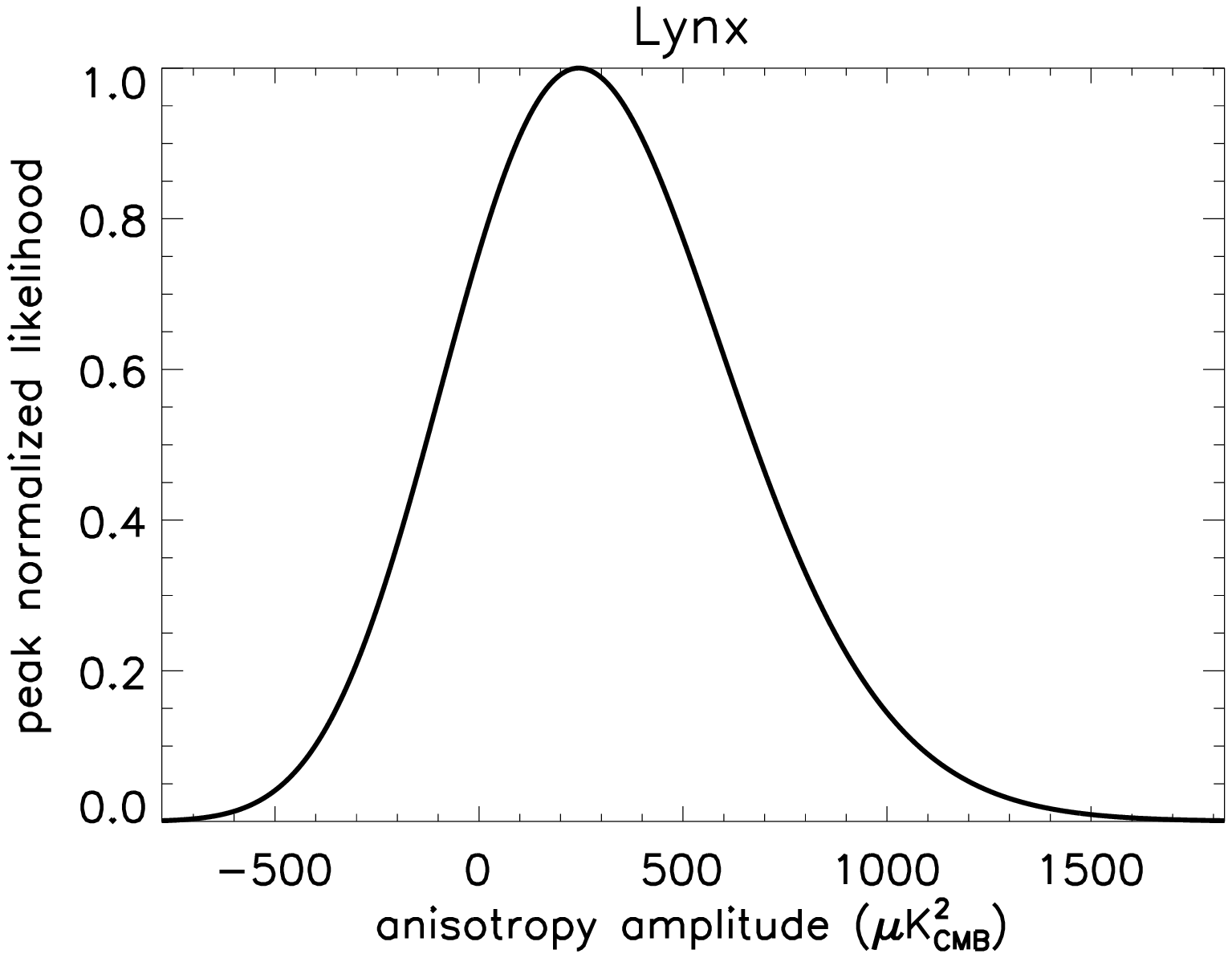}{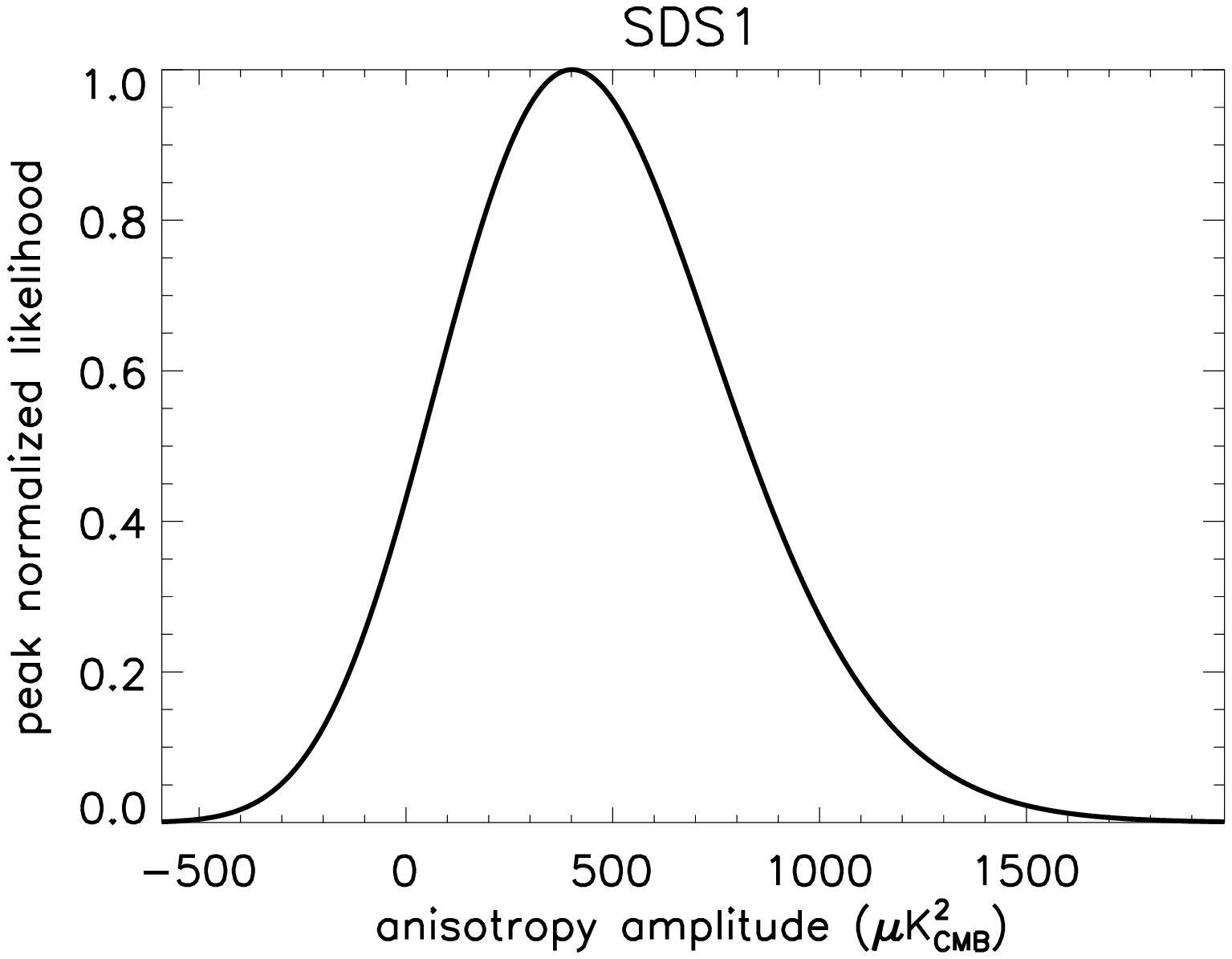}
  \caption{The Bayesian likelihood given by Equation~\ref{eqn:log_l}
    for each science field.
    The likelihoods have all been normalized to one at 
    the peak.
    These plots should only be considered as rough estimates
    for determining confidence intervals
    because the cosmic variance of the CMB
    spectra, correlations among map pixels, and the physical
    boundary that the anisotropy amplitude must be greater 
    than or equal to zero have not been fully accounted for
    in the likelihood function.}
  \label{fig:liklihood}
\end{figure}

\clearpage
\begin{figure}
  \plottwo{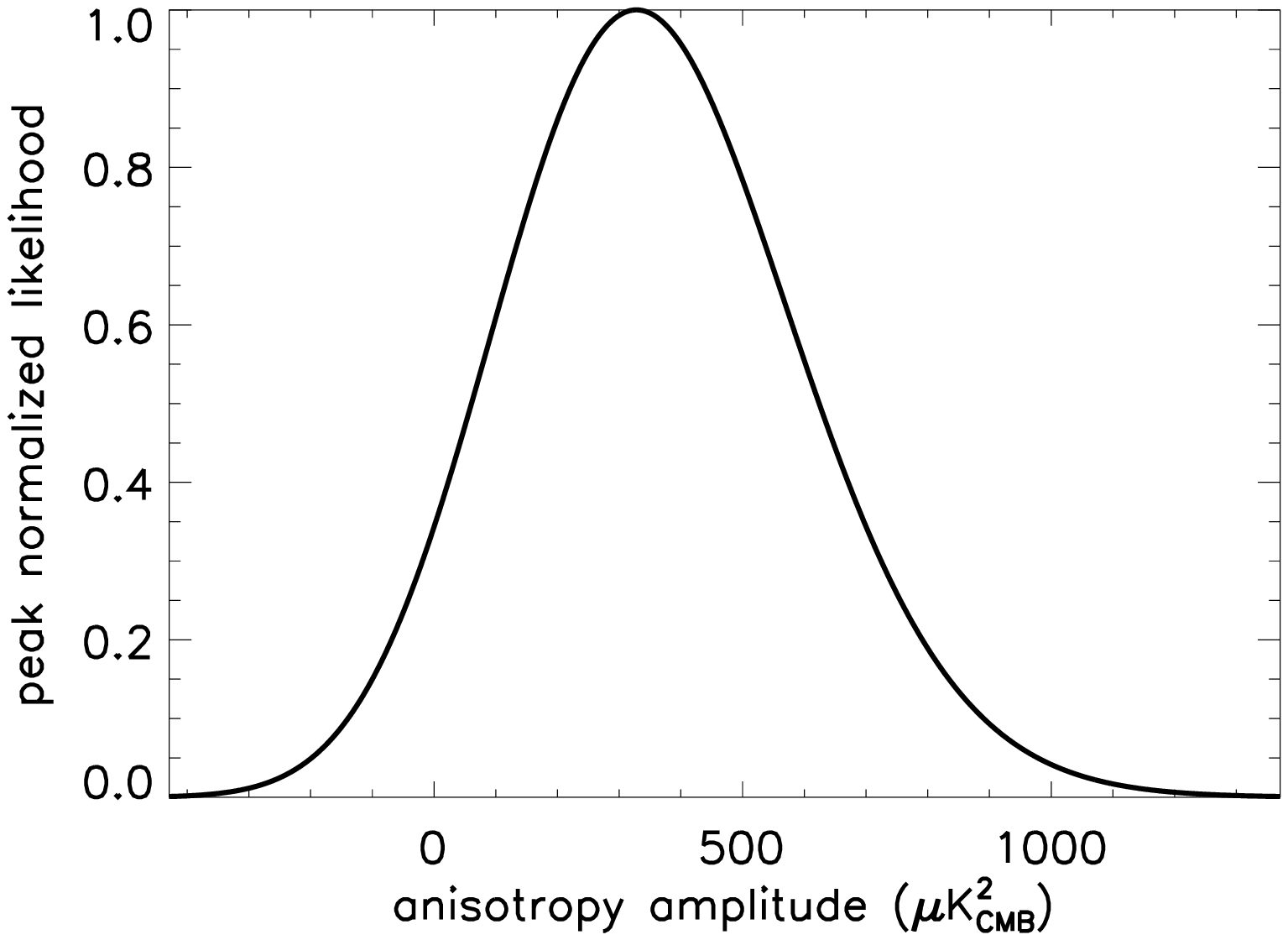}{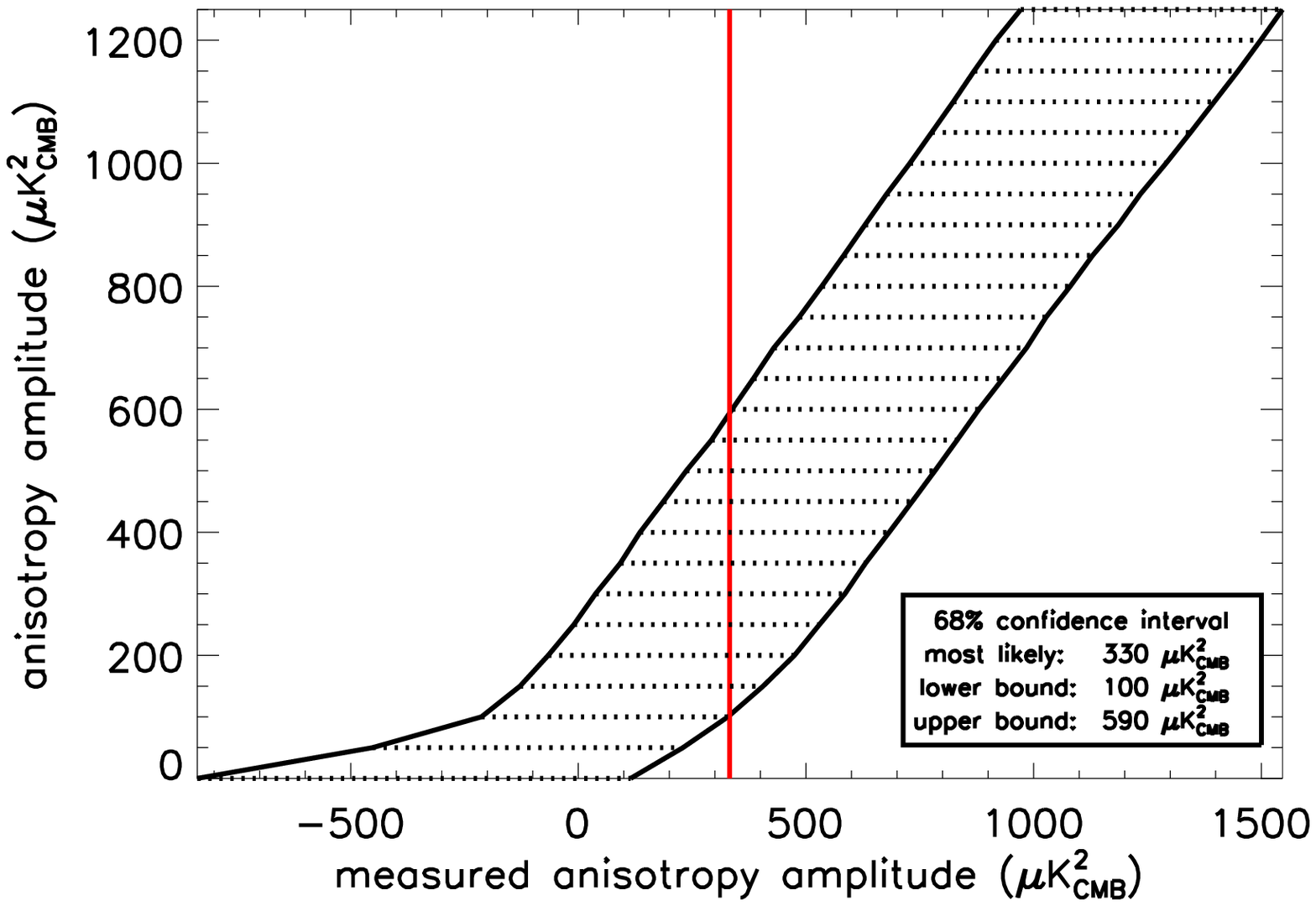}

  \plottwo{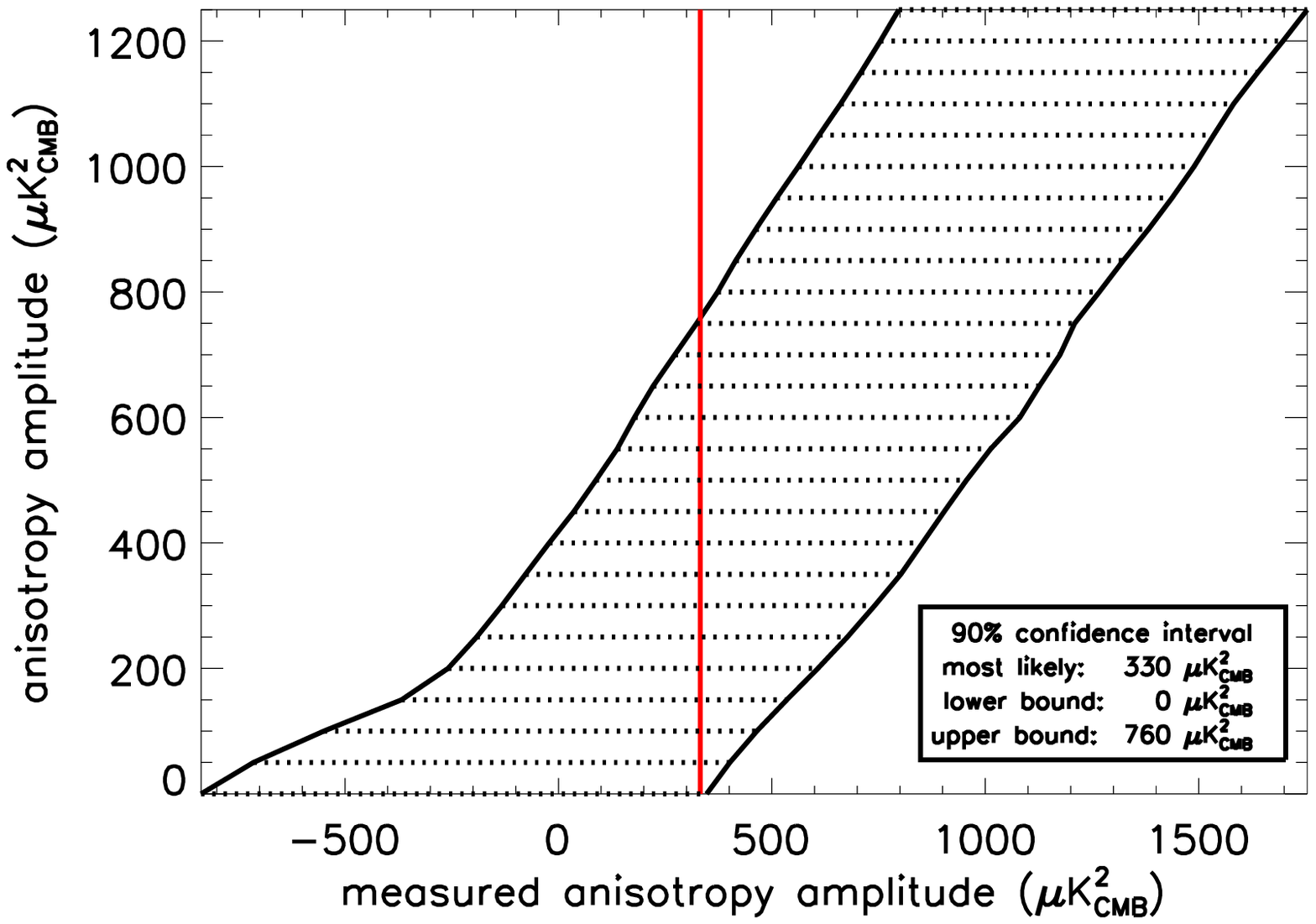}{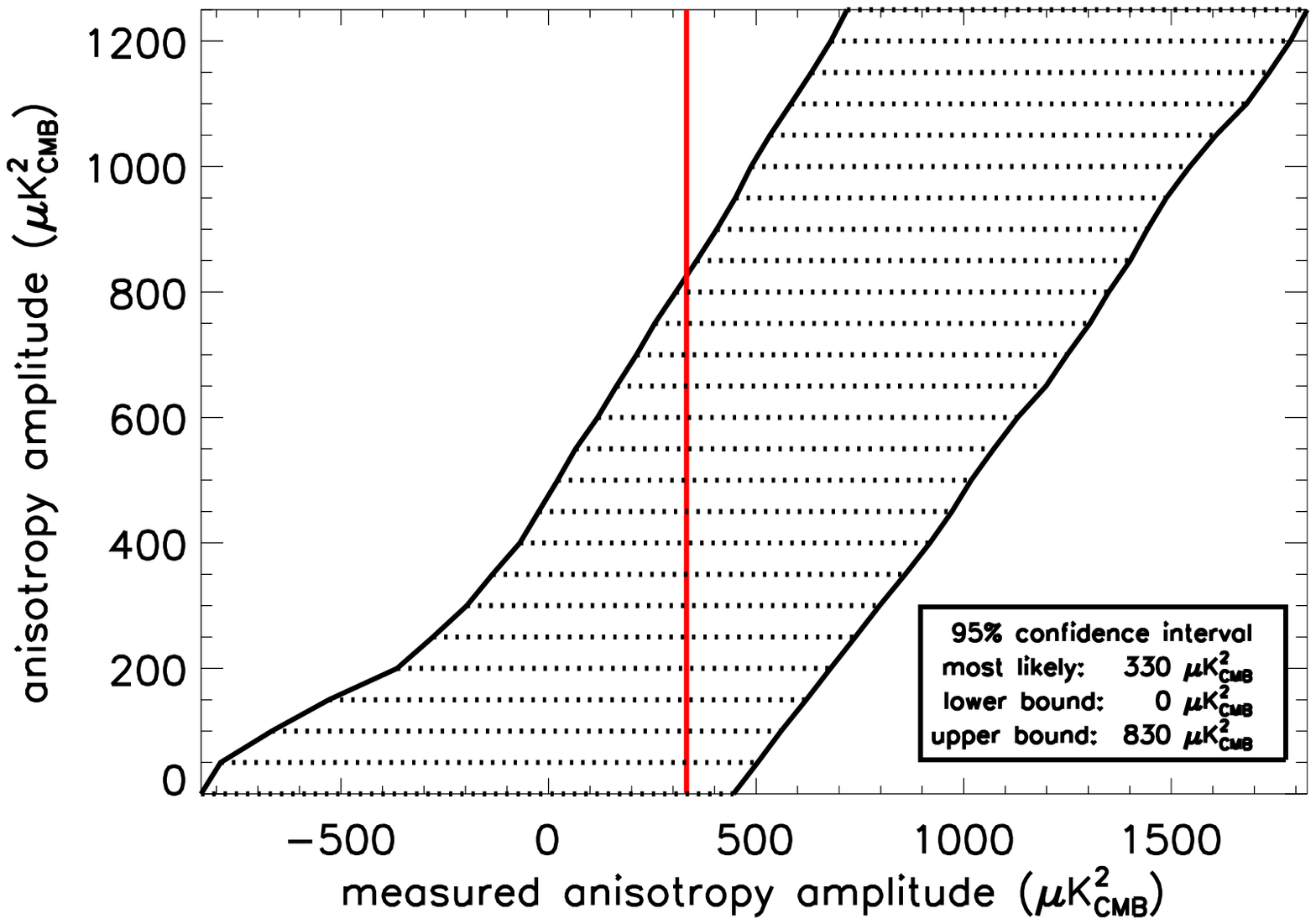}
  \caption{The first plot shows the Bayesian likelihood for 
    a range of anisotropy amplitudes for the full data
    set, which includes all of the observations
    of both science fields.
    The remaining three plots show the frequentist
    confidence belts for the full data set for confidence
    levels of 68\%, 90\%, and 95\%.}
  \label{fig:full_conf_belts}
\end{figure}

\clearpage
\begin{figure}
  \plotone{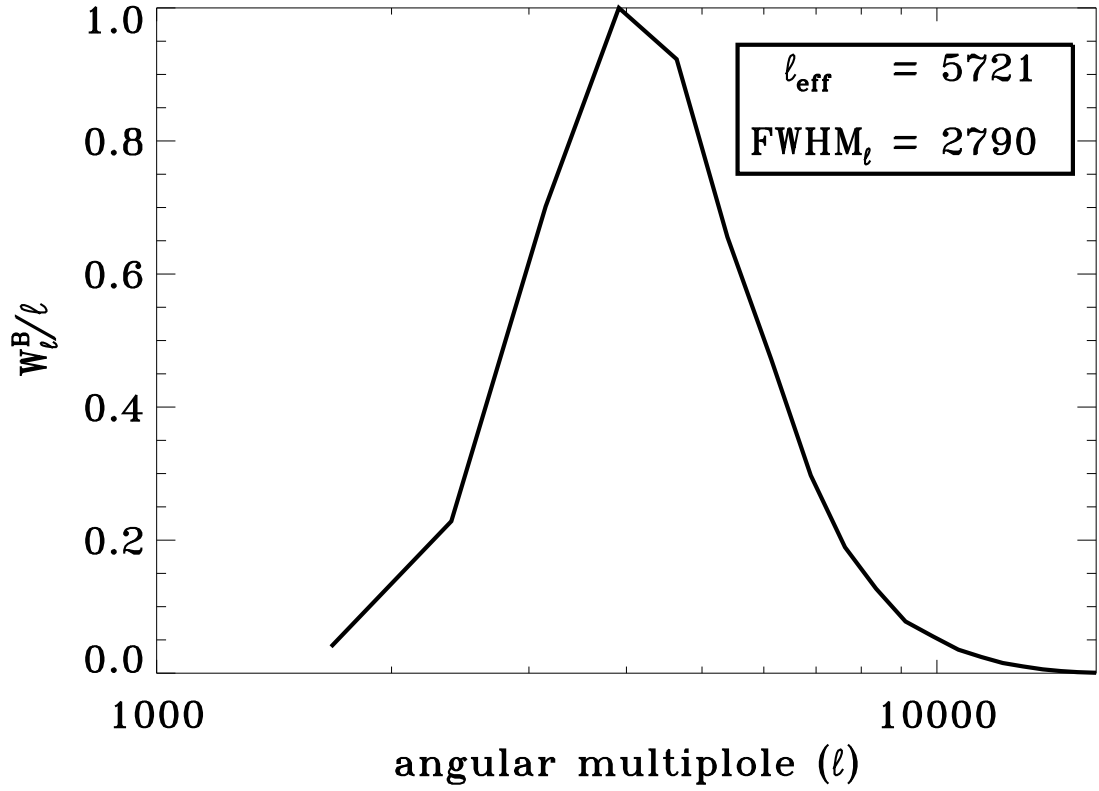}
  \caption{The band power window function, $W_{\ell}^B/\ell$, 
    for the full data set.
    We have arbitrarily peak normalized the window function.}
  \label{fig:knox}
\end{figure}

\clearpage
\begin{figure}
  \plotone{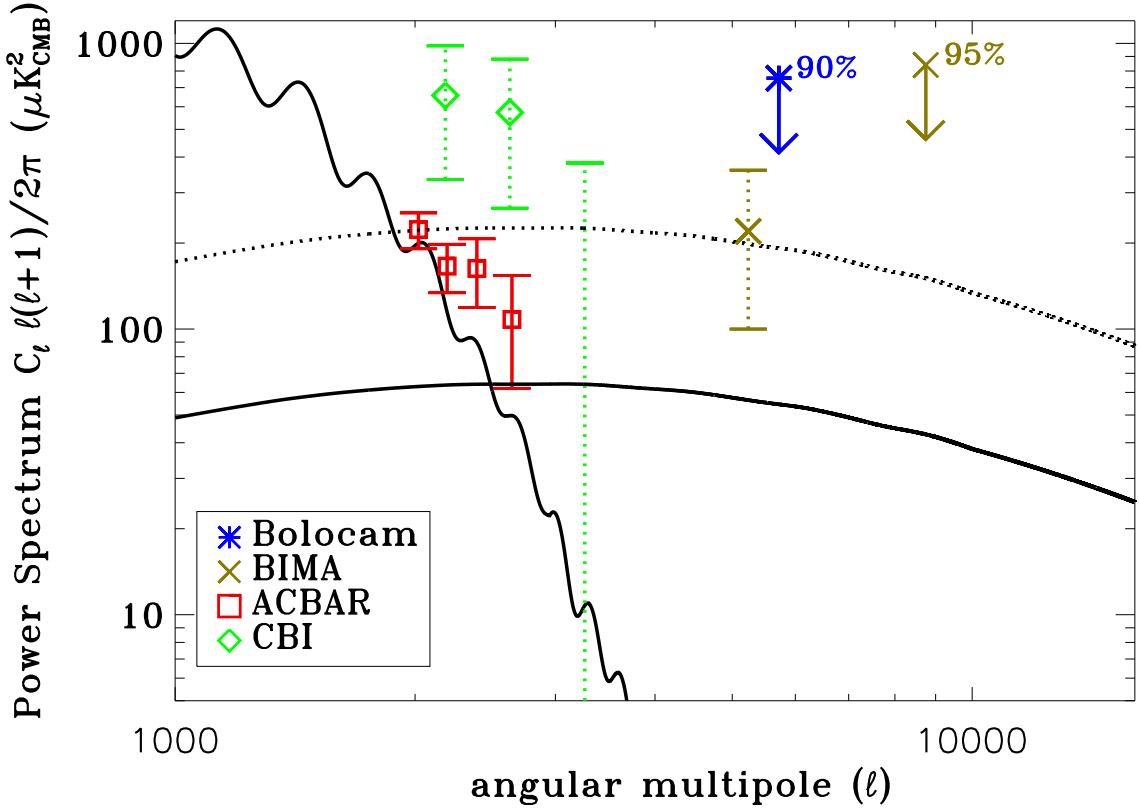}
  \caption{A plot of all of the current CMB anisotropy measurements
    above $\ell = 2000$.
    Solid lines represent observations made near 150~GHz, 
    and dashed lines represent observations made near
    30~GHz.
    The primary CMB anisotropies are
    represented by a solid black line on
    the left side of the plot, and the predicted SZE-induced CMB
    anisotropies are shown as solid (150~GHz) and
    dashed (30~GHz) black lines.
    The primary CMB anisotropies were calculated using
    the parameters given in Section~\ref{sec:astr_noise};
    the analytic model of \citet{komatsu02},
    along with the best estimate of $\sigma_8$ from
    \citet{dawson06}, were used to estimate
    the SZE-induced CMB anisotropies.
    All of the data are plotted with 1$\sigma$ error bars, 
    except for the Bolocam upper limit at $\ell=5700$ and
    the BIMA upper limit at $\ell=8748$, which are given
    as 90\% and 95\% confidence level upper limits, respectively.
    The ACBAR data were taken from \citet{reichardt08},
    the BIMA data were taken from \citet{dawson06},
    and the CBI data were taken from \citet{mason03}.}
  \label{fig:cmb_power}
\end{figure}


\end{document}